\documentclass{aa}  

\usepackage{amsmath,amssymb}
\usepackage[breaklinks=true]{hyperref}
\usepackage{threeparttable}

\usepackage[capitalize]{cleveref}

\crefname{figure}{Fig.}{Figs.}
\Crefname{figure}{Figure}{Figures}

\usepackage{graphicx}
\usepackage{xcolor}
\usepackage[normalem]{ulem}
\usepackage{listings}
\definecolor{dkgreen}{rgb}{0,0.6,0}
\definecolor{gray}{rgb}{0.5,0.5,0.5}
\definecolor{mauve}{rgb}{0.58,0,0.82}
\usepackage{txfonts}
\usepackage{tikz}
\usepackage{booktabs}
\providecommand{\kms}{\ensuremath{\rm \,km\,s^{-1}}\xspace}

\providecommand{\kmskpc}{\ensuremath{\rm \,km\,s^{-1}\,kpc^{-1}\xspace}}

\providecommand{\kms}{\ensuremath{\textrm{km\,s}^{-1}}}

\providecommand{\degree}{\ensuremath{^\circ}}

\newcommand\Gaia{\textit{Gaia}}

\newcommand{\agama}{{\sl AGAMA}}

\newcommand\gaia{\textit{Gaia}}

\newcommand\gdrtwo{\gaia~DR2}

\newcommand\gdrthree{\gaia~DR3}
\newcommand\gdrfour{\gaia~DR4}

\newcommand\apg{\texttt{APOGEE}}

\newcommand\sdss{\texttt{SDSS}}

\newcommand\cmetal{\texttt{C-MetaLL}}

\newcommand\veloce{\texttt{VELOCE}}
\newcommand\ogle{\texttt{OGLE}}
\newcommand\asasn{\texttt{ASAS-SN}}

\newcommand\ecc{\texttt{ecc}}

\newcommand{\gunlim}{\textit{GaiaUnlimited}}

\newcommand{\twomass}{\textit{2MASS}}

\newcommand{\wise}{\textit{WISE}}

\newcommand{\gaussian}{\textit{Gaussian}}

\newcommand{\hvs}{\textit{HVS}}

\newcommand{\feh}{\ensuremath{[\mathrm{Fe/H}]}}

\newcommand{\ruwe}{$\text{RUWE}$}

\newcommand{\rgal}{\ensuremath{R}}

\newcommand{\xgc}{\ensuremath{X}}
\newcommand{\ygc}{\ensuremath{Y}}
\newcommand{\zgc}{\ensuremath{Z}}

\newcommand{\vrgc}{\ensuremath{V_{\text{R}}}}

\newcommand{\vxgc}{\ensuremath{V_{\text{X}}}}
\newcommand{\vygc}{\ensuremath{V_{\text{y}}}}
\newcommand{\vzgc}{\ensuremath{V_{\text{Z}}}}

\newcommand{\vgal}{\ensuremath{V_{\text{Gal}}}}
\newcommand{\vej}{\ensuremath{\Delta V_{\text{ej}}}}

\newcommand{\dcep}{{DCEPs}}
\newcommand{\dcepsing}{{DCEP}}

\newcommand{\gc}{{GC}}
\newcommand{\typetwo}{{T2Cs}}
\newcommand{\dtypetwo}{{$d_{\rm T2C}$}}
\newcommand{\dglob}{{$d_{\rm glob}$}}
\newcommand{\blher}{{BLHer}}
\newcommand{\wvir}{{WVir}}
\newcommand{\rvtau}{{RVtau}}

\newcommand{\tbirth}{{$\tau_{\rm Ceph}$}}
\newcommand{\tdyn}{{$\tau_{\rm dyn}$}}
\newcommand{\tclus}{{$\tau_{\rm clus}$}}

\newcommand{\gamfunc}{{$\gamma_{\rm orb}$}}

\newcommand{\pawel}{{P21}}
\newcommand{\rogue}{{`rogue'}}

\newcommand{\skowceph}{{S25}}

\newcommand{\greatwavepoggio}{{P25}}

\newcommand{\cruzreyes}{{CR25}}

\newcommand{\vtot}{\ensuremath{V_{\text{tot}}}}
\newcommand{\vlos}{{$V_{\rm los}$}}

\newcommand{\orbinc}{{$orb\_inc$}}

\newcommand{\lperpwarp}{{$\sqrt{\Delta L_{x}^{2} + \Delta L_{y}^{2} }$}}

\newcommand{\enunits}{\ensuremath{10^{5} \, \mathrm{(km/s)}^2}}
\newcommand{\angunits}{\ensuremath{10^3 \, \text{kpc km s}^{-1}}}

\newcommand{\angun}{\ensuremath{L_{\text{0}}}}
\newcommand{\enun}{\ensuremath{E_{\text{0}}}}

\newcommand{\lperp}{\ensuremath{L_{\mathrm{\perp}}}}

\newcommand{\lx}{\ensuremath{L_{\text{X}}}}
\newcommand{\ly}{\ensuremath{L_{\text{Y}}}}
\newcommand{\lz}{\ensuremath{L_{\text{Z}}}}

\newcommand{\energy}{\ensuremath{E}}
\newcommand{\jacobi}{\ensuremath{H_{\text{j}}}}

\usepackage{graphicx}
\usepackage{txfonts}

\usepackage{hyperref}
\newcommand{\orcit}[1]{\protect\href{https://orcid.org/#1}{\protect\includegraphics[width=8pt]{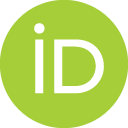}}}

\makeatletter
\renewcommand*\maketitle{%
  \thispagestyle{firstpage}
\begingroup
    \if@wideboxfn
    \setlength\bibindent{1.4\parindent}
    \else
    \setlength\bibindent{\parindent}
    \fi
    \renewcommand*\thefootnote{\@fnsymbol\c@footnote}%
    \renewcommand\@makefntext[1]{%
    \ifaa@longfn\hsize\textwidth\fi
    \noindent
    \hb@xt@\bibindent{\hss\@makefnmark\enspace}##1}
  \ifaa@twocolumn
  \begingroup
    \begin{aa@strip}
          \aa@maketitle
    \end{aa@strip}
    \@thanks	  	
  \endgroup
  \else
    \begingroup
      \let\thanks\footnote
      \aa@maketitle
    \endgroup
  \fi
\endgroup
  \setcounter{footnote}{0}%
}
\makeatother

\DeclareUnicodeCharacter{2212}{-}

\begin{document} 

\title{Rogue Ones \\}
\subtitle{Orbital census of Galactic Cepheids and their Anomalies}

\titlerunning{MW Cepheid orbits}

\author{ 
Shourya Khanna\inst{1} \and
Ronald Drimmel\inst{1} \and
Eloisa Poggio\inst{1} \and
Dorota Skowron \inst{2} \and
Jie Yu\inst{3,4,5}
}
\institute{INAF - Osservatorio Astrofisico di Torino, via Osservatorio 20, 10025 Pino Torinese (TO), Italy\\    \email{shourya.khanna@inaf.it}  
\and 
Astronomical Observatory, University of Warsaw, Al. Ujazdowskie 4, 00-478 Warsaw, Poland
 \and  
  School of Astronomy and Space Science, Nanjing University, Nanjing 210023, People's Republic of China.
 \and  
 Key Laboratory of Modern Astronomy and Astrophysics, Ministry of Education, Nanjing 210023, People's Republic of China.
 \and
 Research School of Astronomy \& Astrophysics, Australian National University, Cotter Rd., Weston, ACT 2611, Australia
}         
       \date{Received ; accepted }

  \abstract{Classical Cepheids (DCEPs) are excellent standard candles expected to trace the spatial and kinematic distribution of the Galaxy's young and dynamically cold stellar disc. Using the most precise mid-infrared DCEP distances to date combined with \gdrthree{} astrometry and line-of-sight velocities, we perform a comprehensive 6D dynamical census of the Milky Way's DCEP population. While the vast majority exhibit the expected disc-like kinematics, we identify 18 kinematically anomalous Cepheids. These `rogue' stars reside on highly inclined orbits (three at $>$70$^{\circ}$), including two in retrograde motion and one with a total velocity of $\sim$480 \kms. Despite their extreme trajectories, their optical light curves are consistent with \dcepsing{} classifications. We explore whether these anomalies originate from possible classification systematics or physical processes and find only three of our sources are likely misclassified. Assuming a runaway scenario we derive dynamical ages for the kinematic anomalies, which we find highly consistent with their Cepheid ages. Spectroscopic follow-up would be insightful as one source in particular is exceptionally metal poor (\feh $\sim$-1.6 dex), which is highly atypical for a \dcepsing{}. Integrating photometric classification with 6D kinematics will help fully characterise the Galaxy's variable star populations.}


   \maketitle

\section{Introduction} 

With their high intrinsic luminosity and well defined Period-Luminosity Relations (PLR) \citep{Leavitt1908}, Classical Cepheids (\dcep) allow us to map the Milky Way's three-dimensional structure and kinematics with high precision out to large distances. Detailed studies of the Galaxy's spiral \citep{drimmel2023gaia,drimmel2025ceph}, warped \citep{Skowron_2019warp_science,Skowron:2019,CabreraGadea:2024}, and perturbed nature \citep[][P25 hereafter]{greatwavepoggio} have been enabled thanks to \dcepsing{} distances. Moreover, they're excellent tracers of abundance gradients in the young Galactic disc \citep[e.g.][]{Genovali:2014,Nunnari:2026}.

Traditionally, \dcep{} are classified by their characteristic `saw-tooth' optical lightcurves \citep{Skowron_2019warp_science,2024A&ARv..32....4B}. A vetted list\footnote{\href{https://www.astrouw.edu.pl/ogle/ogle4/OCVS/allGalCep.listID}{\tiny https://www.astrouw.edu.pl/ogle/ogle4/OCVS/allGalCep.listID}} of bona fide Milky Way \dcep{} is maintained by \citet[][\pawel{} hereafter]{p21ceph}; with their most recent update in March 2026 listing a total of 3640 Galactic \dcep{}. Using a mid-infrared PLR, \citet[][\skowceph{} hereafter]{skowron2025} estimated distances to 3425 Milky Way \dcep{} with distance uncertainties on the order of 6\%.

Because \dcep{} are inherently young stars---typically with ages \tbirth{} $<250$ Myr---they serve as excellent tracers of the Galaxy's recent star-formation history. As such, they are expected to reside within the dynamically cold, thin stellar disc, exhibiting near-circular orbits with minimal vertical excursions from the Galactic mid-plane. However, unlike for older populations of variable stars such as the RR-Lyrae \citep{orazi_rr24}, a comprehensive census of \dcep{} orbits has not yet been performed. 

Therefore, in this paper we conduct the first dynamical census to our knowledge for Galactic \dcep{} by combining their highly precise distances from S25 with proper-motion and line-of-sight velocities (\vlos) from \gdrthree{} \citep{gdr3release}. We consider their orbits and distribution in angular momentum space. In particular, we identify 18 kinematically anomalous `rogue' Cepheids that exhibit highly inclined orbits uncharacteristic of a dynamically cold thin-disc population. We investigate whether the origin of these peculiar objects is systematic (such as due to misclassification) or if their peculiar trajectories result from physical interactions with Galactic Globular clusters or if these are runaway stars \citep{2024Natur.634..809S,Valli:2025}.

The paper is organised as follows: In Sec.~\ref{sec:data} we describe how we construct the phase-space for the various datasets (\dcep{}, \typetwo{}, \gc{} etc) used in this study. In Sec.~\ref{sec:kincensus} we perform the orbital census of Milky Way Cepheids and identify outliers. Sec. ~\ref{sec:sytematics_} considers possible systematics in the data. Finally, in Sec. ~\ref{sec:origin_rogue} we explore physical mechanisms that could explain some of the kinematically anomalous behaviour.

\section{Observational data and phase-space}
\label{sec:data}
The sky distribution ($l,b$) for the various datasets used in this work is shown in the top panel of \autoref{fig:lbplot}. Below we summarise the selection criteria for each dataset, how their distances were estimated and how their 6D phase-space was constructed.


\begin{figure*}
  \includegraphics[width=2.\columnwidth]{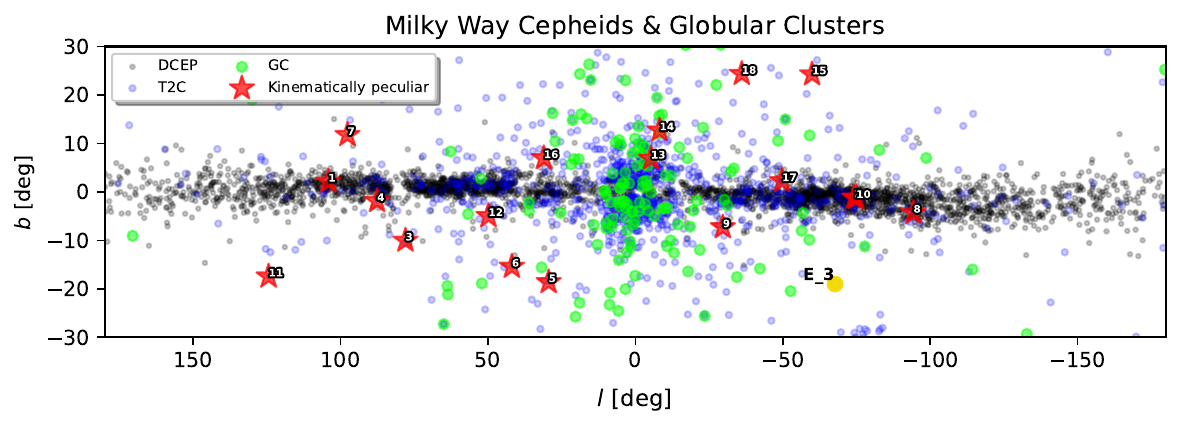}
\caption{Sky distribution of Milky Way Cepheids in Galactic coordinates, with all \dcep{} (black), \typetwo{} (blue), and Galactic globular clusters (lime) shown together. The kinematically anomalous \dcep{} are shown in red and are generally found at $|b|>4^\circ$. The Globular Cluster of interest (E\_3) is indicated in yellow and discussed later in the text.} \label{fig:lbplot}
\end{figure*}

\begin{figure}
\sidecaption
\includegraphics[width=1.\columnwidth]{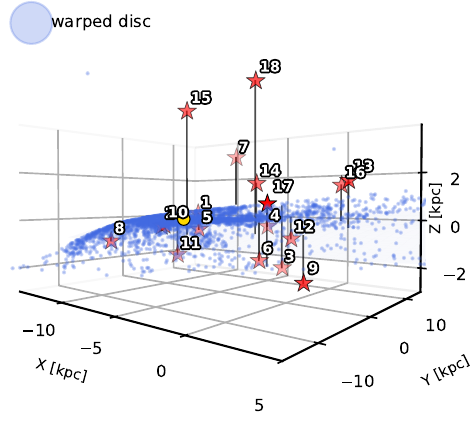}
\caption{Spatial distribution of the peculiar Cepheids (red) shown against the 3D warp model from \greatwavepoggio{}. The vertical lines illustrate the difference between the predicted and the measured positions. The Sun is indicated by the yellow dot.
} \label{fig:lbplot}
\end{figure}

\begin{table*}[ht]
\caption{Kinematically peculiar (\rogue{}) Classical Cepheids in the Milky Way.} 
\label{table:cat}
\centering
\footnotesize 
\setlength{\tabcolsep}{3.5pt} 
\renewcommand{\arraystretch}{1.15}  
\begin{tabular}{r l c c c c r@{\,$\pm$\,}l c r@{\,$\pm$\,}l r@{\,$\pm$\,}l c c c r@{\,$\pm$\,}l c}
\toprule
\# & Survey Source ID & $P$ & $G$ & $Q$ & $Q2$ & \multicolumn{2}{c}{\orbinc} & \ecc & \multicolumn{2}{c}{\vtot} & \multicolumn{2}{c}{\vej} & $D_{S25}$ & $D_{\text{GDR3}}$ & RUWE & \multicolumn{2}{c}{$\tau_{\text{dyn}}$} & $\tau_{\text{Ceph}}$ \\
 & & [d] & [mag] & [mag] & & \multicolumn{2}{c}{[deg]} & & \multicolumn{2}{c}{[\kms]} & \multicolumn{2}{c}{[\kms]} & [kpc] & [kpc] & & \multicolumn{2}{c}{[Myr]} & [Myr] \\
\midrule
1  & J221546.39+591307.3 & 2.65$^{*}$ & 13.6 & 0.83 & 8.80 & 14.1 & 0.3              & 0.24 & 216 & 3              & 110 & 16             & 5.3  & --   & 1.05          & 122 & 5              & 150 \\
2  & OGLE-GD-CEP-1669    & 2.83$^{*}$ & 15.8 & 2.12 & 2.55 & 14.7 & 10.5$^{\dagger}$ & 0.38 & 234 & 35$^{\dagger}$ & 87  & 59             & 7.4  & 8.1  & \textbf{12.61}& 136 & 37             & 132 \\
3  & J210816.15+324443.4 & 2.46       & 14.1 & 1.50 & 3.96 & 15.1 & 0.4              & 0.42 & 165 & 6              & 104 & 28             & 14.7 & 15.8 & 1.06          & 230 & 23             & 228 \\
4  & J211033.19+451848.5 & 2.97$^{*}$ & 14.9 & 2.83 & 5.44 & 17.0 & 0.6              & 0.55 & 160 & 4              & 151 & 5              & 16.7 & 16.0 & 1.08          & 109 & 5              & 190 \\
5  & V1162 Aql           & 5.38       & 7.6  & 0.80 & 18.63& 17.7 & 0.9              & 0.17 & 222 & 2              & 156 & 9              & 1.3  & 1.3  & 0.95          & 121 & 2              & 102 \\
6  & V572 Aql            & 3.77       & 11.0 & 1.57 & 9.27 & 19.4 & 0.5              & 0.27 & 214 & 2              & 118 & 3              & 6.6  & 7.6  & 1.13          & 84  & 2              & 90 \\
7  & J204329.89+614522.8 & 2.10$^{*}$ & 14.3 & 1.07 & 4.12 & 20.9 & 0.8              & 0.47 & 183 & 6              & 181 & 20             & 15.9 & 15.7 & 1.07          & 181 & 13             & 258 \\
8  & AM Vel              & 7.52       & 12.8 & \textbf{0.37} & \textbf{0.01} & 21.5 & 19.0$^{\dagger}$ & 0.57 & 252 & 81$^{\dagger}$ & 135 & 85 & 10.1 & 13.6 & \textbf{41.08}& 16  & 80             & 103 \\
9  & AB Ara              & 5.96       & 12.8 & 0.91 & 4.99 & 25.8 & 0.4              & 0.63 & 301 & 14             & 187 & 37             & 13.3 & 14.2 & 0.91          & 139 & 30             & 96 \\
10 & OGLE-GD-CEP-0507    & 3.69       & 14.6 & 0.78 & 1.14 & 28.4 & 10.0$^{\dagger}$ & 0.47 & 279 & 31$^{\dagger}$ & 124 & 24             & 6.3  & --   & \textbf{26.32}& 137 & 20             & 135 \\
11 & DQ And              & 3.20       & 11.6 & 0.60 & 5.96 & 28.9 & 1.0              & 0.76 & 123 & 2              & 186 & 9              & 6.3  & 6.1  & 1.04          & 173 & 3              & 173 \\
12 & MQ Aql              & 1.48       & 13.8 & 0.41 & 2.88 & 35.1 & 0.9$^{\dagger}$ & 0.97 & \textbf{489} & \textbf{26}$^{\dagger}$ & \multicolumn{2}{c}{--} & 9.6 & 11.3 & \textbf{1.41} & \multicolumn{2}{c}{--} & 215 \\
13 & OGLE-BLG-CEP-171    & 1.67       & 14.5 & 0.80 & 3.11 & 39.9 & 1.7              & 0.79 & 266 & 21             & 232 & 50             & 12.5 & 14.9 & 1.01          & 171 & 22             & 169 \\
14 & V1287 Sco           & 1.96       & 13.1 & 1.15 & 3.93 & 43.5 & 11.3             & 0.99 & 218 & 3              & 331 & 74             & 6.8  & --   & 0.89          & 153 & 5              & 154 \\
15 & V1253 Cen           & 4.32       & 12.1 & 0.70 & 4.10 & 48.7 & 0.7              & 0.32 & 191 & 5              & 220 & 6              & 9.8  & 11.2 & 1.20          & 127 & 9              & 122 \\
16 & DG Ser (\textbf{retro}) & 21.02  & 13.3 & 1.88 & 3.79 & 70.7 & 4.5$^{\dagger}$  & 0.94 & 271 & 4$^{\dagger}$  & 431 & 4              & 11.9 & --   & \textbf{1.55} & 36  & 4              & 46 \\
17 & OGLE-GD-CEP-0955    & 5.47$^{*}$ & 15.6 & \textbf{2.28} & \textbf{0.19} & 77.9 & 1.0 & 0.83 & 316 & 10             & \multicolumn{2}{c}{--} & 19.4 & --   & 0.99          & \multicolumn{2}{c}{--} & 109 \\
18 & V675 Cen (\textbf{retro}) & 4.63  & 12.4 & 0.34 & 4.62 & 89.1 & 0.7              & 0.38 & 206 & 3              & 304 & 10             & 11.6 & 12.8 & 1.28          & 122 & 7              & 106 \\
\bottomrule
\end{tabular}
\tablefoot{
List of \dcep{} satisfying $\gamma_{\rm orb} = |\frac{L_{Z}}{L_{\rm tot}}| < P1$, ranked by orbital inclination. $P$ is the fundamentalised pulsation period, and the pulsators in 1O mode are indicated by an asterisk. \vtot{} is in bold for potential high-velocity candidates. $\dagger$ indicates sources for which proper motion uncertainties were inflated by $\min(\text{RUWE}, 4.0)$. The \gdrthree{} source identifiers are provided in \autoref{tab:ruwe_summary}. The dynamical ages (\tdyn{}) and ejection velocities (\vej{}) are omitted for two cases where either there is no disc crossing or the \tdyn{} is unrealistic (\tdyn{} $>500$Myr).
}
\end{table*}

\subsection{Kinematic tracers}
\subsubsection{Classical Cepheids (\dcep{})}
Our primary sample is based on the 3425 \dcep{} for which \skowceph{} derived distances (D\_S25) using mid-infrared \wise{} photometry. We apply their recommended quality flag ($Q<5$) to remove sources with potentially unreliable distance estimates. Furthermore, we crossmatched this list with the latest version of the \pawel{} catalogue, excluding any sources that have since been removed from the list of bona fide \dcep{}. This vetting results in a final sample of 3343 \dcep{} (\autoref{fig:lbplot}, black points), the vast majority of which are, as expected, confined to the Galactic mid-plane ($|b|<4^\circ$). To propagate distance uncertainties into our kinematic analysis, we adopt the 
S25 catalogue distance modulus ($\mu$) and its associated uncertainty, modeling it as a Gaussian distribution $\mathcal{N}(\texttt{mu-av}, \texttt{e\_mu-av})$. The heliocentric distance in kpc is then given by,
\begin{equation}
\label{eqn:dmod2dist}
 d = 10^{0.2\mu -2}. 
\end{equation} By drawing multiple Monte Carlo realizations from this distance modulus distribution for each DCEP, we compute the uncertainties in their full 6D phase-space coordinates and orbital parameters.

\subsubsection{Old kinematic benchmarks: \typetwo{} \& Globular Clusters }
\label{sec:typ2dist}
To establish a clear kinematic benchmark for older stellar populations ($\geq10$ Gyr) distinct from the younger \dcep{}, we compile a comparison sample of Type II Cepheids (\typetwo{}) and Galactic Globular Clusters.

\emph{Globular Clusters:} We adopt the mean parallaxes, distances, proper motions, and uncertainties in each, from the \gaia{} based catalogue of about 170 \gc{} in the Milky Way of \citet{Baumgardt:2021} and \citet{Vasiliev:2021}, which we will collectively refer to as BV21 hereafter. \autoref{fig:lbplot} shows that the GCs (lime dots) are mostly distributed about the Galactic center away from the disc, extending out to $|b|\sim$30$^\circ$ as would be expected of a generally older halo-like population. 

\emph{Type II Cepheids (\typetwo{}):} From the \gdrthree{} variable star catalogue \citep{Ripepi:2023}, we select all sources with the classification “T2CEP" and retain fundamental-mode pulsators (“p1\_o is null"). \typetwo{} follow a very tight PLR, and mostly pulsate in the fundamental mode \citep{Soszynski:2008anom,Groenewegen:2017}. Restricting to the fundamental mode allows us to minimise the uncertainty in their distance estimates. \typetwo{} consist of three main subclasses each in a different stage of stellar evolution that can broadly be separated by their period-range; we restrict the sample to the BL Herculis (\blher{}, period < 5 days) and W Virginis (\wvir{}, period < 20 days) subclasses ignoring the \rvtau{} type (period > 20 days) which do not follow a well defined PLR \citep{Sicignano_t2c_MC:2024}. The final sample of 1194 \typetwo{} is shown as blue points in \autoref{fig:lbplot} which are heavily concentrated within 10 degrees of the Galactic center.

The absolute magnitude for the \typetwo{} in the \twomass{} $K_{s}$ band \citep{twomass}, is obtained using the 
near-infrared PLR relations derived by \citet[][W21 hereafter]{wielg_typ2}, to minimise the effects of extinction.
\begin{equation}
    M_{K_{s}} = \alpha (log_{10}P_{F} - log_{10}P_{0}) +  \beta
\end{equation} where $P_{F}$ is the fundamental-mode period, $log_{10}P_{0}=0.3$, and the  coefficients $(\alpha, \beta)$ are defined in \autoref{tab:plr_type2} based on the subclass. The distance modulus $\mu$ is then given by,
\begin{equation}
\label{eqn:absmag_cal}
{\rm \mu} = m_{Ks} - A_{Ks} -  M_{Ks}. 
\end{equation} which is converted to distance using \autoref{eqn:dmod2dist}. The \twomass{} photometry ($m_{Ks}$) was obtained from the \Gaia{} archive \citep[\texttt{tmass\_psc\_xsc\_best\_neighbour}\footnote{\href{https://gea.esac.esa.int/archive/documentation/GDR3/Catalogue_consolidation/chap_crossmatch/}{Gaia crossmatch documentation}}][]{dr2Marrese:2019}. Since the extinction $A_{Ks} (l,b,d)$ here is a function of the sky coordinates and distance itself, the Type II distances (\dtypetwo{}) are estimated in an iterative manner, where we first derive the maximum possible $\mu$ (and \dtypetwo{}) assuming zero extinction ($A_{Ks}=0$) along the line of sight, and then update $A_{Ks} (l,b,d)$ and recalculate the $\mu$ (and \dtypetwo{}) until the distance converges (typically after 5 iterations). We estimate the extinction following the procedure set out in \cite{khanna2025rc} by combining publicly available 2D/3D dust maps \citep{Schlegel1998,Green:2019,vergely22} and adopting $A_{Ks}/E(B-V)$=0.304 \citep{Yu:2026}.

In order to validate our distance estimates, we use the catalog by \citet[][\cruzreyes]{reyes_type2} of \typetwo{} associated with Galactic Globular clusters. This allows us to compare our values directly against the independent distances of the host \gc{}. From \cruzreyes{}, we retain sources with a membership posterior $> 0.5$, that are classified as fundamental-mode pulsators and “T2CEP" in the \gdrthree{} variable source catalog. The final sample consists of the \blher{} and \wvir{} subclasses, along with a few lacking a specific subclassification. Using their \twomass{} $K_{s}$ photometry (as described earlier) and the PLR from \autoref{tab:plr_type2} we estimate the distances for these cluster  \typetwo{} stars. \autoref{fig:dgc_comparison} shows that our estimates are largely consistent with the literature values from BV21, except for a notable offset such that \dtypetwo{}$<$ \dglob{} by about 9\%. This offset is still present (albeit reduced to 5\%) if we instead adopt the PLR relations from \citet[][B17]{Bhardwaj2017}; a similar discrepancy was also noted by \citet[][their Figure 18]{Sicignano_t2c_MC:2024}. Because our broader goal is to compare the ensemble distribution of all \typetwo{} with \dcep{} rather than individual sources, we choose to use the more recent PLR calibration of W21, as this small offset does not qualitatively affect our conclusions.

\subsection{Gaia DR3 astrometry \& velocities}
\label{sec:kindata}
For all objects we obtain their proper motions ($\mu_{\alpha^*}, \mu_{\delta}$) and line-of-sight velocities (\texttt{radial\_velocity}) from the \gdrthree{} \href{https://gea.esac.esa.int/archive/}{archive}. \gaia{} provides goodness-of-fit statistics for astrometry, such as the \texttt{astrometric\_excess\_noise} and the Renormalised Unit Weight Error (\textit{RUWE}). 

A high \ruwe{} value typically indicates non-single star behaviour or compromised astrometry. Strict thresholds (e.g., \ruwe$<1.25$) are often used to exclude binaries 
\citep{Penoyre:2022}. However, given that the binary fraction among Classical Cepheids is estimated to be at least 80\% \citep{Kervella:2019_binary_anomaly}, excluding these would heavily bias our sample. Furthermore, \cite{El-Badry:2025} demonstrated that Gaia parallaxes for high-RUWE sources can still yield robust distances if uncertainties are appropriately inflated (by upto a factor of 4). Following this logic, rather than applying an arbitrary \ruwe{} cut, we inflate the proper motion uncertainties ($\sigma_{\mu,Gaia}$) for high-\ruwe{} sources (see \autoref{tab:ruwe_summary}). We cap the inflation factor at 4 to prevent the sampling of unphysical velocity space, given that the radial velocity semi-amplitude in Cepheid binaries is typically between 10--30 \kms{} \citep{Shetye:2024}.

Finally, to convert heliocentric data to galactocentric coordinates we assume the Sun's coordinates are \rgal,\zgc{}=(8.277,0) kpc \citep{GravityCollaboration:2021} and a Solar motion with respect to the Galactic centre of,
\begin{equation}
{\bf{v}}_\odot =
\begin{pmatrix}
9.3 \pm 1.3  \\
251.5 \pm 1.0  \\
8.59 \pm 0.28 
\end{pmatrix} \kms
\end{equation} assuming that Sgr A$^{*}$ is stationary with respect to the Galactic centre \citep{Reid2020,drimmel2023gaia}. The cylindrical coordinate angle $\phi={\rm tan}^{-1}(Y/X)$ increases in the anti-clockwise direction, while the rotation of the Galaxy (\vgal{}) is clockwise. 
\subsection{Compilation of \dcepsing{} metallicities}
Because only a small fraction of Galactic Cepheids have publicly available high-resolution spectroscopy, we compile metallicity estimates for our \dcep{} sample from several major large-scale and Cepheid dedicated surveys: a) We include data from the \href{https://dr19.sdss.org/sas/dr19/spectro/astra/0.6.0/summary/}{\texttt{astraMWMLite}} catalog from the recent SDSS-DR19 release \citep{sdss19}. We filter this sample to retain sources satisfying: \texttt{spectrum\_flags == 0}, \texttt{snr} $> 50$, \texttt{e\_v\_rad} $< 1$, and \texttt{std\_v\_rad} $< 1$; b) We extract data from the \texttt{galah\_dr4\_allstar} catalog \citep{galahdr4}, selecting sources satisying:\texttt{snr\_px\_ccd3} $> 30$ and \texttt{e\_fe\_h} $< 0.2$; c) We include stellar parameters derived from \gaia{} RVS spectra via the \texttt{GSP-Spec} pipeline \citep{gspspecdr3}. We select sources satisfying: \texttt{[vbroadM, KMgiantPar]} ($\le 1$) and \texttt{[vradM, fluxNoise, extrapol]} ($\le 2$). Additionally, we restrict the sample to sources with well constrained metallicities (\texttt{mh\_gspspec\_upper} $-$ \texttt{mh\_gspspec\_lower} $< 0.4$) and effective temperatures \texttt{teff\_gspspec} $> 3750$~K ; d) Finally, we include the measurements from the dedicated \href{https://sites.google.com/inaf.it/c-metall/data?authuser=0}{C-MetaLL} survey \citep{cmetal2021,cmetall2024,trentin2026a} which provides high resolution \feh{} for approximately 300 \dcep{}. Specifically, we used values from Table 4 of their latest release \texttt{(VizieR: }\href{https://vizier.cds.unistra.fr/viz-bin/VizieR-3?-source=J/A%2bA/707/A142/tablea4}{J/A+A/707/A142}).

\subsection{Orbit integration and quasi-integrals of motion}
We compute the total 3D Galactocentric velocity \vtot{} = $\sqrt{V_{X}^{2} + V_{Y}^{2} + V_{Z}^{2}}$, and integrate orbits with the \agama{} package \citep{agamapaper}. We adopt the non-axisymmetric Galactic potential $\Phi$ from \cite{hunter24}. Alongside the galactic disc and dark matter halo, this model includes the Supermassive Black Hole, and a rotating Galactic Bar with constant pattern speed $\Omega_{b}$=37.5 \kmskpc and orientation angle of 25\degree. Additionally, we compute three quasi-integrals-of-motion (\energy{}, \lz{}, \lperp{}) which, although not strictly conserved in a non-axisymmetric rotating potential, are still useful indicators of orbital dynamics. These are defined as, 
\begin{gather}
    \energy = \frac{1}{2} \vtot^{2} + \Phi \,, \\
    \lz = \rgal \vgal \quad ; \quad \lperp = \sqrt{L_{X}^{2} + L_{Y}^{2}} \,, \\
    \lx = \ygc \vzgc - \zgc \vygc \quad ; \quad \ly = \zgc \vxgc - \xgc \vzgc \,.
\end{gather} 
In addition we calculate the Jacobi integral  \jacobi{}:
\begin{equation}
     \jacobi = \energy - \Omega_{b}\lz  \,,
\end{equation}
that is conserved for a rotating non-axisymmetric potential, such as in the case of a bar with a fixed pattern speed \citep{dillamore_orbits25}. For brevity we express energy and angular momenta in terms of \enun{}=\enunits{} and \angun{}=\angunits, respectively. We adopt \lz{} $>0$ (prograde) and \lz{} $<0$ (retrograde) for the sense of rotation.  We generate 50 realisations per object which allows us to capture the spread in the aforementioned quantities. For each realisation, the orbits are integrated for a total of 1 Gyr in the forward/backward directions in time in steps of 1 Myr. We note that the conclusions in the paper are not sensitive to the choice of Galactic potential adopted given that we limit to 1 Gyr integration time. S25 provide age estimates (\tbirth{}) for the \dcep{} in their sample with nearly 90\% having \tbirth{}$<500$ Myr, hence integration for 1 Gyr seems a reasonable upper limit. The eccentricity is then computed using the apocenter ($r_{apo}$) and pericenter ($r_{peri}$) as,
\begin{equation}
    \texttt{ecc} = \frac{r_{apo} - r_{peri}}{r_{apo} + r_{peri}}
\end{equation} such that perfectly circular orbits have \ecc{}=0 while those on perfectly radial orbits have \ecc{}=1. 
\begin{figure}
\includegraphics[width=1.\columnwidth]{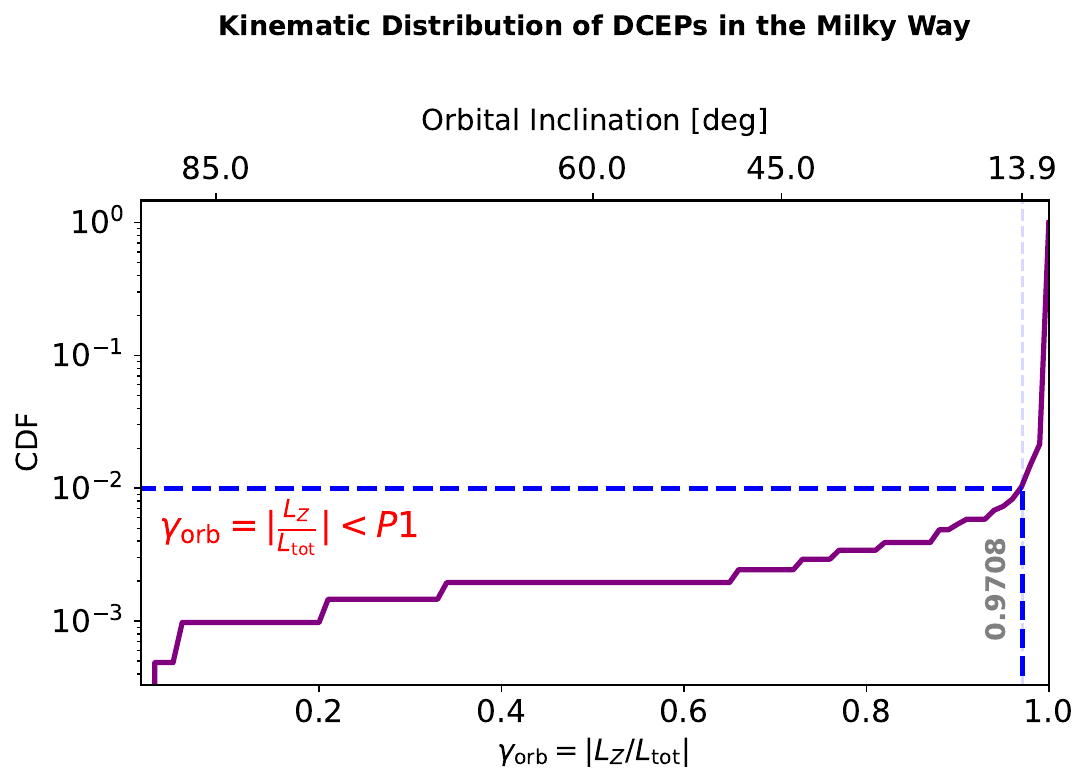} 
\caption{Cumulative distribution function of the \gamfunc{} for all MW \dcep{} shown by the purple curve. The region below the blue dashed line represents outliers that are selected since they fall below the one percentile level which corresponds to an orbital inclination threshold of $P1=13.9^{\circ}$.} \label{fig:cdf_gamma}
\end{figure}
\section{The 6D Dynamical census and rogue Cepheids}
\label{sec:kincensus}
\subsection{Identification of rogue Cepheids}
We characterise the orbital plane of each object in terms of the ratio of its angular momentum about the \zgc{} axis to the total angular momentum,
\begin{equation}
    \gamma_{\rm orb} = \left|\frac{L_{z}}{L_{tot}}\right| =\cos (\rm orb\_inc) .
\end{equation} \gamfunc{} ranges between 0 ($orb\_inc$=90$^\circ$ i.e polar orbit) and 1 ($orb\_inc$=0$^\circ$ i.e. disc-like orbit) and thus defines the orientation of the orbital plane with respect to the plane of the Galaxy. \cite{Ye:2024} recently characterised the orbital profile for several stellar streams and accreted structures in the Galactic halo using this parameter, finding for example for the Helmi streams \gamfunc{}=0.68, for Wukong \gamfunc{}=0.33, etc., showing that many accreted substructures exhibit an orbital inclination in the range between 30$^\circ<$\orbinc$<$ 75$^\circ$. In \autoref{fig:cdf_gamma} we show the cumulative distribution (CDF) of \gamfunc{} for our \dcep{}, which turns out to be highly skewed towards \gamfunc{} =1, showing clearly that nearly all Milky Way \dcep{} are on very disc-like orbits as expected. We select kinematic anomalies (\rogue{} Cepheids) by picking out the \dcep{} in the bottom one-percentile of the CDF i.e.
\begin{equation}
\label{eqn:gamma_cond}
\gamma_{\rm orb} = \left|\frac{L_{Z}}{L_{\rm tot}}\right| < P1(=0.9708) , 
\end{equation} which corresponds to orbital inclination $\geq13.9^{\circ}$. This criterion selects 18 \rogue{} stars, their properties are listed in \autoref{table:cat}, where these are ranked by \orbinc{} going from the most disc-like (\orbinc{}=14.1$^\circ$) to the most polar orbit (\orbinc{}=89.3$^\circ$).

\subsection{Sample validation}
The optical light curves of the \rogue{} stars are compiled in \autoref{app:light_curves}). In the following we summarise their general characteristics that point to similarities with the overall \dcep{} population. 

\begin{figure}
\includegraphics[width=1.\columnwidth]{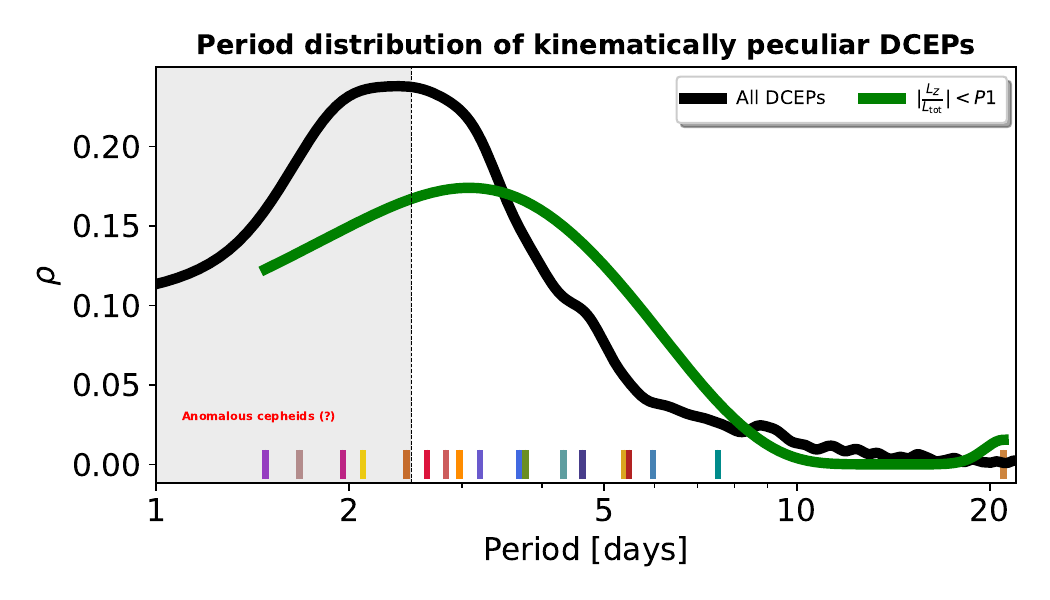} 
\caption{Period distribution of all Milky Way \dcep{} (black curve) and those with $\gamma_{\rm orb} = |\frac{L_{Z}}{L_{\rm tot}}| < P1 $ are indicated as vertical lines with the green curve showing the Kernel density estimate for their distribution. Also marked is the typical region (Period < 2.5 days) where Anomalous Cepheids are found.} \label{fig:period_hist}
\end{figure}
\begin{figure}
\includegraphics[width=.99\columnwidth]{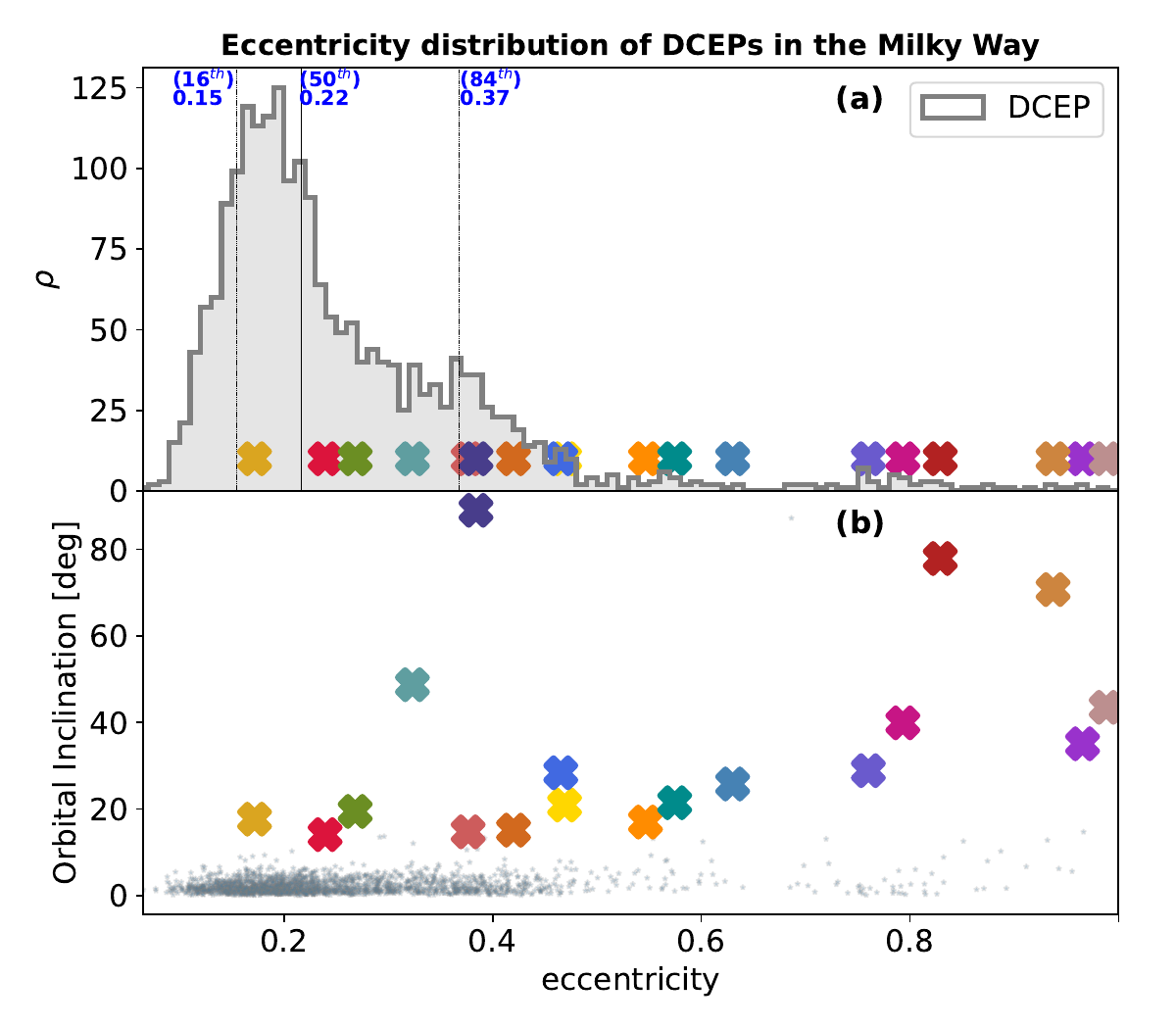} 
\caption{The Top panel shows the distribution of orbital eccentricity (\ecc{}) for all Milky Way \dcep{} (gray histogram) with the (16$^{th}$,50$^{th}$,84$^{th}$) percentiles indicated. The kinematically peculiar \dcep{} are shown as coloured crosses. The bottom panel compares the orbital eccentricity against the orbital inclination (\orbinc{}) for the same two samples.} \label{fig:eccentricity}
\end{figure}
\begin{figure}
\centering
\includegraphics[width=1.\columnwidth]{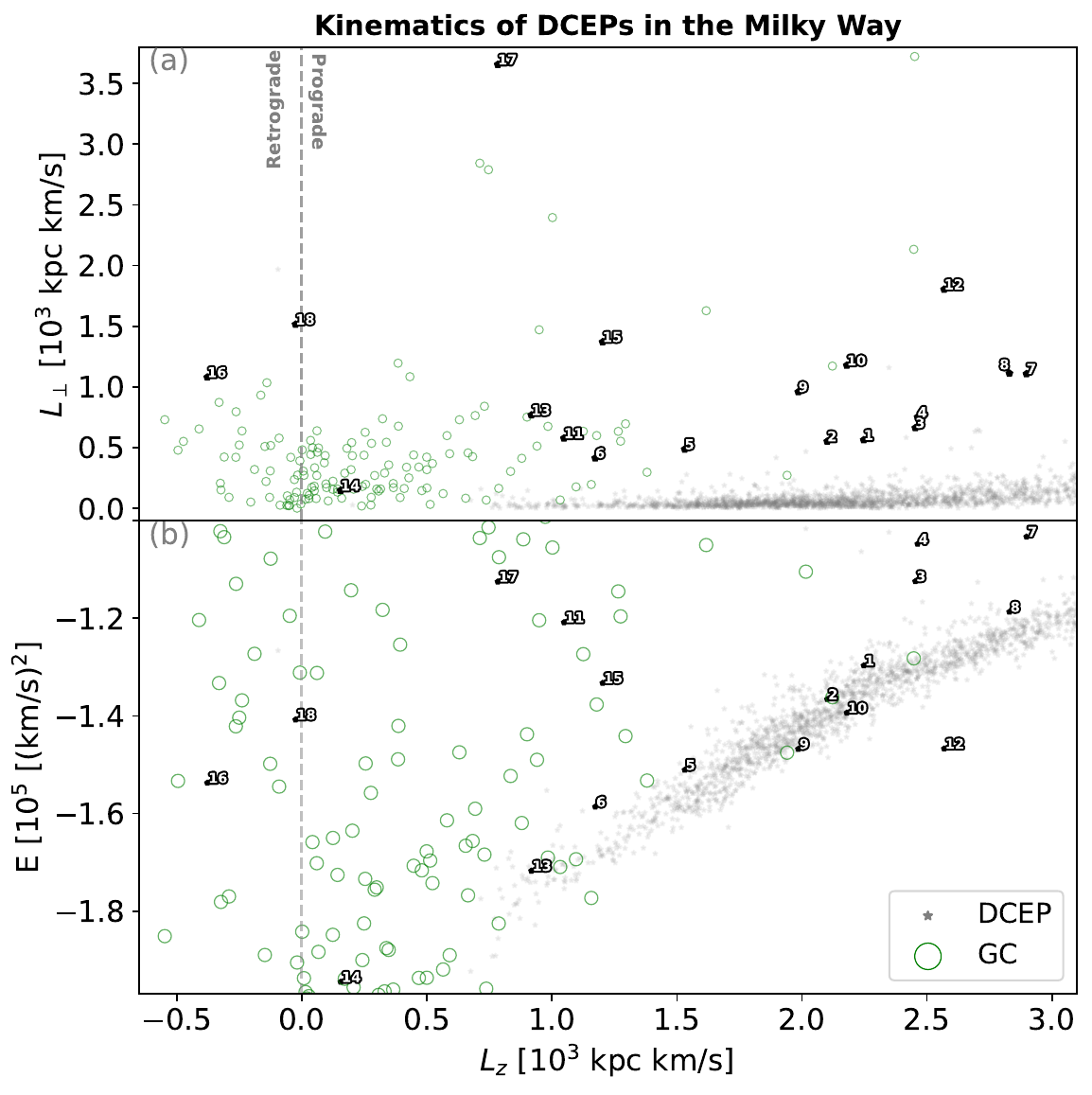} 
\caption{Panel (a) shows the distribution of all \dcep{} (gray) in the \lz{} vs. \lperp{} plane. The dashed vertical line marks the boundary between retrograde and prograde stars. The \dcep{} satisying the condition $\gamma_{\rm orb} = |\frac{L_{Z}}{L_{\rm tot}}| < P1 $ are labelled and we consider these as kinematic outliers (\rogue{} stars). The positions of the Galactic Globular Clusters (green) are also shown for perspective. Panel (b) shows the distribution for the three samples (\dcep{}, \gc{}, \rogue{}) in the \lz{}-\energy{} space.}\label{fig:lzlperp}
\end{figure}
\begin{figure*}
\centering
\includegraphics[width=1.9\columnwidth]{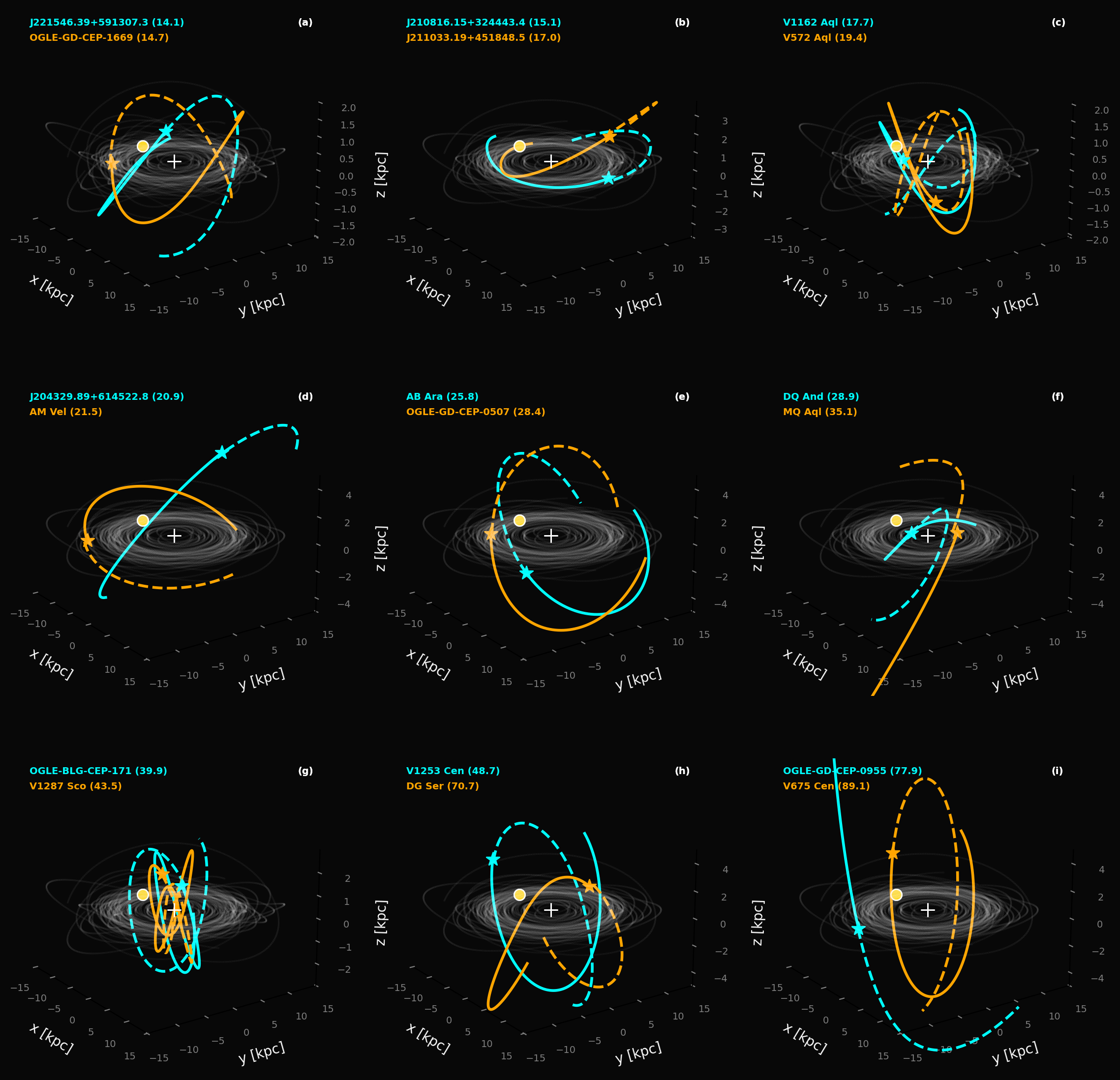}
\caption{Gallery of orbits for the kinematically peculiar \rogue{} stars listed in \autoref{table:cat}. Each panel plots two objects, with increasing \orbinc{} (in degrees indicated in parenthesis) from top left to bottom right. For each, the trajectory is shown for 150 Myr in both the backwards (solid line) and the forwards (dashed line) direction, and the current position is indicated by a star symbol. The orbits for a random subset of all \dcep{} are shown in gray in the background, the Galactic center and the Sun are marked by the '+' and 'O' symbols respectively. The top-down perspective of the orbits is shown in \autoref{fig:orbgal_top}.}\label{fig:orbgal}
\end{figure*}
\textit{Periods:} The period distribution of this sample shown in \autoref{fig:period_hist}(green curve) compares well with that of all \dcep{} (black curve), with several sources found with a period between 2 and 5 days. We note that there are five sources with a period $<2.5$ days, a threshold below which Anomalous Cepheids (AC) are typically found \citep{Ripepi:2024}. ACs are thought to be either intermediate-age stars with exceptionally low metallicity, or coalesced old binary stars \citep{Soszynski:2008anom}. However, as \autoref{fig:period_hist} shows, approximately 20\% of all \dcep{} ($\sim$700 stars) oscillating in the fundamental (F) or first-overtone (1O) mode are also found in the period $<2.5$ day regime. In that respect then, the period distribution of the \rogue{} stars is not so unusual.

\textit{Distance estimates:} The \dcep{} sample analysed in this work relies on the \pawel{} classifications, with distances (D\_S25) derived from mid-IR PLR. Independently, \cite{Ripepi:2023} classified \dcep{} using \gdrthree{} optical photometry and estimated their distances (D\_GDR3) via Wesenheit magnitudes. \autoref{table:cat} shows there are 13 of our 18 \rogue{} stars common to both catalogs, for which D\_S25 and D\_GDR3 are in good agreement. Thus our adopted distance estimates are robust despite the differences in the adopted methodology.

\subsection{Orbital and kinematic properties}
\label{sec:orbkin}
We summarise below the notable kinematic characteristics of the \rogue{} stars.

\textit{Eccentricity:} \autoref{fig:eccentricity}(a) compares the orbital eccentricity (\ecc{}) distribution of the full \dcep{} sample (gray) against the \rogue{} sources (crosses). The full sample exhibits a median eccentricity of \ecc{}=0.22$^{+0.15}_{-0.13}$ reflecting cold near-circular orbits. Instead, the \rogue{} stars are more uniformly distributed; all but one of the sources lie above the median eccentricity of the full sample and 13 sources exceed the 84$^{th}$ percentile (\ecc{} $>0.37$). The most eccentric of these are  \texttt{DG Ser},  \texttt{MQ Aql}, and \texttt{V1287 Sco} (\ecc{}=0.94, 0.97, 0.99). 

\autoref{fig:eccentricity}(b) shows the \ecc{} and the \orbinc{} together. The full \dcep{} sample clusters tightly at low orbital inclination and eccentricities characteristic of a dynamically cold population. The \rogue{} stars instead exhibit a spread with generally higher \ecc{} and the \orbinc{} values. A direct correlation however is not expected in this space as a source could both be on a near-circular orbit but with its orbital plane misaligned with the disc, as is apparent in the case of V675 Cen where the orbit is at a near 90$^{\circ}$ polarity. 

\textit{High velocity stars (\hvs):} For the entire \dcep{} sample, the median total 3D velocity \vtot{}$=$232$\pm13$ \kms, is about the same as the normalisation of the Milky Way's rotation curve \citep{drimmelkhanna}, while the 99$^{th}$ percentile is at \vtot{}$\sim$274 \kms. The \rogue{} stars have median \vtot{}=224\kms, but there are five sources between 271$<$ \vtot{} $<$ 489\kms i.e. these are moving much faster than expected for \dcep{}. While the exact escape velocity of the Milky Way depends on the Galactocentric distance (lower at higher \rgal{}), in general objects moving above a total velocity of 300-400 \kms can be considered as candidates of hyper-velocity stars \citep{2022MNRAS.515..767M,2025A&A...700A.172C}. In the literature \hvs{} have been found in a great many variety of environments/scenarios such as; escaping the central black hole of the Galaxy/LMC, runaway stars in White-dwarf binaries, or in the Galactic halo \citep{2020MNRAS.491.2465K,elbadry_wdhyper,2025A&A...700A.172C,2025ApJ...982..188H}. Here, we report a potential \hvs{} candidate seen in \dcep{}: \texttt{MQ Aql} with \vtot=489$\pm26$ \kms. However, this star also has the shortest period (1.48 days) in our sample, and may be a misclassified Anomalous Cepheid. 

\textit{Dynamical properties:} From their distribution in \lz{}-\lperp{} space (\autoref{fig:lzlperp}a) it is clear that Milky Way \dcep{} (gray) have near zero \lperp{} and are on predominantly prograde motion ($1<$ \lz{}/\angun{} $<3$), as expected of a young disc population. Their large span in \lz{} is a reflection of \lz{}$\propto$\rgal{} since the rotation curve is nearly flat (near constant \vgal{}) for \dcep{} \citep{drimmelkhanna}. Instead, the 18 \rogue{} stars (overplotted as labelled points) are populated well away from the background population, with three sources (14,16,18) placed close to the boundary between retrograde and prograde orbits. Finally, for perspective, we have also included the Milky Way's \gc{} population from BV21 (green). These are a very distinct population evident from their near random distribution about the \lz{}$=0$ boundary, well away from the discy \dcep{} background population. The \gc{} also show a spread in their \lperp{} values, with a few in close proximity to specific \rogue{} stars. 

\autoref{fig:lzlperp}b) shows the distribution of the three samples in \lz{}-\energy{} space. The majority of \dcep{} follow a distinct tight and monotonic sequence where the total energy increases proportionally with \lz{}. Stars along this track are part of the dynamically cold population tightly bound to the disc. The \rogue{} stars are located either at the upper or lower boundaries of this track or in other cases well above at higher energies indicative of less-bound orbits. This would then suggest that many of the \rogue{} stars have a disc-like origin and were perturbed on high \lperp{} orbits. The \gc{} show a large spread in \energy{}; a few with energies and \lz{} similar to the disc, while others are in close proximity to specific rogue stars. This could suggest the possibility of past dynamical interactions between \gc{} and a few of the \rogue{} stars.

\emph{Orbits:} \autoref{fig:orbgal} shows the orbits for the peculiar sources. For clarity, each panel plots only two sources at a time, following the sequence in \autoref{table:cat} (i.e. from top-left to bottom-right the orbits are arranged in order of ascending \orbinc{}). For each source we show their orbits over 150 Myr in both the backward (solid line) and forward (dashed line) directions. The current positions of the source, the Galactic center, and the Sun are indicated ( `*',`+', `O' respectively). For perspective, we also plot the orbital tracks for a random subset of all the \dcep{} (white) which clearly exhibit highly disc-like trajectories compared to the \rogue{} stars. As shown in \autoref{fig:orbgal} these peculiar objects are distributed across the sky, varying in their positions with respect to both the Galactic center and the Sun, though are in general found at higher latitudes compared to the background \dcep{} (\autoref{fig:lbplot} top panel). The trajectory of the \hvs{} candidate MQ Aql (\tbirth{}= 215 Myr) is shown in \autoref{fig:orbgal}(f), and suggests that it originates from well beyond \xgc=15 kpc and from far below the disc. Similarly, the trajectory of the other fast moving candidate OGLE-GD-CEP-0955 (\tbirth{}= 109 Myr) in \autoref{fig:orbgal}(i) suggests an origin well above the disc and well beyond \ygc=-15 kpc. For added clarity, the top-down perspective of all these orbits is provided in \autoref{fig:orbgal_top}.

\section{Ruling out systematics}
\label{sec:sytematics_}
Before exploring physical origins of the kinematically \rogue{} stars, we consider possible systematics that could give rise to such anomalies. 
\subsection{Classification systematics: Are these miclassified \typetwo{}?}
If the \rogue{} stars were misclassified as \dcep{} instead of \typetwo{}, their distance estimates would need to be re-derived using the appropriate T2C PLR (Section ~\ref{sec:typ2dist}). At a given period, reclassification from \dcep{} to \typetwo{} lowers a star's predicted luminosity, assigning it a smaller heliocentric distance, consequently also lowering the inferred 3D \vtot{}. For the 18 \rogue{} stars their \typetwo{} distances are derived---assuming no knowledge about the subclassification---using the coefficients in \autoref{tab:plr_type2}(`unclassified'). In the following we anaylse if reclassifying the \rogue{} stars would be consistent with their astrometry, metallicities, and the dynamical expectation of a T2C population. We do not consider other potential misclassifications (such as Young Stellar Objects or spotted stars) as evaluating that would likely require improved light curves.

\emph{Consistency with Astrometry:} In order to compare our T2C photometric distances to \gaia{} astrometric parallaxes, following S25 we define a quality parameter ($Q2$) that accounts for uncertainties in photometric distances, parallax, as well as the \gaia{} zero point variance (see \autoref{app:q2}). A lower $Q2$ value indicates better agreement with the astrometric distance. Essentially we check if the quality parameter for a \dcepsing{} classification ($Q$) improves when the source is instead assumed to be T2C ($Q2$). \autoref{table:cat} highlights the two sources where $Q2< Q$; \textit{AM Vel} ($Q=0.37$ vs $Q2=0.001$) and \textit{OGLE-GD-CEP-0955} ($Q=2.28$ vs $Q2=0.19$). Based on their astrometry, while these two sources are likely misclassifications, for the remaining 16 \rogue{} stars, the quality parameter did not improve with reclassification.
\begin{figure}
\centering
\includegraphics[width=.95\columnwidth]{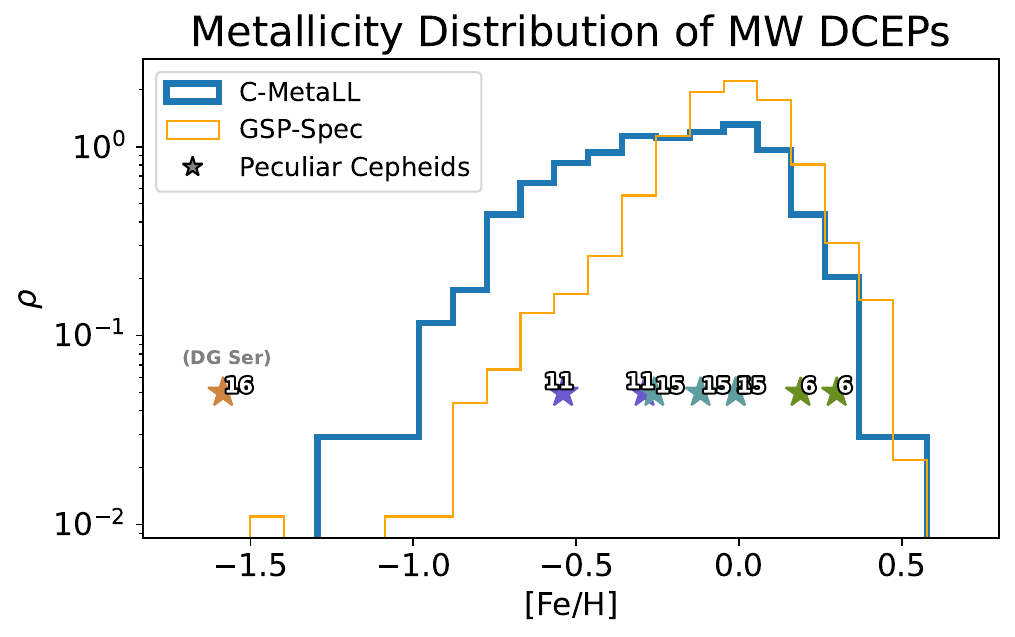} 
\includegraphics[width=.95\columnwidth]{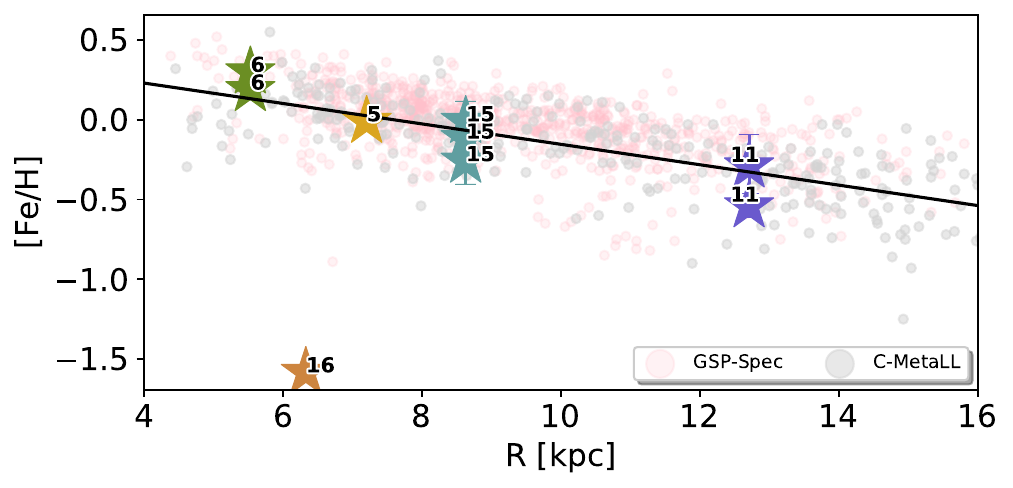} 
\caption{Metallicity distribution of the kinematically peculiar sources compared with that of all MW \dcep{}. The top panel shows the distribution of spectroscopic \feh{} from the \cmetal{} and \texttt{GSP-spec} in the background, while for the peculiar sources these are compiled from a variety of surveys as available (see \autoref{tab:feh_summary}). The bottom panel shows the metallicity gradient of all MW \dcep{} with the \rogue{} sources overplotted. For all sources here we adopt distances from S25. The black solid line is adopted from \cite{trentin2026a}.} \label{fig:feh_dist}
\end{figure}

\emph{Metallicity distribution:} \dcep{} are young stars (\tbirth{}$\ll$ 1 Gyr) with typical metallicities between $-1<$\feh{}$<0.5$ \citep[e.g.][]{Genovali:2014,trentin2026a}. The top panel of \autoref{fig:feh_dist} presents the metallicity distribution from the \gdrthree{} \texttt{GSP-spec} (N$\sim 850$) and \cmetal{} (N$\sim 300$) datasets, which together form the largest homogeneous compilation of metallicities for Galactic \dcep{}. Despite differences in their selection functions, the two datasets are largely consistent within their respective uncertainties ($\sim0.2$ dex). Even with these two datasets combined, nearly 65\% of all known MW \dcep{} still lack reliable spectroscopy. Indeed, across the major surveys we could only retrieve metallicities for 6 \rogue{} stars (Table \ref{tab:feh_summary}).

As shown in \autoref{fig:feh_dist}(top panel) the \rogue{} stars generally align with this background distribution. The sole exception is \texttt{DG Ser} (labelled 16) which is unusually metal poor, with two available estimates: \feh{} = $-1.58$ (\apg) and \feh{} = $-1.71$. (\texttt{GSP-spec} omitted from the figure because it fails recommended quality flags.) A value of \feh{} = $-1.58$ is significantly lower than the most metal poor \dcepsing{} in \cmetal{} (\texttt{OGLE-GD-CEP-128}, \feh{}=$-1.25$ at \rgal{}=14.9 kpc). The bottom panel in \autoref{fig:feh_dist} presents the metallicity as a function of \rgal{}. The background samples closely trace the well known metallicity gradient in \dcep{} (solid black line) adopted from \cite{trentin2026a}. Notably our \rogue{} stars with reliable \feh{} also follow this gradient, suggesting they are genuine classical Cepheids. Again, \texttt{DG Ser} (labelled 16) is a clear exception as it is too metal poor for a \dcepsing{} located at \rgal{}=6.3 kpc. In contrast to Classical Cepheids, \typetwo{} show a broader, metal-poorer distribution--ranging between $-2.4<$ \feh{} $<-0.4$ in globular clusters and between $-2<$ \feh{} $<$ 0.4 in the Galactic field \citep[][their Figure A.1]{bono2020_t2c_met}. So while on the one hand based on its metallicity alone \texttt{DG Ser} would appear to be \typetwo{}-like, since its light curve in \ref{fig:light_curves_1} is consistent with \dcep{}, we suggest re-deriving its metallicity using templates suited to variable stars. We will also present further corroborating evidence in Section \ref{sec:origin_rogue} that this star is indeed a \dcepsing{}.
\begin{figure}
\includegraphics[width=1.\columnwidth]{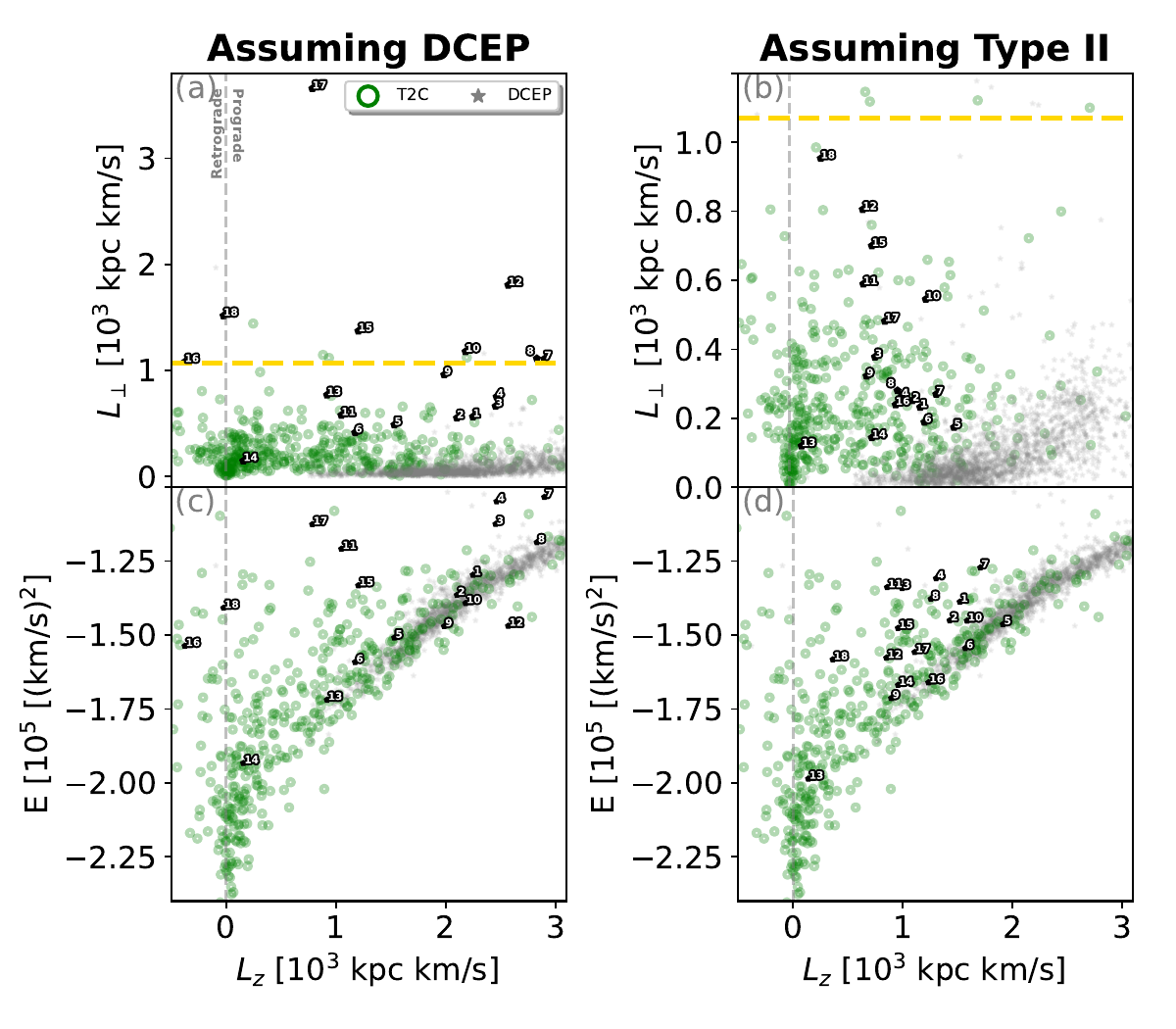} 
\caption{Distribution of all MW \dcep{} (gray), all MW \typetwo{} (green) and the peculiar sources (labelled) in the \lz{}-\lperp{} and  \lz{}-\energy{} planes. This is similar to \autoref{fig:lzlperp}, except here we assume two different classifications for the peculiar sources: in panels (a,c) we assume the peculiar sources are of the \dcep{} type, while in panels (b,d) we assume these are of the \typetwo{} type. The horizontal yellow line is included as a visual aid.} \label{fig:assumedcep_t2}
\end{figure}

\emph{Dynamical comparison:} Although we find only two of our 18 rogue \dcep{} to be likely misclassified T2Cs, we still evaluate how their dynamics would shift under a manual reclassification to T2Cs. \autoref{fig:assumedcep_t2} illustrates how reclassification affects the distribution of the \rogue{} stars on the \lz{}-\lperp{} and \lz{}-\energy{} dynamical planes. 
Panels (a) and (c) show their distribution assuming a \dcepsing{} classification as in \autoref{fig:lzlperp}, while panels (b) and (d) plot the dynamical quantities derived using the \typetwo{} distances. For context, we include background populations of Milky Way \dcep{} (gray), and \typetwo{} (green).

\autoref{fig:assumedcep_t2}(b) reveals that the background \typetwo{} and \dcep{} are distinct. In particular the \typetwo{} mirror the \gc{} distribution from \autoref{fig:lzlperp}, with a dense concentration around \lz{}=0 with \lperp{}$<$1 \angun{}. Additionally, a sparse tail of  \typetwo{} extends toward higher \lz{}, overlapping with the disc-like \dcep{} population.

 In the \lz-\lperp{} plane (\autoref{fig:assumedcep_t2}a,b), reclassification shifts all  peculiar sources below the \lperp{}=1.07\angun{} reference line. This happens because  \typetwo{} are intrinsically fainter than their classical Cepheid counterparts, which reduces both their inferred distances and angular momenta. With the exception of \texttt{OGLE-BLG-CEP-171} (labelled 13), which shifts into the core \typetwo{} distribution around \lz{}=0, the peculiar sources remain within the sparse tail of the background \typetwo{} population.
 Similarly, on the \lz-\energy{} plane (\autoref{fig:assumedcep_t2}c,d), reclassification concentrates most of the peculiar sources between $-1.75<$\energy{}/\enun{}$< -1.25$. Again \texttt{OGLE-BLG-CEP-171} is the exception, shifting to an energy (\energy{}/\enun{}$\sim$-2) more typical of the background \typetwo{} core. However, we note that reclassifying this Cepheid as a T2C would give it a photometric distance less consistent with the astrometry as its $Q2 (3.11)> Q(0.80)$. 
 
 While the dynamical distribution of all \rogue{} stars shifts significantly under a \typetwo{} assumption, the majority still reside in the sparse tail of the background population rather than its core. To investigate whether or not this makes the rogue stars any less likely to have originated from T2C, we reverse the experiment - essentially we now derive \dcep{} distances for the known T2C, which has the effect of inflating their angular momenta (compared to the original T2C distribution). Then we select fake `rogue' stars by applying the same orbital inclination threshold as for \dcep{} and compare the location of these with respect to their original T2C distribution as well as with respect to our \dcep{} rogue stars. Specifically we find that those in the original core-T2C are generally inflated well above the yellow horizontal line in (\autoref{fig:assumedcep_t2}a,b), instead those in the tail of the original T2C distribution are inflated below this reference line, but nevertheless overlap with the distribution of our \dcep{} rogue stars. So, even though our rogue stars only end up in the dynamical tail of the background T2C upon reclassification, some of these could well have originated from a true T2C population and based on this alone we cannot thus rule out misclassification. On the other hand, those rogue stars that do end up in the core of the background T2C distribution are very likely to have been misclassified.

In summary, we find that reclassification (from \dcep{} to \typetwo{}) improves the consistency between photometric and astrometric parallaxes for only 2 \rogue{} stars. From a dynamical and astrometric standpoint alone, it is not possible to quantify how many of the rogue sources warrant reclassification. We anticipate that the improved astrometry from the upcoming \gaia{} DR4 will further help in this regard.

\subsection{Astrometric systematics: signatures of unresolved binaries}
Given the high binary fraction of about 80\% among Classical Cepheids \citep{Kervella:2019_binary_anomaly}, in the following we check which of the \rogue{} stars may suffer from compromised single-source astrometric solutions due to the presence of (un)resolved companions. 

\emph{Known binaries \& proper motion anomalies:} The \rogue{} star \texttt{V1162 Aql} is flagged as a potential Spectroscopic Binary (SB1) in the \veloce{} survey \citep{Shetye:2024}, based on trends in its pulsation residuals. For this source the \vlos{} from \veloce{} ($12.6\pm0.1$ \kms) and \gdrthree{} ($15.8\pm2.7$ \kms) are comparable. Interestingly, \texttt{V1162 Aql} is also listed in the catalogue of proper motion anomalies (PMa) by \citet{Kervella:2019_binary_anomaly}. This catalog identifies anomalies by measuring the discrepancy between the short-term proper motions from Hipparcos \citep{hipparcos97,hipparcos2007} and \gdrtwo{}, against the long-term average velocity derived from the positional difference over the $\sim$24-year epoch baseline. However, for \texttt{V1162 Aql}, the PMa detection is at a very low signal-to-noise ratio ($S/N = 0.5$), falling well below their formal threshold for suspected binaries ($S/N > 2.0$).

No other \rogue{} stars were found in the publicly available \gaia{} NSS catalogs \citep{2023A&A...674A..34G}, the resolved binary catalog by \cite{El-Badry:2021}, or the \href{https://webarch.konkoly.hu/cep/intro.html}{Cepheid binary database} maintained by \citet{szabados2003}

\emph{Sources with high \ruwe{}:} \cref{tab:ruwe_summary} lists the \gdrthree{} \ruwe{} for each of the \rogue{} stars alongside its sky-dependent threshold used to identify potential signatures of binarity. We find five sources that exceed this threshold (flagged as $\text{RUWE}_{\text{cond}}$=`high'). While two of these are only moderately above their threshold  (\ruwe{}=1.41 and 1.55), the remaining three sources have rather extreme values: \texttt{OGLE-GD-CEP-1669} (\ruwe{}=12.6), \texttt{OGLE-GD-CEP-0507} (\ruwe{}=26), and \texttt{AM Vel} (\ruwe{}=41). In general the \ruwe{} is most sensitive to binary systems with a high astrometric signal (photocentric wobble), so while a high \ruwe{} by itself does not guarantee binarity, these three extreme cases likely have unresolved companions that could be confirmed by future radial velocity monitoring.
\begin{figure}
\includegraphics[width=.95\columnwidth]{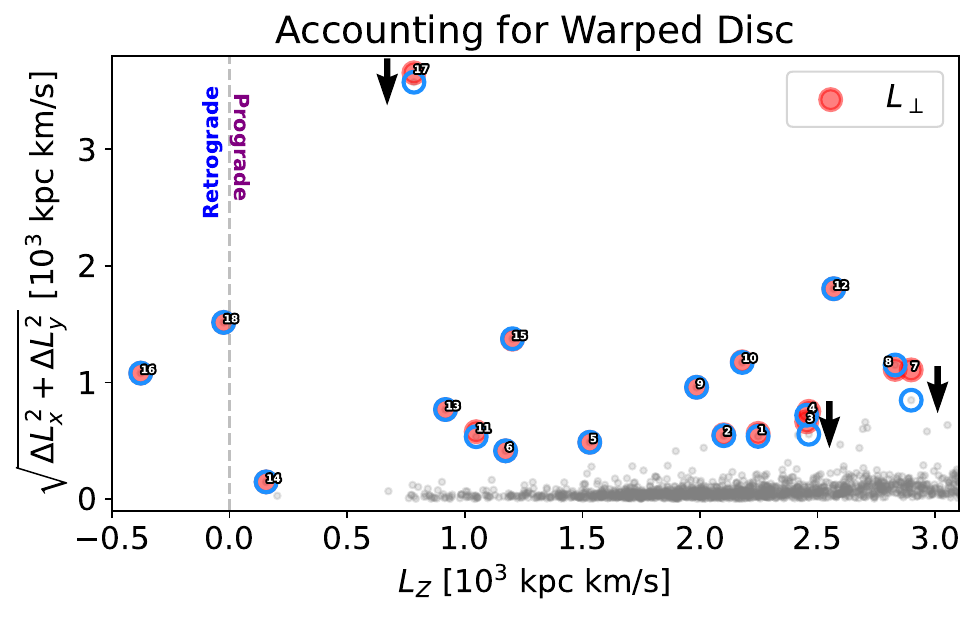} 
\caption{Comparison with a warp model. Same as \autoref{fig:lzlperp} but here the y-axis instead plots \lperpwarp{} i.e. \lperp `corrected' for a warp model. All \dcep{} are shown as grey dots, while the kinematic peculiar ones are marked as blue circles. For comparison with \autoref{fig:lzlperp} also plotted are the \lperp{} values marked as red circles. The black arrows indicate where there is a notable difference going from \lperp{} to \lperpwarp{}.} \label{fig:warpmodel}
\end{figure}
Given our small sample size, rather than discarding the \rogue{} stars with high \ruwe{} we adopt a conservative approach and inflate their proper motion uncertainties by up to a factor of four (see Sec.~\ref{sec:kindata}). This is likely a very generous overestimate compared to the expected proper motion bias induced by the wobble of a companion, which we estimate as 
\begin{equation}
    \mu_{\mathrm{bias}} \, \left[\mathrm{mas}\,\mathrm{yr}^{-1}\right] = \frac{V_{\mathrm{wobble}}\,\left[\mathrm{km}\,\mathrm{s}^{-1}\right]}{4.74 \times d\,\left[\mathrm{kpc}\right]} \,.
\end{equation} For example assuming an upper-limit radial velocity semi-amplitude of $V_{\rm wobble}$=30 \kms, for \texttt{OGLE-GD-CEP-0507} ($d=6.27$ kpc, $\mu=8$ mas/yr, $\sigma_{\mu}=0.5$ mas/yr), we estimate $\mu_{\rm bias}\sim1$ mas/yr which is about twice its nominal proper motion uncertainty. With the over-inflated uncertainties, the orbital inclinations (lower limit) of \texttt{OGLE-GD-CEP-1669} and \texttt{AM Vel} fall below the threshold for \rogue{} stars, while \texttt{OGLE-GD-CEP-0507} is still above the threshold (\autoref{table:cat}).

\section{The origin of the rogue Cepheids}
\label{sec:origin_rogue}
In the following we consider a few physical scenarios which might explain the kinematically anomalous \dcep{}.

\subsection{Can some \rogue{} stars be explained by the warp?}
The Galactic disc is warped, a feature especially conspicuous in the spatial distribution of young stars such as \dcep{} \citep{Skowron_2019warp_science,Skowron:2019,Chen_2019warp}. In the bottom panel of \autoref{fig:lbplot} we present the 3D (\xgc,\ygc,\zgc) warp model (blue) from \greatwavepoggio{}, with the positions of the \rogue{} stars overlaid. Several of these are at larger \zgc{} than predicted by the warp model, and their displacement is indicated by the vertical lines. Conversely, a few others show minimal offsets, suggesting their kinematics may closely follow that of the Galactic warp. 

\begin{figure}
\centering
\includegraphics[width=1.\columnwidth]{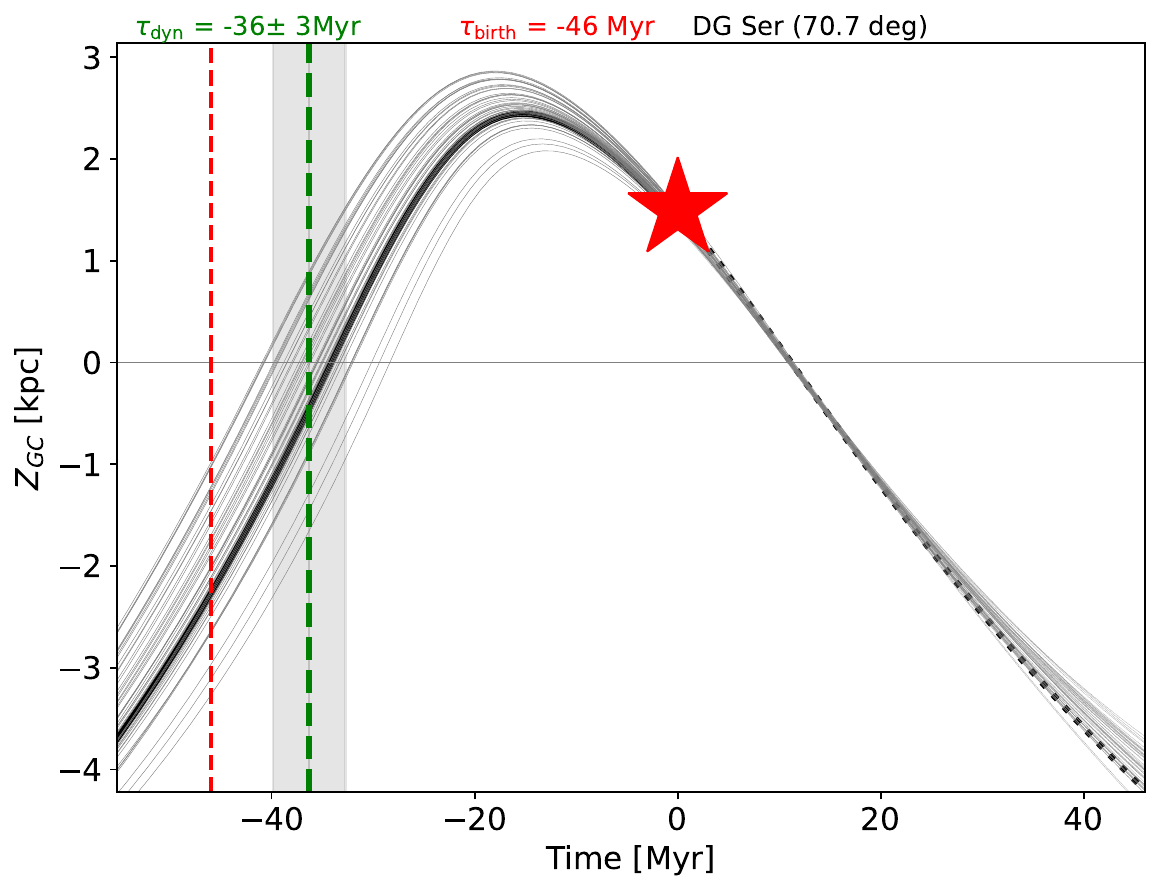}
\caption{Example Orbital tracks for one of the \rogue{} stars (\texttt{DG\_Ser}) showing its  displacement from the Galactic plane as a function of time. The trajectory based on present day initial conditions is shown by the black track (solid:past, dotted:future), its present day position is marked in red, and its Cepheid age estimate is marked by the red dashed line. Its dynamical age estimate is marked by the green dashed line with its uncertainty illustrated by the gray region around it. Tracks for the remaining \rogue{} stars are presented in \autoref{fig:tracks}.} \label{fig:example_ztrack}
\end{figure}

We compute the expected contribution of the warp in angular momenta, by adopting a model with $(\zgc, \vzgc)_{\rm warp}$ from \greatwavepoggio{} and additionally $(\vgal, \vrgc) = (-231.5, 0)~\mathrm{km~s^{-1}}$ for a galactic model with rotation and zero streaming motion in the plane for simplicity. The warp-corrected $L_{\perp, \rm corr} = \sqrt{\Delta L_x^2 + \Delta L_y^2}$, is plotted against \lz{} for the \rogue{} stars in \autoref{fig:warpmodel} (blue circles), overlaid with their observed \lperp{} from \autoref{fig:lzlperp} (filled red circles). For most of the sources the blue and red circles coincide such that the warp's contribution to their \lperp{} is negligible. Only in three cases (labels 17,4,7) is the corrected \lperp{} lower than the observed (indicated by black arrows), but this revision is still not enough to reconcile their kinematics with the background disc (grey). In summary,  
the warp alone is not sufficient to explain the \rogue{} orbits.
\begin{figure}
\includegraphics[width=1.\columnwidth]{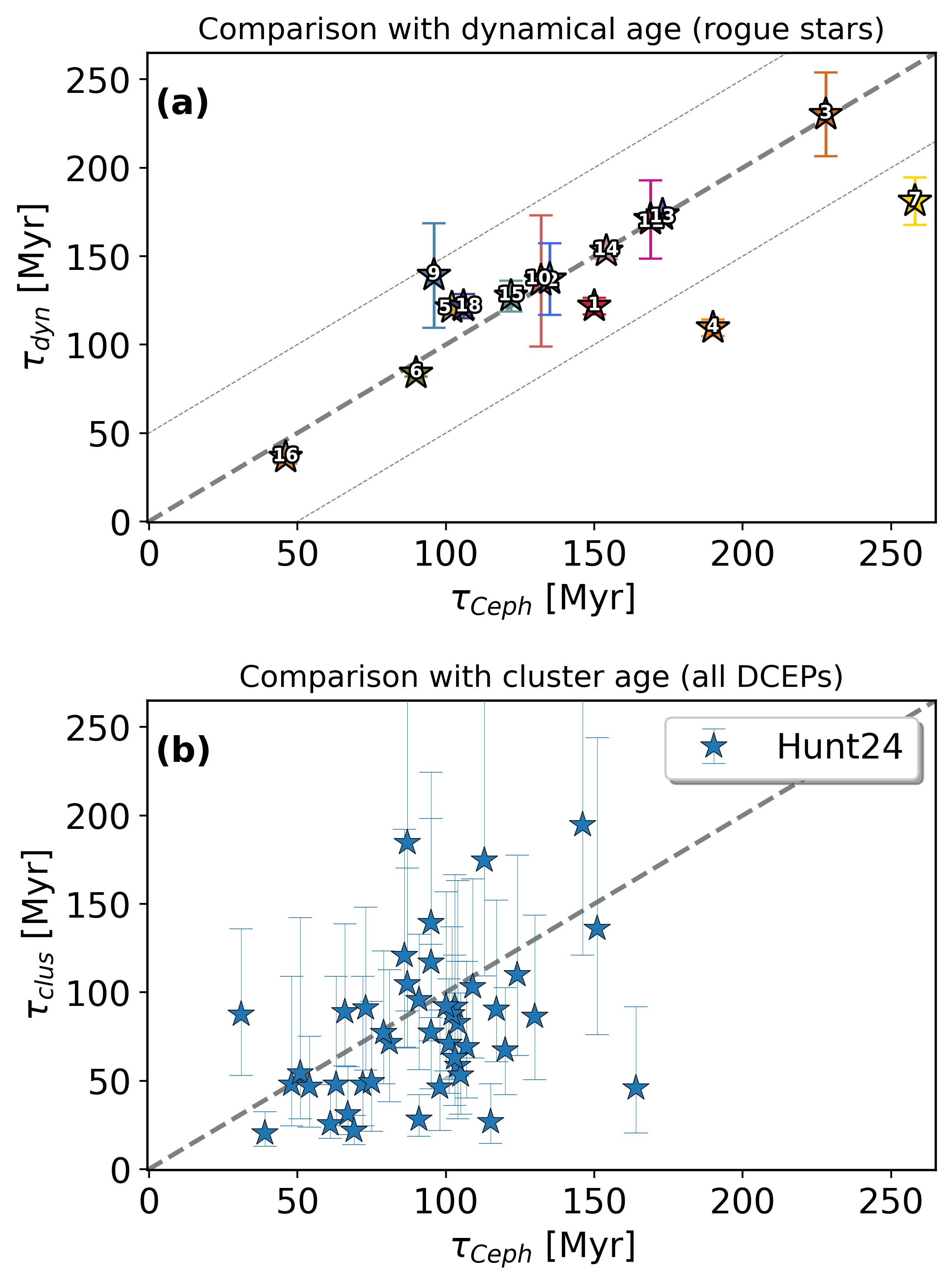}
\caption{Comparing age estimates for \dcep{} from three independent methods. Panel (a) compares the estimates from standard period-age relations (\tbirth{}) with dynamical ages (\tdyn{}) estimated in this work for the kinematically peculiar \rogue{} stars. The dotted black lines indicate the $|$\tdyn{} -  \tbirth{}$|< 50$ Myr region. Panel (b) compares \tbirth{} of all \dcep{} found in clusters with their cluster-based ages \tclus{} from \cite{Hunt:2024}. The 1:1 relation is shown as a grey dashed line in both panels.} \label{fig:dynage_comp}
\end{figure}
\subsection{Dynamical scattering by globular clusters: the case of OGLE-GD-CEP-0507 and E3}
\label{sec:globclus_assoc}
In Section \ref{sec:orbkin} we noted the close proximity between some of the \rogue{} stars and \gc{} in $(\lz, \lperp, \energy)$ space (Fig. ~\ref{fig:lzlperp}). 
We used three methods to measure the dynamical distance of the DCEP orbits with those of the Globular Clusters (see appendix...)  

\emph{Separation metric:}  In order to compute the separation in dynamical space between each \gc{}---\rogue{} star pair (i,j) we use the following metric,
\begin{equation}
\label{eqn:ijmetric}
        \Delta I_{ij}^{2} = (L_{Z,i} - L_{Z,j})^{2} + (L_{\perp,i} - L_{\perp,j})^{2} + (H_{i} - H_{j} )^{2} .
\end{equation} In practice, we rank the \gc{}---star association by the median separation computed over multiple realisations for both objects in order to account for uncertainties in the dynamical quantities. In \autoref{tab:dcep_gc} we list the associations using (\jacobi,\lz,\lperp). However since \jacobi{} and \lz{} are not entirely independent, we also list the case using only two components (\jacobi,\lperp). For 50\% of the \rogue{} stars the associations remain unchanged whether we use two or all three components.

\emph{Orbital overlap:} Additionally we compute the orbital overlap for each \gc{}---\rogue{} star pair. With a Kernel density estimator, each object's discrete 3D trajectory $\mathbf{x}$ (over all realisations) is converted to a continuous distribution function $\hat{f}_i$, using a \gaussian{} kernel and a bandwidth determined following \citet{Scott:1992}. The spatial overlap between two objects is quantified using the coefficient \citep{Bhattacharyya1943},
\begin{equation}
\label{eqn:bhattacharyya}
    \mathit{BC} = \int \sqrt{\hat{f}_1(\mathbf{x}) \hat{f}_2(\mathbf{x})} \, \mathrm{d}^3x,
\end{equation} which returns a value between 0 (no overlap) and 1 (100\%  overlap). The associations with the highest overlap (above 80\%) are included in \autoref{tab:dcep_gc}

In particular only one rogue Cepheid, \texttt{OGLE-GD-CEP-0507} (labelled 10), is strongly associated with the same cluster (\texttt{E\_3}) in all three methods and with a particularly high (89\%) orbital overlap. To explore this further, in \autoref{fig:gc_dcep_0507}\footnote{The separation and orbital overlap were computed using nominal uncertainties, but as an exercise in \autoref{fig:gc_dcep_0507_inf} we show the orbital projections for \texttt{E\_3} \& \texttt{OGLE-GD-CEP-0507} assuming over-inflated uncertainties. The overlap drops to 78\%.} we present the orbital projections (integrated backwards) for both objects, with their present-day locations at $\tau=0$ (now) indicated in red. The high orbital overlap noted earlier is now also visually striking. At $\tau=0$ the two objects have very similar face-on (\xgc,\ygc) positions, and are separated mostly along \zgc{} by about 3 kpc. In panel(d) we plot their 3D separation $\Delta d_{ij}$=$\sqrt{\Delta X_{ij}^{2} + \Delta Y_{ij}^{2}+ \Delta Z_{ij}^{2}}$, computed over multiple realisations, as a function of time from $\tau$=0 up to the Cepheid's assumed age (\tbirth{}=135 Myr). The median separation (cyan curve) evolves from $\Delta d_{ij}\sim3$ kpc at present to a minimum $\Delta d_{ij}\sim1.5$ kpc approximately 50 Myr ago, before rising back to $\Delta d_{ij}\sim3$ kpc at the time of \texttt{OGLE-GD-CEP-0507}'s birth. It would be natural then to speculate whether \texttt{E\_3} passed through the disc shortly after the birth of \texttt{OGLE-GD-CEP-0507}, potentially perturbing its orbit. To illustrate this, in panels (a-c) we highlight two key epochs: \tbirth{}=135 Myr (gray markers), and their closest approach at $\tau$=51 Myr ago (purple markers). These markers correspond to the shaded regions in panel (d). As can be clearly seen, the position of closest approach is already 5 kpc outside the disc, so that the GC could not have been responsible for perturbing \texttt{OGLE-GD-CEP-0507} outside the disc. 

Furthermore, we note that for a Globular Cluster such as \texttt{E\_3} to impart a significant kick velocity exceeding several km/s would require a sub-parsec separation between the star and the cluster as well as a relative velocity of $<$5 \kms (Eugene Vasiliev, private communication). Based on this consideration it is unlikely for \texttt{E\_3} to have gravitationally perturbed the orbit of \texttt{OGLE-GD-CEP-0507}.

\subsection{Runaway scenario \& dynamical age estimates}
\emph{Dynamical ages:} \dcep{} are young stars likely born in high-density environments such as spiral arms. We consider a scenario where our \rogue{} stars were kinematically ejected shortly after birth via dynamical interactions with one or more massive companions in their birth cluster. 
Assuming each star was born in the Galactic plane we trace its orbit backward in time to determine its dynamical age (\tdyn{}), which we define as the epoch at which it crosses the midplane (\zgc{}=0), closest to its estimated birth age (\tbirth{}), as provided in S25.

As an example, \autoref{fig:example_ztrack} shows the orbital tracks over multiple realisations for \texttt{DG Ser} charting its vertical displacement (\zgc{}) from the Galactic plane as a function of time. Its current position is marked in red, and its Cepheid age \tbirth{} is marked as a dashed red line. The tracks for the remaining stars are compiled in \autoref{fig:tracks}. To identify the likely ejection event from these tracks we minimise the absolute difference $|$\tdyn{} -  \tbirth{}$|$, as the Cepheid age itself is an uncertain quantity which means the mid-plane crossing can occur on either side of the nominal birth age (see for example \texttt{DG Ser} vs. \texttt{V675 Cen} ). We estimate the median \tdyn{} and its uncertainty by repeating the procedure over multiple realisations, resampling the observation uncertainties. The estimated \tdyn{} is included as a green dashed line with its uncertainty illustrated by the grey region around it and also included in \autoref{table:cat}.

Fig.~\ref{fig:dynage_comp}(a) shows that the estimated dynamical ages for the majority of the \rogue{} stars are in excellent agreement with their Cepheid ages, closely following the 1:1 relation. 
From the plot we have excluded the two likely misclassified T2C stars \texttt{AM Vel}(8) and \texttt{OGLE-GD-CEP-0955}(17). Also \texttt{MQ-Aql}(12) never crosses the disc during backward integration, providing further evidence that it may well be misclassified, leading to an erroneous distance and velocity estimate. Of the remaining 15 stars, only three sources, \texttt{J221546.39+591307.3}(1), \texttt{J211033.19+451848.5}(4) and \texttt{J204329.89+614522.8}(7), have dynamical ages significantly shorter that their Cepheid ages, showing clear evidence that they were perturbed out of the disc well after their birth.

The \tbirth{} estimates from S25 are derived using standard period--age relations for \dcep{}. For a subset of Milky Way \dcep{} associated with open clusters, independent cluster-based age estimates are available from \cite{Hunt:2024}. In \autoref{fig:dynage_comp}(b) we compare these cluster ages with the \tbirth{}. In contrast to the tight correlation in  \autoref{fig:dynage_comp}(a), there is substantial scatter between these two age estimates with median uncertainties in \tclus{} on the order of 50\%. This likely indicates that the ages for the youngest clusters are highly uncertain. Conversely, the close agreement between our derived dynamical ages and \tbirth{} is an independent validation of the \dcep{} period-age relation and an additional verification that our sources are indeed Classical Cepheids, including the Cepheid 
\texttt{DG Ser} (\autoref{fig:example_ztrack}) with an anomalous metallicity, and \texttt{OGLE-BLG-CEP-171} which is found in the dynamical core of Type II population.
To our knowledge this is a first such estimate of dynamical ages for Classical Cepheids in the Milky Way.

\emph{Ejection velocity \& mechanisms:} At least 30\% of progenitors of core-collapse supernovae---that mark the end of massive stars---typically have initial masses upward of 8 M$_{\odot}$ \citep{Jerkstrand:2018}. A significant fraction of these explode between 50-200 Myr after birth, with their pre-explosion lifespans extended especially in binary systems \citep{Zapartas:2017}. Since binary components are coeval, the temporal gap between the Cepheid's birth and ejection (\tdyn{} -  \tbirth{}) could thus be informative about the possible ejection mechanisms. Certainly, \autoref{fig:dynage_comp}(a) shows that all but three sources fall within the $|$\tdyn{} -  \tbirth{}$|< 50$ Myr region (dotted lines), hinting at an ejection not long after the star's birth, most likely involving three-body encounters with other birth cluster members \citep{Gvaramadze2009}, or possibly a massive companion going supernova \citep{Evans2020}.

\begin{figure}
\centering
\includegraphics[width=1.\columnwidth]{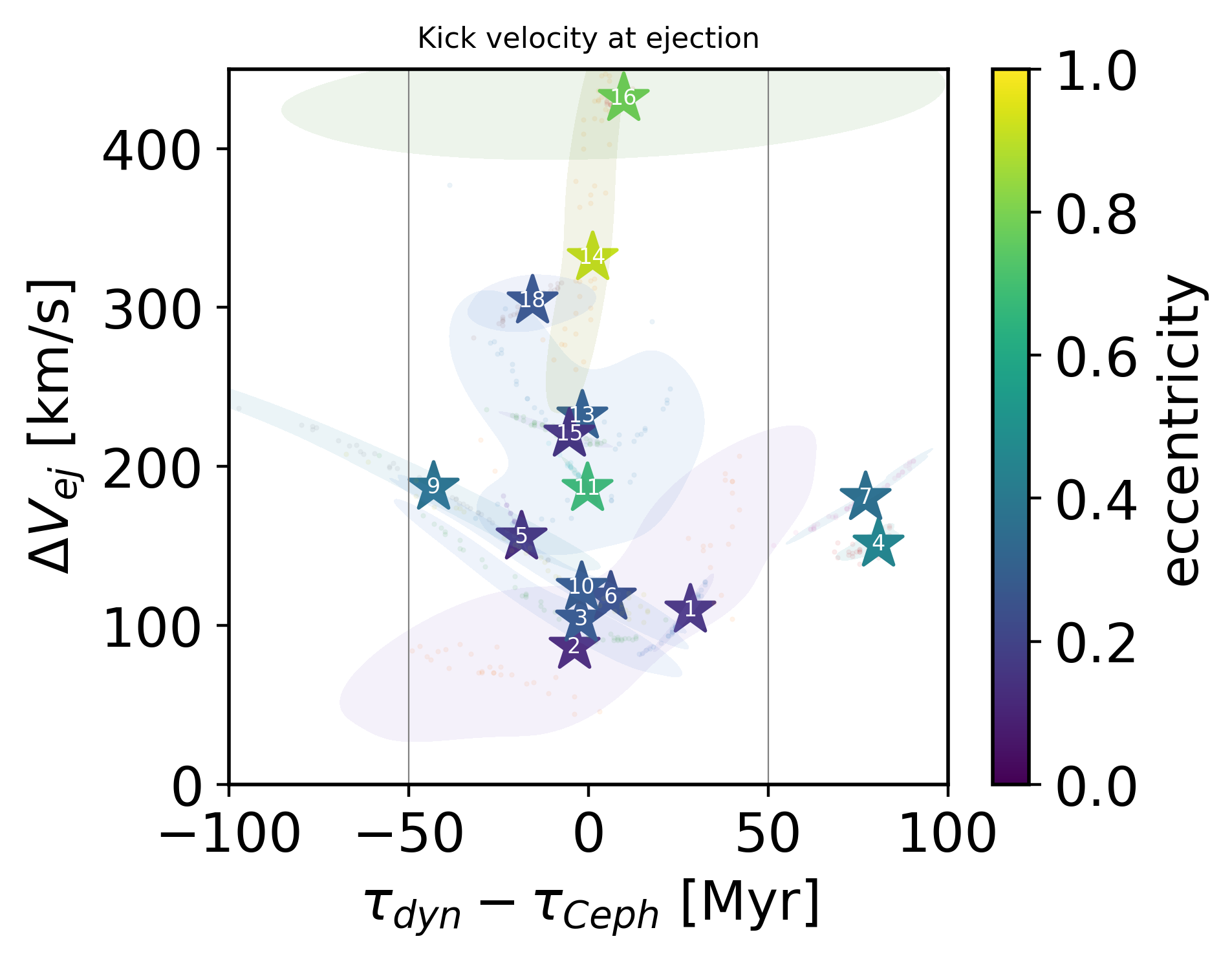}
\caption{Estimated ejection velocity for the \rogue{} stars plotted against the time gap between assumed birth and ejection epochs. The median values are shown as stars while the uncertainties over many realisations are displayed as clouds in the background. These are also available in \autoref{table:cat}. The data are colour-coded by orbital eccentricity.} \label{fig:kickvel}
\end{figure}

Finally, regardless of the nature of the ejection, we compute the kick-velocity \vej{} of each star at its \tdyn{} epoch. We define this as the difference between the 3D velocity vector of the star and a background model assuming only Galactocentric rotation and no radial or vertical streaming motion i.e.,
\begin{equation}
\Delta V_{ej} \, [\kms] = (\vgal ,\vrgc,\vzgc{})-(-231.5,0,0) \, .   
\end{equation} These kick-velocities are listed in \autoref{table:cat} and in \autoref{fig:kickvel} we plot \vej{} against (\tdyn{} -  \tbirth{}). Not surprisingly, the \rogue{} stars on the most polar orbits (those with higher labels) also have a very high ejection velocity such as \texttt{DG\_Ser} with the highest \vej{}$\sim 400$ \kms. Furthermore, the data points are colour coded by their orbital eccentricity, and we note that in general sources with \ecc{}$<0.2$ (highly circular orbits) have \vej{} below 150 \kms{}. However, this method of estimating the dynamical ages assumes both the reliability of the stellar orbits as well as a scenario where each source was born in the Galactic midplane. Indeed in a few cases the uncertainties on \vej{} and \tdyn{} are quite large, also reflected in the background clouds in \autoref{fig:kickvel}.

\section{Summary and outlook}
\label{sec:summary}
We carried out the first orbital census of Galactic \dcep{}, mapping their distribution in orbital and dynamical  space and comparing with the older Type II Cepheids and Globular Clusters. 
\begin{itemize}
    \item While generally we find \dcep{} to be consistent with a dynamically cold population  confined to the Galactic midplane, we identified 18 candidate \rogue{} stars with high orbital inclinations (between $14^\circ$ to $90^\circ$), several of which are present at high Galactic latitudes. 
    \item After considering possible misclassification and data quality systematics, we find that 15 out of 18 of these sources are consistent with being of a \dcepsing{} origin. Furthermore misclassification of Cepheid type or compromised astrometry due to binarity is not enough to explain the observed kinematics of all these stars. For stars that are likely Type II based on the inconsistency between astrometric and photometric data, we plan to re-evaluate their classifications once improved light curves are available. The Gaia source identifiers for the \rogue{} stars are made available and we suggest high resolution spectroscopic follow-up could help resolve ambiguity in their classification. Alternatively, where spectra are already available, we suggest re-deriving their metallicities using templates suitable for variable stars, such as in the case of \texttt{DG Ser} with measurements of \feh{} $=-1.58$ from \sdss{}-DR19. We summarise our verdict for the individual \rogue{} stars in \autoref{tab:remarks}.
    \item By tracing back their orbits, we also estimated the dynamical ages for these stars assuming a disc-birth origin, finding that their standard Cepheid ages agree very well with our dynamical age estimates. This provides an independent validation of the Cepheid period-age relationship. 
    \item As a physical mechanism, dynamic scattering of Cepheids by Galactic Globular Clusters was explored given their apparent proximity in (\jacobi,\lz,\lperp) space, in particular for the case of \texttt{OGLE-GD-CEP-0507} and \texttt{E\_3} with a $\sim$90\% orbital overlap. However, any significant momentum change to the Cepheid in this system would require a subparsec (rather than a sub kpc) separation and as such scattering by \texttt{E\_3} is highly unlikely. 
    Also, we can exclude the possibility that \texttt{E\_3} scattered this Cepheid from the Galactic midplane, as the dynamical age of \texttt{OGLE-GD-CEP-0507}(10) is in good agreement with its Cepheid age, while the close encounter with \texttt{E\_3} happened sometime later when the star was already well above the Galactic midplane. 
    \item Assuming a runaway ejection scenario we then also estimated a kick velocity imparted on the star at the dynamical age epoch, finding extreme values of \vej{} $>200$ \kms{} in several cases.
   \end{itemize}
\begin{table}
\caption{Final remarks and verdict for the individual \rogue{} stars identified in this work.}
\label{tab:remarks}
\resizebox{1.\columnwidth}{!}{ \footnotesize
\begin{tabular}{l|l|l}
\hline\hline
 survey\_source\_id & Remarks & Verdict \\
\hline
1 J221546.39+591307.3   & \tdyn{} $<$ \tbirth{} & \textcolor{blue}{DCEP}  \\
2  OGLE-GD-CEP-1669     & High RUWE (unresolved binary?)   & \textcolor{blue}{DCEP}\\
3 J210816.15+324443.4   &   &  \textcolor{blue}{DCEP}\\
4  J211033.19+451848.5  &  \tdyn{} $<$ \tbirth{} &  \textcolor{blue}{DCEP}  \\
5           V1162 Aql   & Known SB1 \citep{Shetye:2024}   &  \textcolor{blue}{DCEP}  \\
6           V572 Aql    &   & \textcolor{blue}{DCEP} \\
7   J204329.89+614522.8 & \tdyn{} $<$ \tbirth{} & \textcolor{blue}{DCEP} \\
8           AM Vel      & $Q2 < Q$ | High RUWE (unresolved binary?) &  \textcolor{red}{T2C} \\
9          AB Ara       &  & \textcolor{blue}{DCEP} \\
10    OGLE-GD-CEP-0507  & High RUWE (unresolved binary?)  & \textcolor{blue}{DCEP} \\ 
11           DQ And     &  &  \textcolor{blue}{DCEP}  \\
12          MQ Aql      &  HVS candidate (?),  but low period (1.48 days) &  \textcolor{red}{ACEP}  \\
13   OGLE-BLG-CEP-171   &  Dynamically similar to T2C population, but &  \textcolor{blue}{DCEP}  \\
       &  \tdyn{} $\sim$ \tbirth{} and $Q2 > Q$.   &  \\
14           V1287 Sco  &  &    \textcolor{blue}{DCEP} \\
15           V1253 Cen  &  &  \textcolor{blue}{DCEP}  \\
16         DG Ser       &  Atypically metal poor (\feh{}=$-1.58$) | \tdyn{} $\sim$ \tbirth{} & \textcolor{blue}{DCEP}   \\
17    OGLE-GD-CEP-0955  & $Q2 < Q$ & \textcolor{red}{T2C} \\
18         V675 Cen     &  & \textcolor{blue}{DCEP}  \\
\hline
\end{tabular}}
\end{table}
The distribution of the difference between our dynamical ages and the Cepheid ages also reveals something about the origin of runaway stars in the initial mass range of Cepheids, which were born as O or B stars. Indeed, that the dynamical ages agree with the majority of Cepheid ages confirms that in this mass range the dominant channel for forming runaway stars is the dynamical ejection scenario, as those ejected by a binary companion requires some delay between birth of the binary and the supernova. This is in agreement with the recent finding of \citet{Carretero-Castrillo2023}. 

Probing dynamical processes in the Galaxy is often limited to the availability of high confidence distance tracers. Classical Cepheids (despite their low numbers $\sim$3600) offer a unique dataset with highly precise distances and kinematics, allowing us to trace the dynamical history of the Galactic disc out to its extremities. Looking ahead to the upcoming \gdrfour{}, we can expect improved proper motions (by about a factor of 3), and line-of-sight velocities for an additional $\sim$ 500 \dcep{} (down to $G_{RVS}$=16.5) thus allowing us to expand the orbital census to an even larger sample \citep{Brown:2025}. Currently \gaia{} line-of-sight velocities are only available for about 57\% of all Galactic \dcep{}. Furthermore, we also expect a more complete and vetted catalogue of \dcep{} with \gaia{} time-series data products. 

\begin{acknowledgements} We thank the anonymous referee for a critical read and helpful suggestions.
SK, RD and EP were supported in part by the Italian Space Agency (ASI) through contract ASI-INAF 2025-10-HH.0 to the National Institute for Astrophysics (INAF). SK \& RD also acknowledge support from the European Union's Horizon 2020 research and innovation program under the GaiaUnlimited project (grant agreement No 101004110). DMS acknowledges support from the European Union (ERC, LSP-MIST, 101040160). SK acknowledges use of the INAF PLEIADI@IRA computing resources. We thank Alfred Castro-Ginard, Vincenzo Ripepi, Richard Anderson, Alexandre Gallene and Eugene Vasiliev for useful discussions. This work has made use of data from the European Space Agency (ESA) mission \gaia\ (\url{https://www.cosmos.esa.int/gaia}), processed by the \gaia\
Data Processing and Analysis Consortium (DPAC, \url{https://www.cosmos.esa.int/web/gaia/dpac/consortium}). Funding for the DPAC has been provided by national institutions, in particular the institutions participating in the \gaia\ Multilateral Agreement. This work has used the following additional software products:
\href{http://www.starlink.ac.uk/topcat/}{TOPCAT}, and \href{http://www.starlink.ac.uk/stilts}{STILTS} \citep{Taylor:2005};
Matplotlib \citep{Hunter:2007};
IPython \citep{PER-GRA:2007};  
Pandas \citep{reback2020pandas}; 
Astropy, a community-developed core Python package for Astronomy \citep{AstropyCollaboration:2018}; NumPy \citep{harris2020array}; 
\end{acknowledgements}

   \bibliographystyle{aa} 
   \bibliography{mybib} 


\begin{appendix}
\label{sec:app}
\section{Type II distance validation}
\label{app:typeII_val}
\begin{table}
\caption{Period-luminosity relation coefficients adopted for \typetwo{}. }
\label{tab:plr_type2}
\resizebox{\columnwidth}{!}{ \footnotesize
\begin{tabular}{c|c|c|c|c|c}
\hline\hline
 & Population & $\alpha$ &  $\beta$ & $\sigma$ & Note \\
\hline
& & & & & \\
(1)  & \blher{} & -2.251 &  -1.840 & 0.054 & W21 Table 4 (case 3)  \\
(2)  & \wvir{}/unclassified  & -2.225 & -1.836 & 0.065 & W21 Table 10 (L21)\\
\hline
\end{tabular}}
\tablefoot{The coefficients ($\alpha,\beta$) and the intrinsic scatter in the PLR adopted from W21 for the \typetwo{} subclassifications \blher{} and \wvir{}.} 
\end{table}
\begin{figure}
\includegraphics[width=1.\columnwidth]{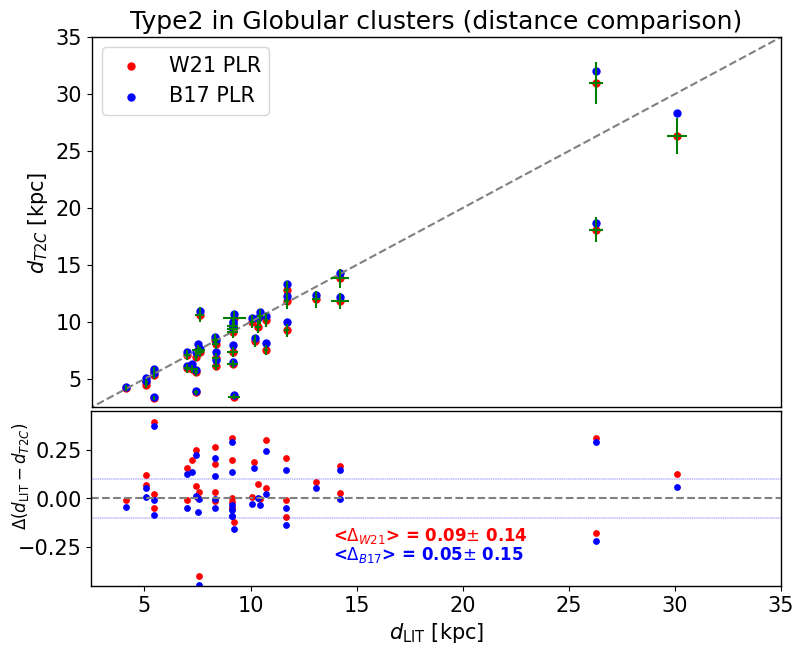} 
\caption{Validation of distance estimates for the \typetwo{} in Globular clusters derived using the PLR from W21, against the values for the parent cluster taken from \cite{Baumgardt:2021}. The top panel compares 1:1 distribution while the residuals are shown in the bottom panel. For reference the comparison is repeated using an older PLR calibration by B17.} \label{fig:dgc_comparison}
\end{figure}
We validate our distance estimation for \typetwo{} using the subset that are associated with Galactic Globular Clusters by \cite{reyes_type2}. \autoref{fig:dgc_comparison} compares our distance estimates against that of the parent clusters taken from \citet[][$d_{\rm LIT}$]{Baumgardt:2021}. The top panel shows that the distances estimated using the W21 PLR adopted in this work, are largely consistent with Globular cluster distances although we note an offset (bottom panel) of approximately 9\% such that \dtypetwo{}$<d_{\rm LIT}$. For reference also shown is the comparison with an older PLR calibration by B17 which also shows an offset of about 5\%. We choose to use the newer calibration in this work.

\section{Light curves for individual sources}
\label{app:light_curves}

\begin{figure*}
\includegraphics[width=.5\columnwidth]{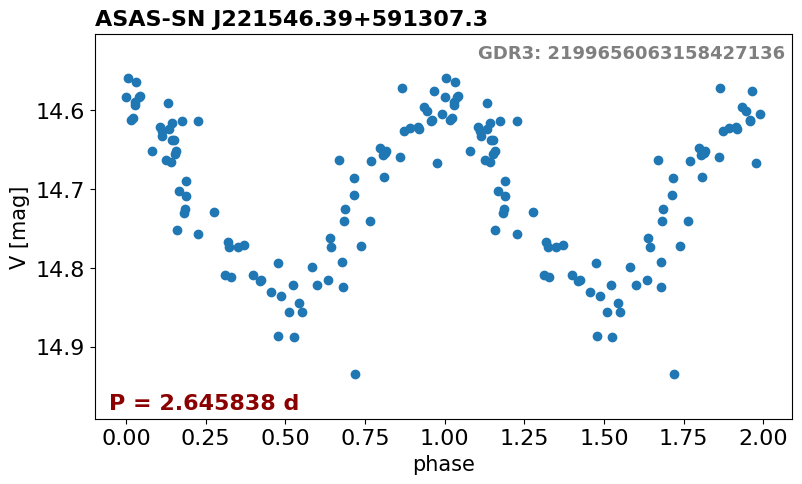}
\includegraphics[width=.5\columnwidth]{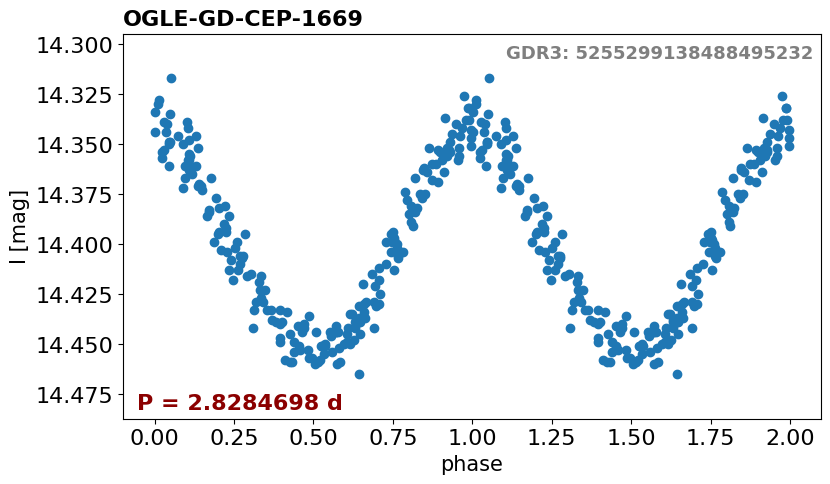}
\includegraphics[width=.5\columnwidth]{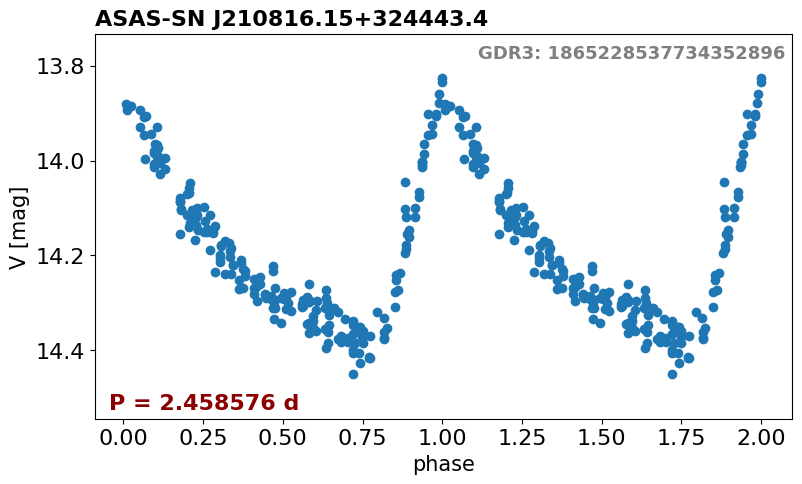}
\includegraphics[width=.5\columnwidth]{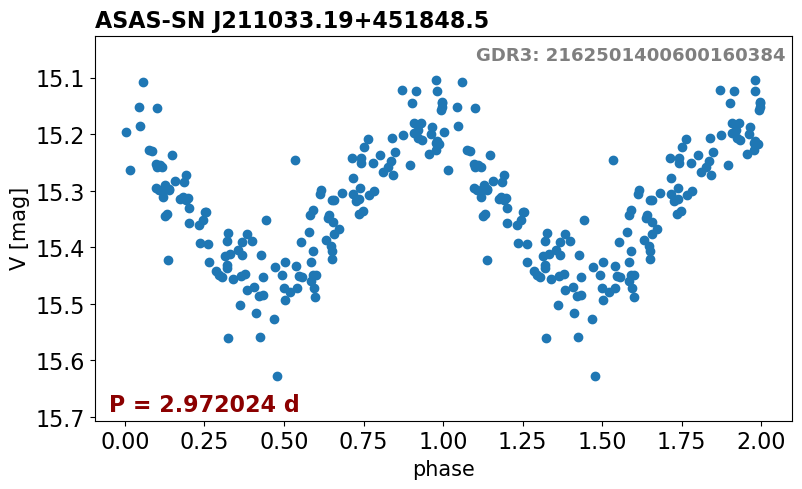}
\includegraphics[width=.5\columnwidth]{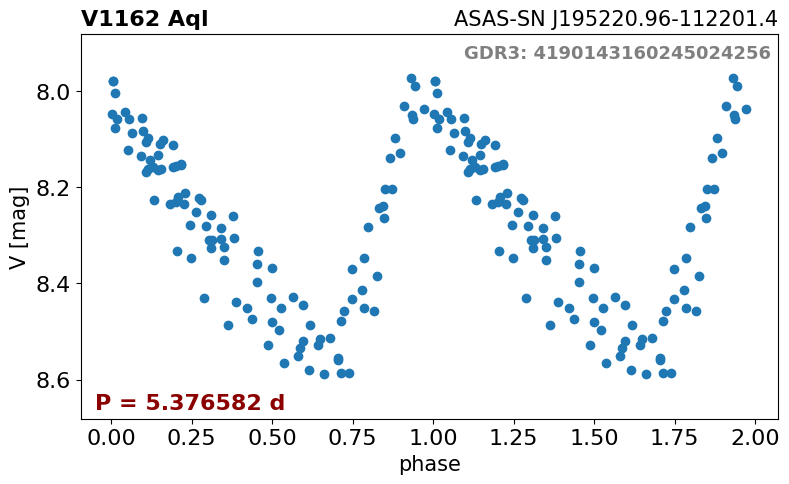}
\includegraphics[width=.5\columnwidth]{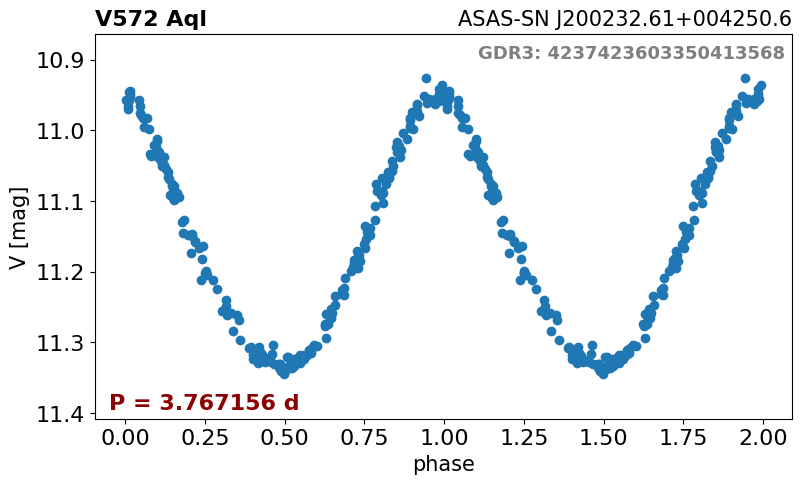}
\includegraphics[width=.5\columnwidth]{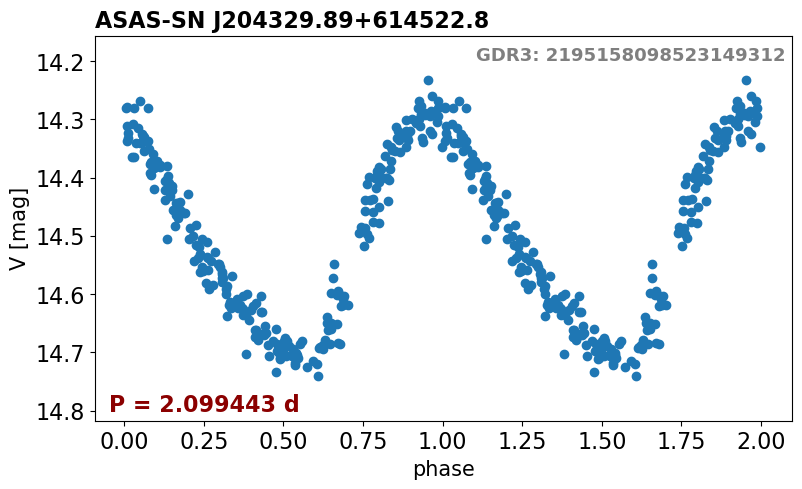}
\includegraphics[width=.5\columnwidth]{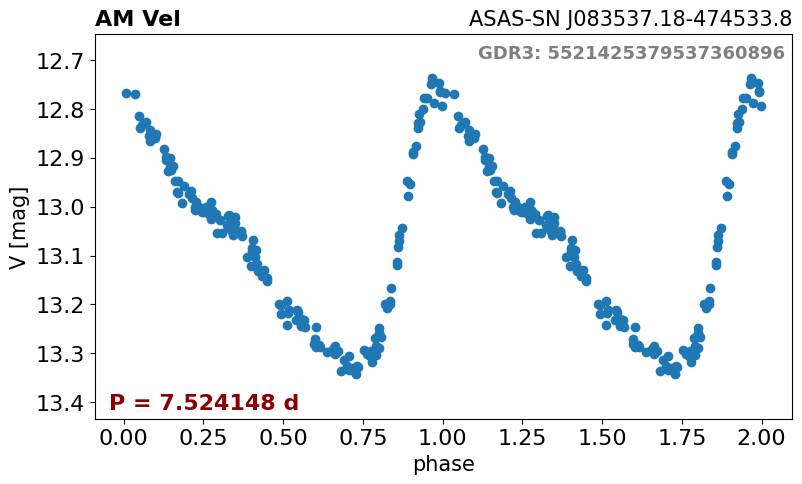}
\includegraphics[width=.5\columnwidth]{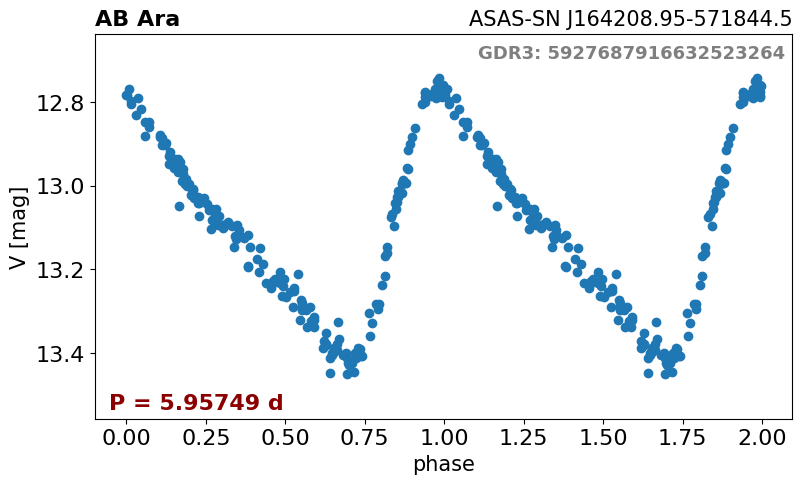}
\includegraphics[width=.5\columnwidth]{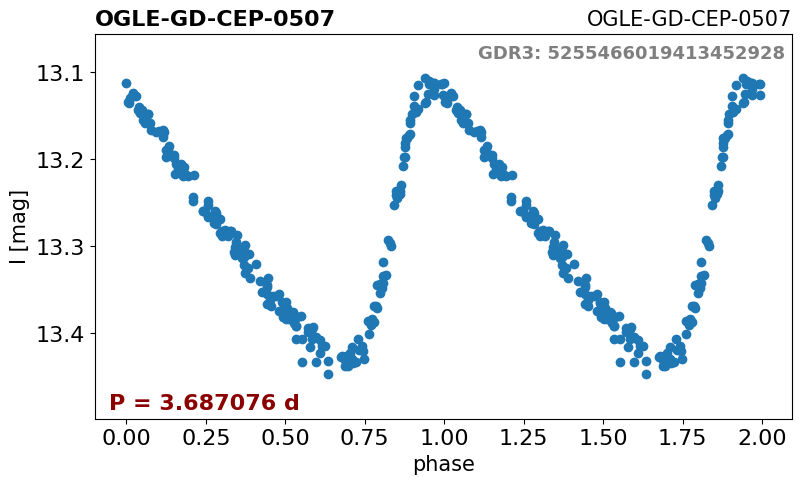}
\includegraphics[width=.5\columnwidth]{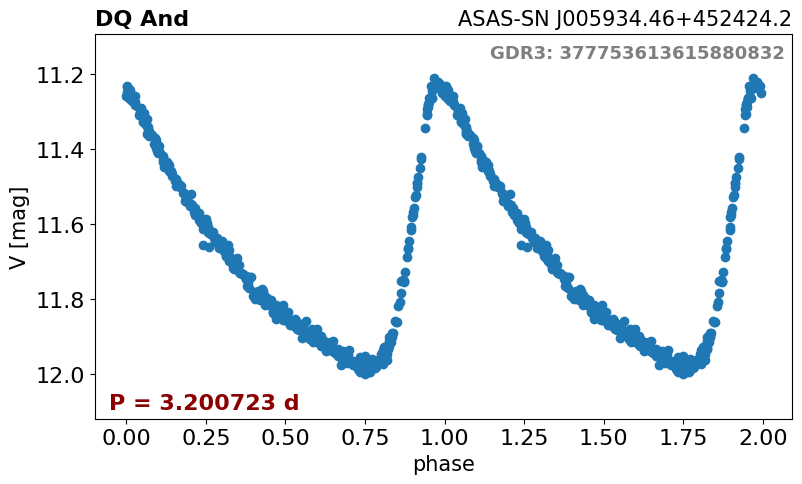}
\includegraphics[width=.5\columnwidth]{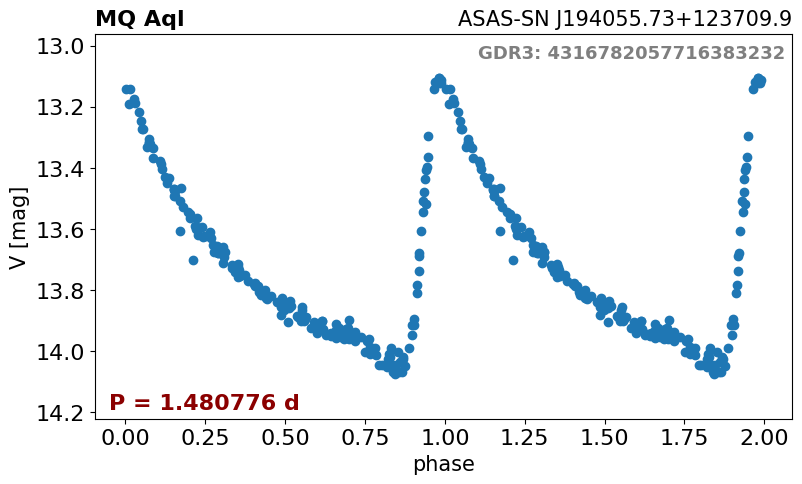}
\includegraphics[width=.5\columnwidth]{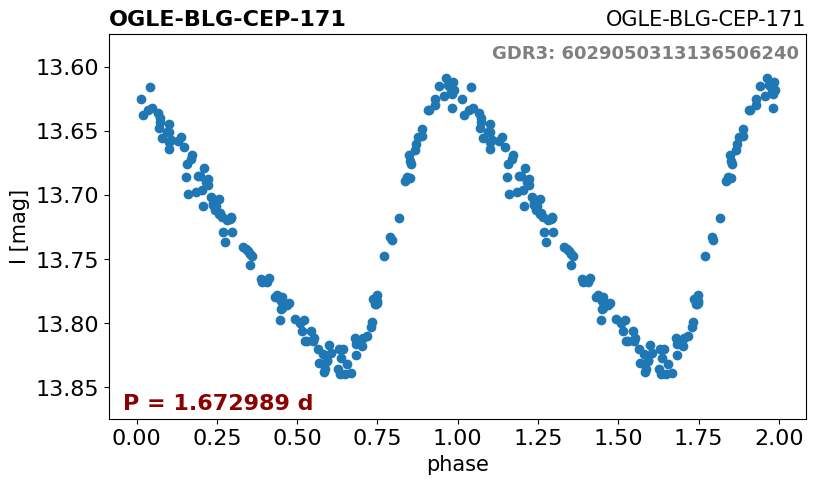}
\includegraphics[width=.5\columnwidth]{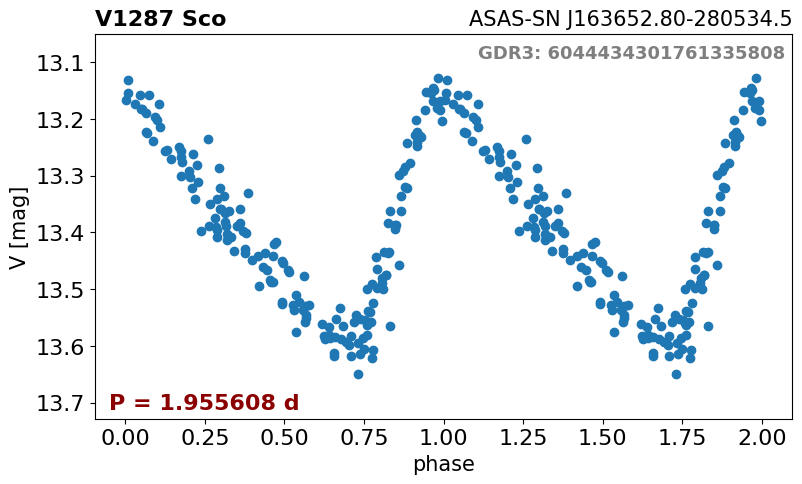}
\includegraphics[width=.5\columnwidth]{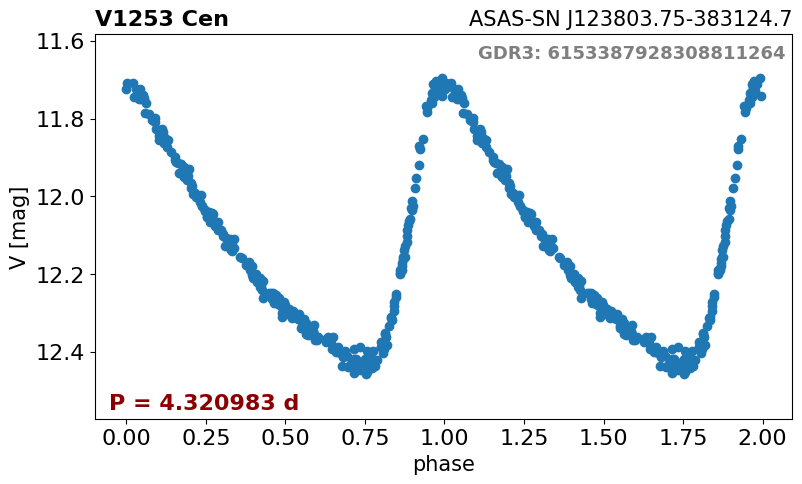}
\includegraphics[width=.5\columnwidth]{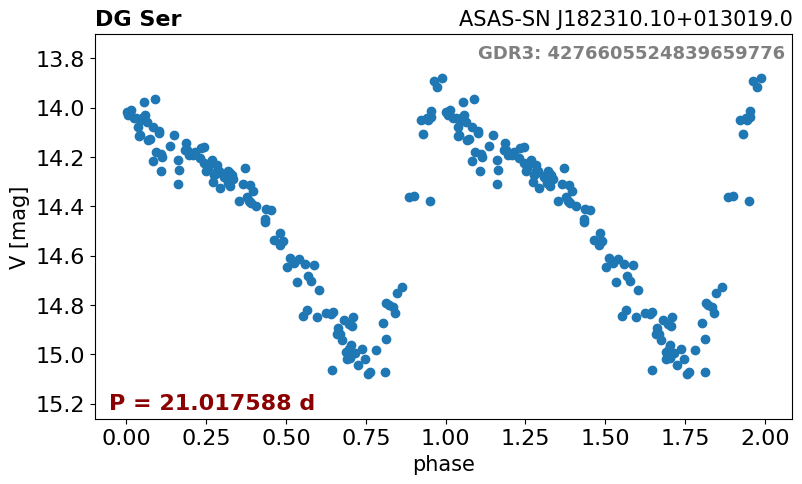}
\includegraphics[width=.5\columnwidth]{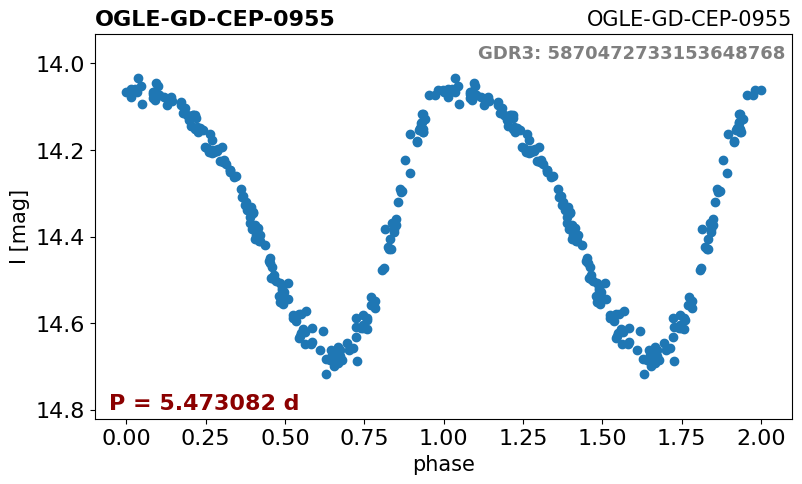}
\includegraphics[width=.5\columnwidth]{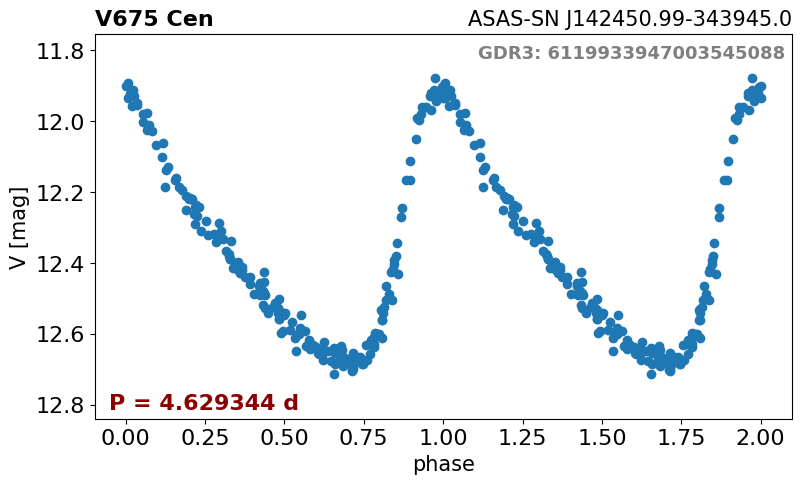}
\caption{Optical light curves of the kinematically peculiar \dcep{}. For each source we list their identifier pointing to the survey where it was discovered (in bold in the top left of each panel). The top right identifier points to the source of photometry ($V$-band for ASAS-SN and $I$-band for OGLE), if different from the discovery survey. We also provide the Gaia Source ID of each star and its pulsation period in days (in red). } 
\label{fig:light_curves_1}
\end{figure*}

In \Cref{fig:light_curves_1} we present the optical light curves for each of the 18 \rogue{} sources. Photometric $V$-band data for 14 sources were downloaded from the \href{https://asas-sn.osu.edu/variables}{\asasn{}} Variable Stars Database \citep{Jayasinghe:2018}, while the $I$-band photometry for the remaining four sources was obtained from the \href{https://ogledb.astrouw.edu.pl/~ogle/OCVS/ceph_query.php}{\ogle{} Collection of Variable Stars (OCVS)} \citep{Udalski:2018}. The retrieved light curves were cleaned of outliers and a Fourier analysis of each light curve was performed to verify the published pulsation periods and, where necessary, refine them. Classical Cepheids exhibit characteristic light-curve morphologies that depend on the pulsation period. To assess whether the variability patterns of our sources are consistent with those expected for Cepheids, we compared the phased light curves with the reference examples at similar pulsation periods, provided in the \href{https://ogle.astrouw.edu.pl/atlas/classical_Cepheids.html}{\ogle{}
Atlas of Variable Star Light Curves}. While the data quality varies from one source to another, in general these light curve shapes seem consistent when compared with those of bona fide \dcep{} variables.

\section{Orbit gallery}
\label{app:orbit_gallery}
\begin{figure*}
\centering
\includegraphics[width=2\columnwidth]{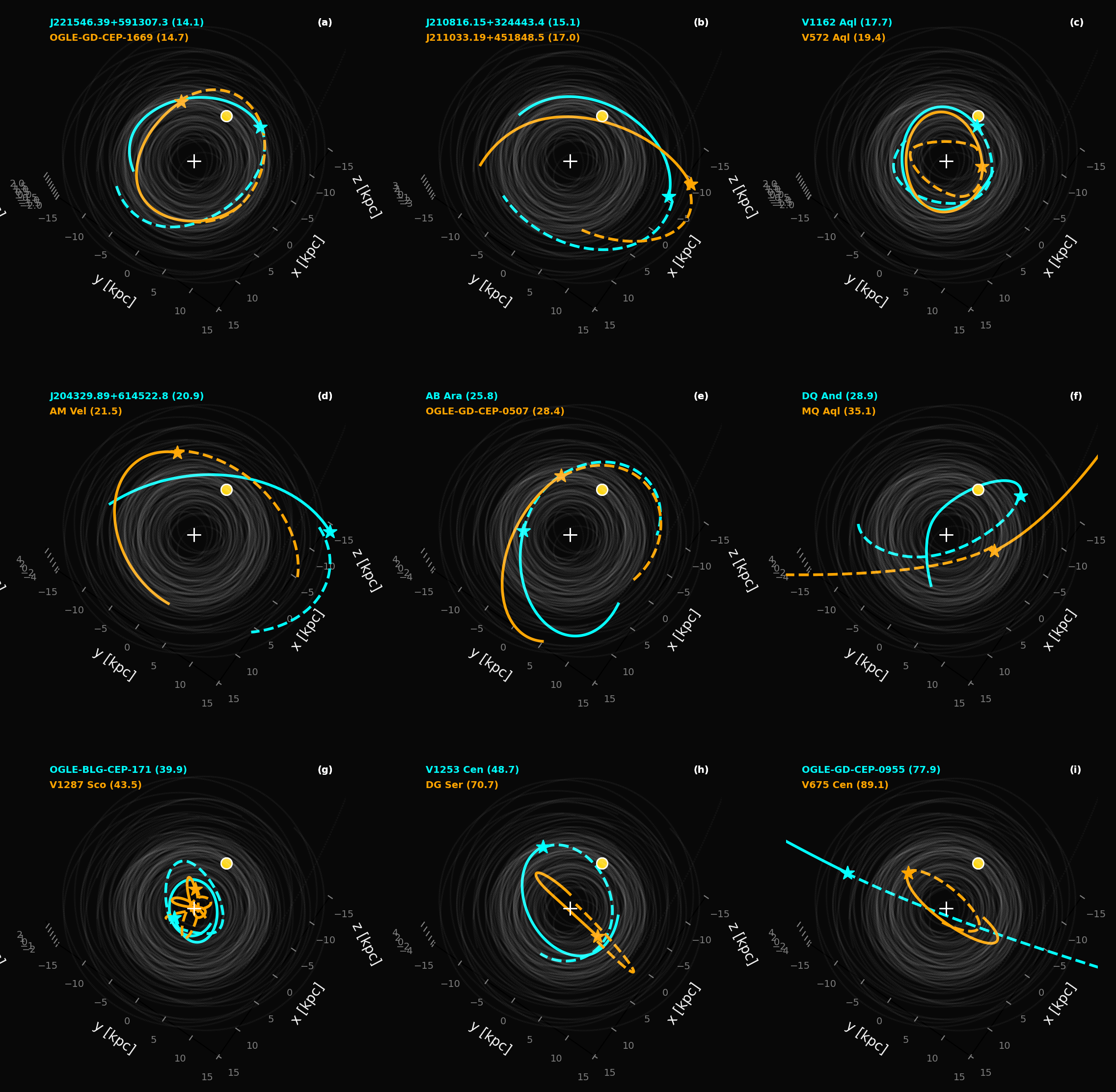}
\caption{Same as \autoref{fig:orbgal} but with a top-down perspective instead.}\label{fig:orbgal_top}
\end{figure*}

The orbits of the peculiar sources are presented in \autoref{fig:orbgal} and in top-down perspective in \autoref{fig:orbgal_top}.

\section{Metallicity compilation}
\label{app:feh_summary}
The spectroscopic metallicity estimates for the \rogue{} stars are compiled in Table~\ref{tab:feh_summary}. For a few sources there are multiple estimates available. Additionally, given the small sample size, we choose to include (in parenthesis) the metallicity estimates that are available but do not pass their respective quality flags.
\begin{table*}
\caption{Compilation of spectroscopic metallicities for the \rogue{} stars.}
\label{tab:feh_summary}
\resizebox{2\columnwidth}{!}{ \footnotesize
\begin{tabular}{c|c|c|c|c|c|c|c|c}
&  Gaia & Survey\_Source\_id & C-MetaLL & SDSS DR19 (apogee) & Lamost LRS  & GALAH dr4 & Gaia GspSpec \\
\hline
\hline
1 & 2199656063158427136 & J221546.39+591307.3 & - & - & - & - & (-0.73$^{+1.73}_{-0.22}$) \\ 
2 &  5255299138488495232 & OGLE-GD-CEP-1669 & - & - & - & - & - \\ 
3 &  1865228537734352896 & J210816.15+324443.4 & - & - & - & - & - \\
4 &  2162501400600160384 & J211033.19+451848.5 & - & - & - & - & - \\
5 &  4190143160245024256 & V1162 Aql & - & - & - & - & -0.01$^{+0.01}_{-0.00}$ \\
6 &  4237423603350413568 & V572 Aql & - & - & 0.19$\pm0.03$ & - & 0.3$^{+0.02}_{-0.05}$ \\
7 &  2195158098523149312 & J204329.89+614522.8 & - & - & - & - & - \\ 
8 &  5521425379537360896 & AM Vel & - & - & - & - & (0.68$^{+0.05}_{-0.71}$) \\
9 &  5927687916632523264 & AB Ara & - & - & - & - & - \\
10 &  5255466019413452928 & OGLE-GD-CEP-0507 & - & - & - & - & - \\
11 &  377753613615880832 & DQ And & -0.29$\pm$0.20 & - & - & - & -0.54$^{+0.08}_{-0.07}$ \\
12 &  4316782057716383232 & MQ Aql & - & - & - & - & - \\
13 &  6029050313136506240 & OGLE-BLG-CEP-171 & - & - & - & - & - \\
14 &  6044434301761335808 & V1287 Sco & - & - & - & - & - \\
15 &  6153387928308811264 & V1253 Cen & -0.26$\pm$0.14 & - & - & -0.12$\pm0.06$ & -0.01$^{+0.13}_{-0.06}$ \\
16 &  4276605524839659776 & DG Ser & - & -1.58$\pm0.01$ & - & - & (-1.71$^{+0.42}_{-0.38}$) \\
17 &  5870472733153648768 & OGLE-GD-CEP-0955 & - & - & - & - & - \\
18 &  6119933947003545088 & V675 Cen & - & - & - & - & - \\
\end{tabular}}
\tablefoot{Metallicities compiled from the major publicly available spectroscopic surveys. Values in parenthesis indicate entries that were found but did not pass quality cuts of the respective survey.} 
\end{table*}

\section{Q2 estimation}
\label{app:q2}
We follow the recipe in S25 to derive an equivalent quality parameter to their `Q', except here it is done for \typetwo{} distances. Specifically, we compare the \gaia{} parallax $\varpi$ with $\varpi_{\mu}$ which is derived by converting the T2C distance modulus estimate $\mu$, with its uncertainty adopted from \autoref{tab:plr_type2} as $\sigma_{mu}$=0.065. We also adopt the parallax zero-point variance from S25 to be 23.7 $\mu$as. Putting it together, we obtain
\begin{gather}
Q2 = \left|\frac{\varpi - \varpi_{\mu}}{ \sqrt{C_{1}^{2} + \sigma_{\varpi}^{2} + \frac{23.7}{1000}^{2} } }\right| \,, \\
C_{1} = \frac{1}{2}(\mu_{+} - \mu_{-}) \,, \\
\mu_{\pm} = 10 ^{- (\mu \mp \sigma_{\mu} - 10)/5} \,.
\end{gather} 
\section{Ruwe summary}
\label{app:ruwe_summary}
We use the \gunlim{} code \citep{2024A&A...688A...1C} to compute the \ruwe{} threshold for potential binaries as a function of Galactic coordinates ($l,b$). The threshold values range between $1.15 < \ruwe_{\rm gunlim} < 1.36$. \cref{tab:ruwe_summary} lists the \gdrthree{} \ruwe{} along the binarity threshold for each of the 18 candidate \rogue{} \dcep{}. Sources with an observed \ruwe{} above this threshold are assigned a flag \ruwe$_{cond}$=`high' while those below are flagged as `ok'. 

\begin{table}
\caption{RUWE of the high \lperp{} Cepheids. }
\label{tab:ruwe_summary}
\resizebox{\columnwidth}{!}{ \footnotesize
\begin{tabular}{l|c|c|c|c|c|c}
\hline\hline
 & Gaia ID & survey\_source\_id & AF & \ruwe & \ruwe$_{gunlim}$ & \ruwe$_{cond}$ \\
\hline
1 & 2199656063158427136 & J221546.39+591307.3  & 1.00 &  1.05  &      1.22  &      ok \\
2 &5255299138488495232 &  OGLE-GD-CEP-1669    & 0.25 & 12.61   &     1.25   &   \textbf{high} \\
3 &1865228537734352896 &  J210816.15+324443.4 & 0.99 &  1.06   &     1.19   &     ok  \\
4 &2162501400600160384 &  J211033.19+451848.5 & 0.99 &  1.08   &     1.23  &      ok  \\
5 &4190143160245024256 &          V1162 Aql   & 1.00 &  0.95   &     1.23   &     ok  \\
6 &4237423603350413568 &          V572 Aql    & 0.79 & 1.13    &    1.21   &     ok   \\
7 &2195158098523149312 &  J204329.89+614522.8 & 0.99 &  1.07  &      1.23   &     ok  \\
8 &5521425379537360896 &          AM Vel      & 0.87 & 41.08  &    1.23   &   \textbf{high}  \\
9 &5927687916632523264 &          AB Ara      & 1.00  &  0.91 &     1.26    &    ok  \\
10 &5255466019413452928 &   OGLE-GD-CEP-0507  & 0.23 &  26.32 &      1.25  &    \textbf{high}  \\
11 & 377753613615880832 &          DQ And     & 0.99 &  1.04 &    1.22   &     ok  \\
12 &4316782057716383232 &          MQ Aql     & 0.61 &  1.41 &     1.26  &    \textbf{high}  \\
13 &6044434301761335808 &          V1287 Sco  & 1.00 &  1.01 &      1.26  &      ok  \\
14 &6029050313136506240 &   OGLE-BLG-CEP-171  & 1.00 &  0.89 &      1.29  &      ok   \\
15 &6153387928308811264 &          V1253 Cen  & 0.97 &  1.2  &      1.22   &     ok  \\
16 &4276605524839659776 &          DG Ser     & 0.71 & 1.55  &    1.29    &  \textbf{high}   \\
17 &5870472733153648768 &   OGLE-GD-CEP-0955  & 0.93 &  0.99  &      1.24  &      ok  \\
18 &6119933947003545088 &         V675 Cen    & 0.99 & 1.28   &     1.29   &     ok   \\
\hline
\end{tabular}}
\tablefoot{The \ruwe{} from the \gaia{} archive is compared against the threshold predicted by \gunlim{} for potential binary candidates. The last column indicates whether \ruwe{} $<$ \ruwe$_{gunlim}$ (ok), or a source might have poor astrometry due to a companion. } 
\end{table}

\section{\dcep{}-globular cluster association}
\label{sec:dcep_gc_table}

\autoref{tab:dcep_gc} lists the potential dynamic association between pairs of Galactic Globular Clusters and the \rogue{} stars, considering their proximity in the (\jacobi,\lperp,\lz) and (\jacobi,\lperp) spaces as well as their 3D orbital overlap. Only cases with an orbital overlap exceeding 80\% are included.

\begin{table}
\caption{Exploring potential dynamic association between \gc{} and the kinematically peculiar \dcep{}. }
\label{tab:dcep_gc}
\centering
\resizebox{1.\columnwidth}{!}{ \footnotesize
\begin{tabular}{cc|ccc}
\hline
& & & Ranking method & \\
\hline\hline
& \dcep{} & $\Delta I_{ij}$(\jacobi,\lperp,\lz) & $\Delta I_{ij}$(\jacobi,\lperp) & Orbital overlap \\
\hline
1& J221546.39+591307.3 & Pal 10 & NGC 6333 & - \\
2&OGLE-GD-CEP-1669 & Pal 10 & NGC 6333 & - \\
3&J210816.15+324443.4 & E 3 & NGC 6656 & - \\
4&J211033.19+451848.5 & E 3 & NGC 104 &  -  \\
5&V1162 Aql & NGC 6838 & NGC 6287 & - \\
6&V572 Aql & NGC 6838 & FSR 1716 & \textbf{NGC 6752 (82 \% )} \\
7&J204329.89+614522.8 & E 3 & E 3 &  -  \\
8&AM Vel & Pal 1 & Ter 10 & - \\
9&AB Ara & E 3 & E 3 & - \\
10&OGLE-GD-CEP-0507 & E 3 & E 3 & \textbf{E 3 (89 \% )} \\
11&DQ And & NGC 6426 & NGC 6426 & - \\
12&MQ Aql & NGC 4590 & NGC 4590 &  -  \\
13&OGLE-BLG-CEP-171 & NGC 6656 & NGC 6656 & \textbf{NGC 6235 (83 \% )} \\
14&V1287 Sco & NGC 6717 & NGC 6638 & - \\
15&V1253 Cen & Rup 106 & Rup 106 & - \\
16&DG Ser & NGC 6779 & ESO 280-SC06 & - \\
17&OGLE-GD-CEP-0955 & NGC 5053 & NGC 5053 &  -  \\
18&V675 Cen & NGC 5904 & NGC 5272 &  -  \\
\hline
\end{tabular}}
\tablefoot{For each peculiar source, the \gc{} with the lowest $\Delta I_{ij}$ metric is listed 
by the use of all three components (\jacobi,\lperp,\lz), or just two components (\jacobi,\lperp), or instead using their orbital overlap. Only those cases with orbital overlap $>80$\% are included.}
\end{table}
\section{\texttt{OGLE-GD-CEP-0507} \& \texttt{E3}}
\begin{figure}
\centering
\includegraphics[width=1.\columnwidth]{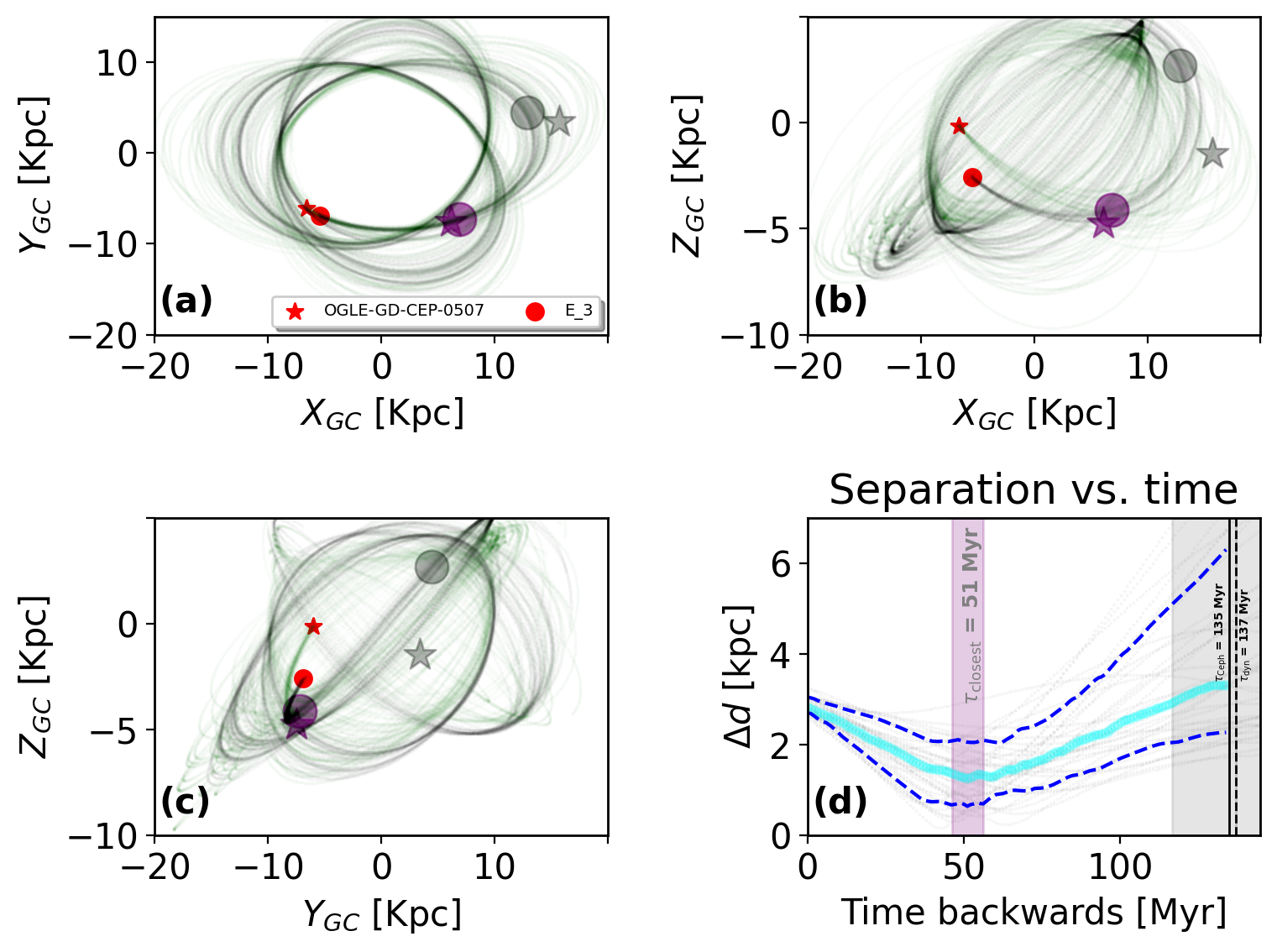}
\caption{Orbital comparison over multiple realisations in different projections (a-c) between \texttt{OGLE-GD-CEP-0507} (green tracks) and the Globular cluster \texttt{E\_3} (black tracks). Panel (d) shows the separation (median as cyan curve) between the two objects from $\tau=0$ back to the Cepheid's estimated age \tbirth{}.  Panels (a-c) also show the positions of the two objects (based on present day initial conditions) at three different epochs: in red at $\tau=0$, in gray at \tbirth{}, and in purple at their closest separation. } \label{fig:gc_dcep_0507}
\end{figure}
\begin{figure}
\includegraphics[width=1.\columnwidth]{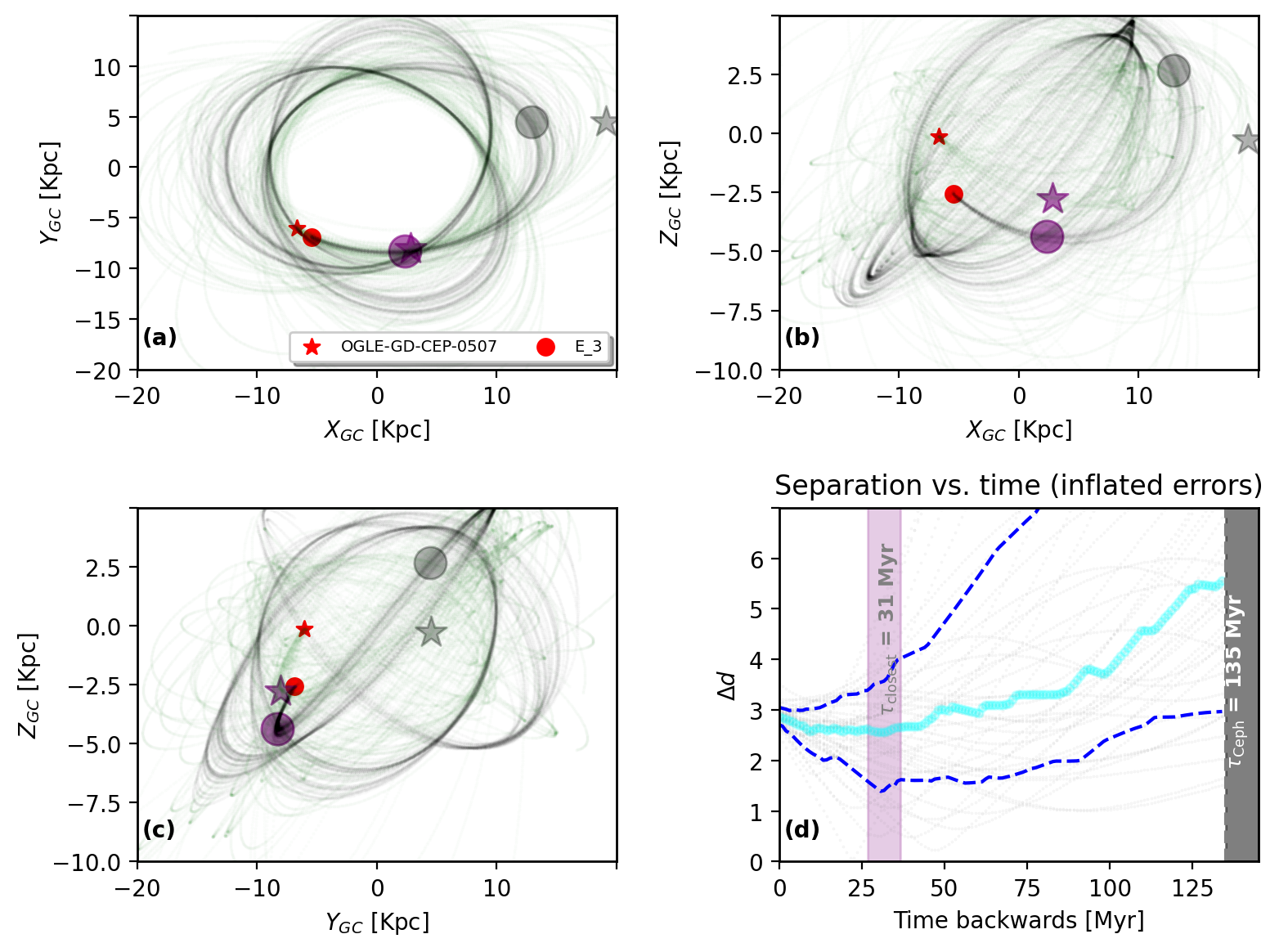}
\caption{Orbit comparison between \dcep{} \texttt{OGLE-GD-CEP-0507} and globular cluster E3, same as \autoref{fig:gc_dcep_0507} but for the case assuming proper motion uncertainty inflation by a factor of 4. The orbital overlap in this case drops to 78\% compared to the case with nominal uncertainties.} \label{fig:gc_dcep_0507_inf}
\end{figure}
\autoref{fig:gc_dcep_0507_inf} compares the orbital tracks of \texttt{OGLE-GD-CEP-0507} and \texttt{E3} when proper motion uncertainties are (over)inflated by a factor of 4. We find that the orbital overlap in this scenario drops to 78\% compared to  89\% for the nominal uncertainties case presented in \autoref{fig:gc_dcep_0507}. 

\section{Kinematic tracks}
\begin{figure*}
\includegraphics[width=.5\columnwidth]{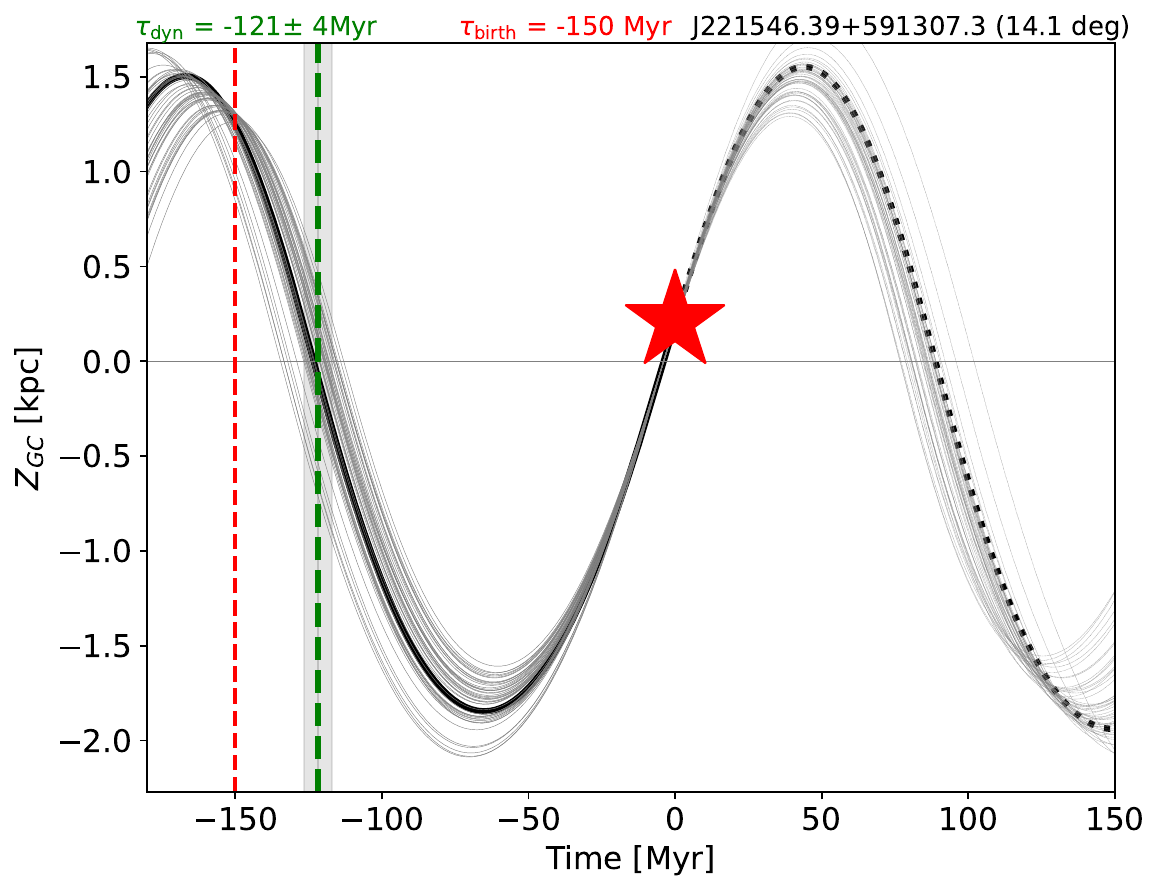}
\includegraphics[width=.5\columnwidth]{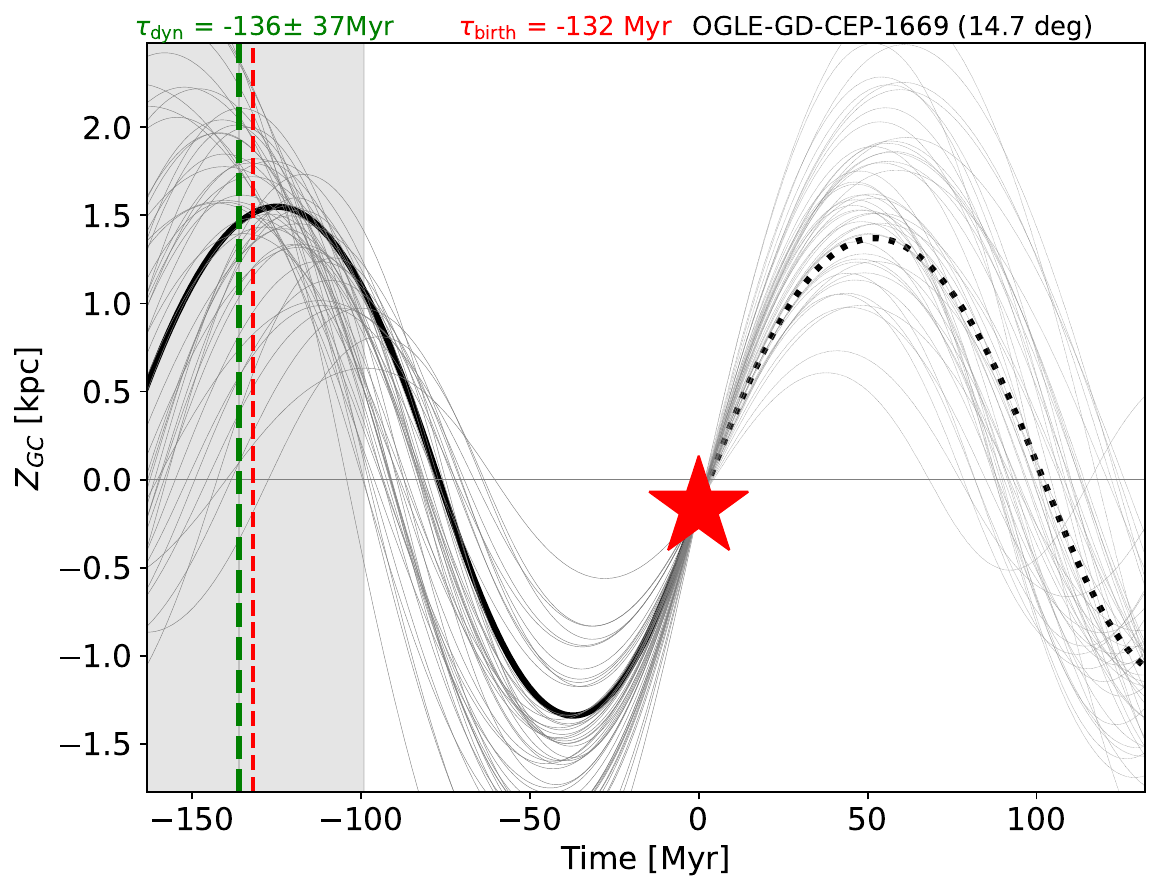}
\includegraphics[width=.5\columnwidth]{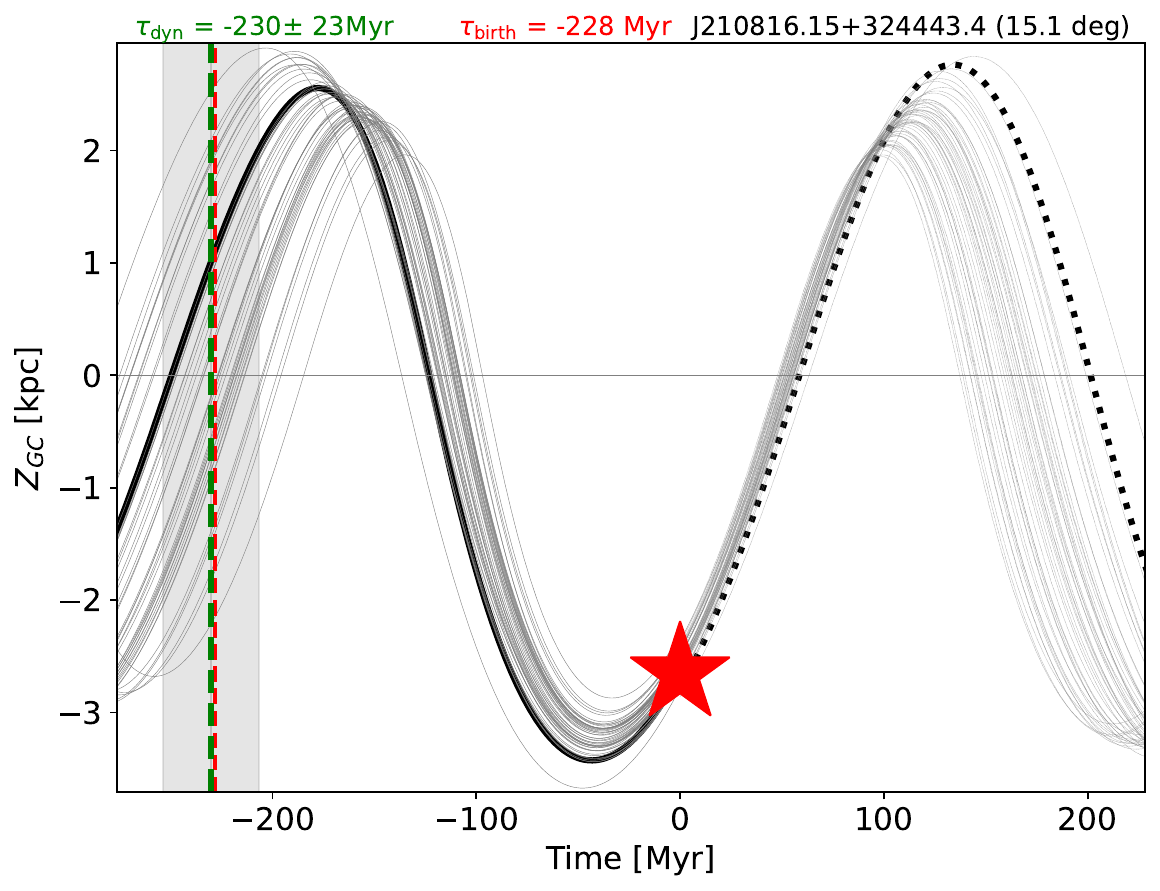}
\includegraphics[width=.5\columnwidth]{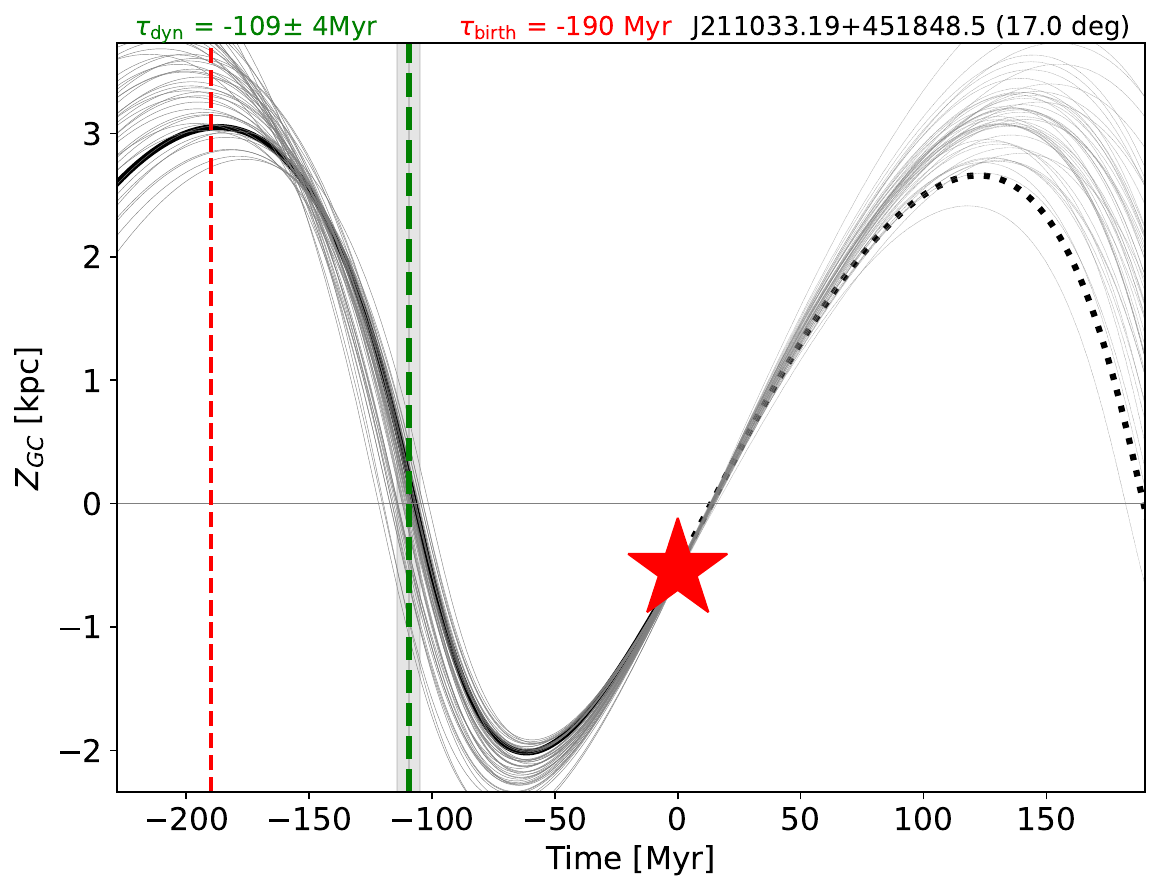}
\includegraphics[width=.5\columnwidth]{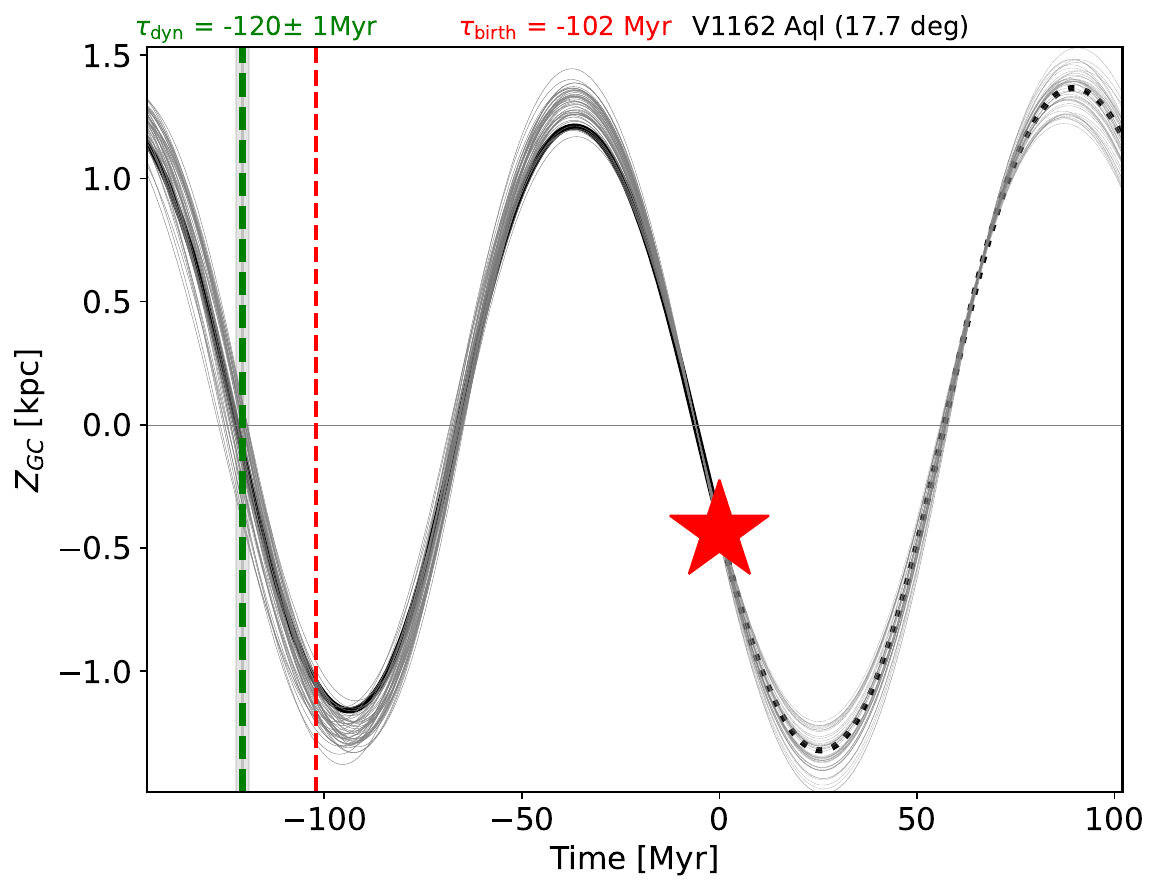}
\includegraphics[width=.5\columnwidth]{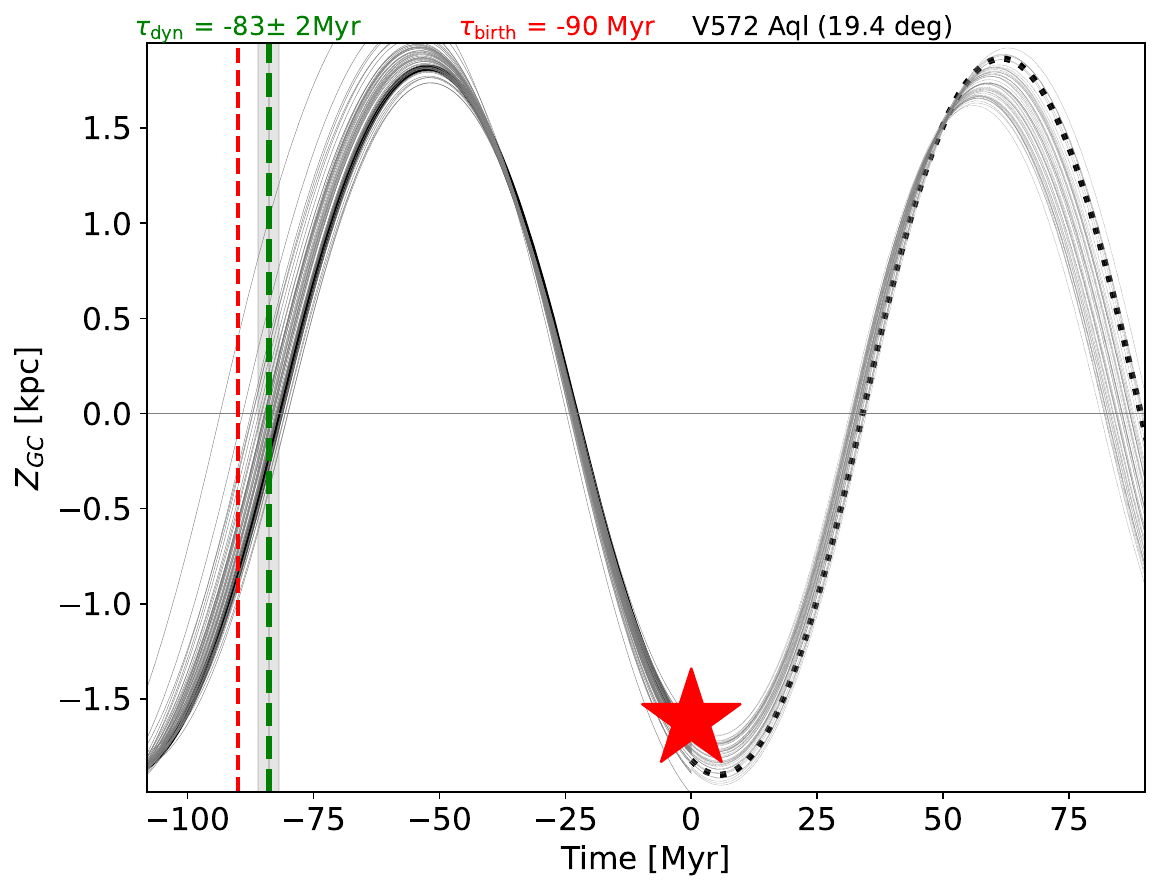}
\includegraphics[width=.5\columnwidth]{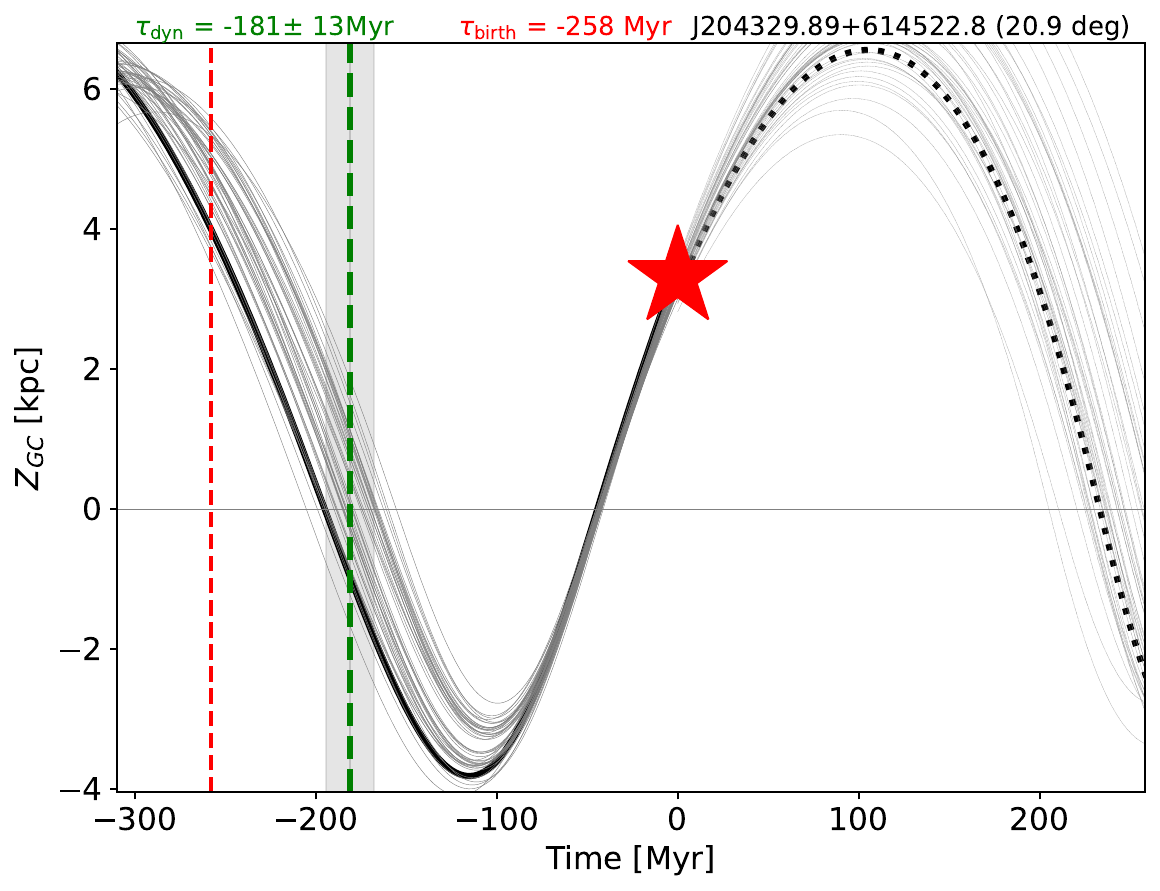}
\includegraphics[width=.5\columnwidth]{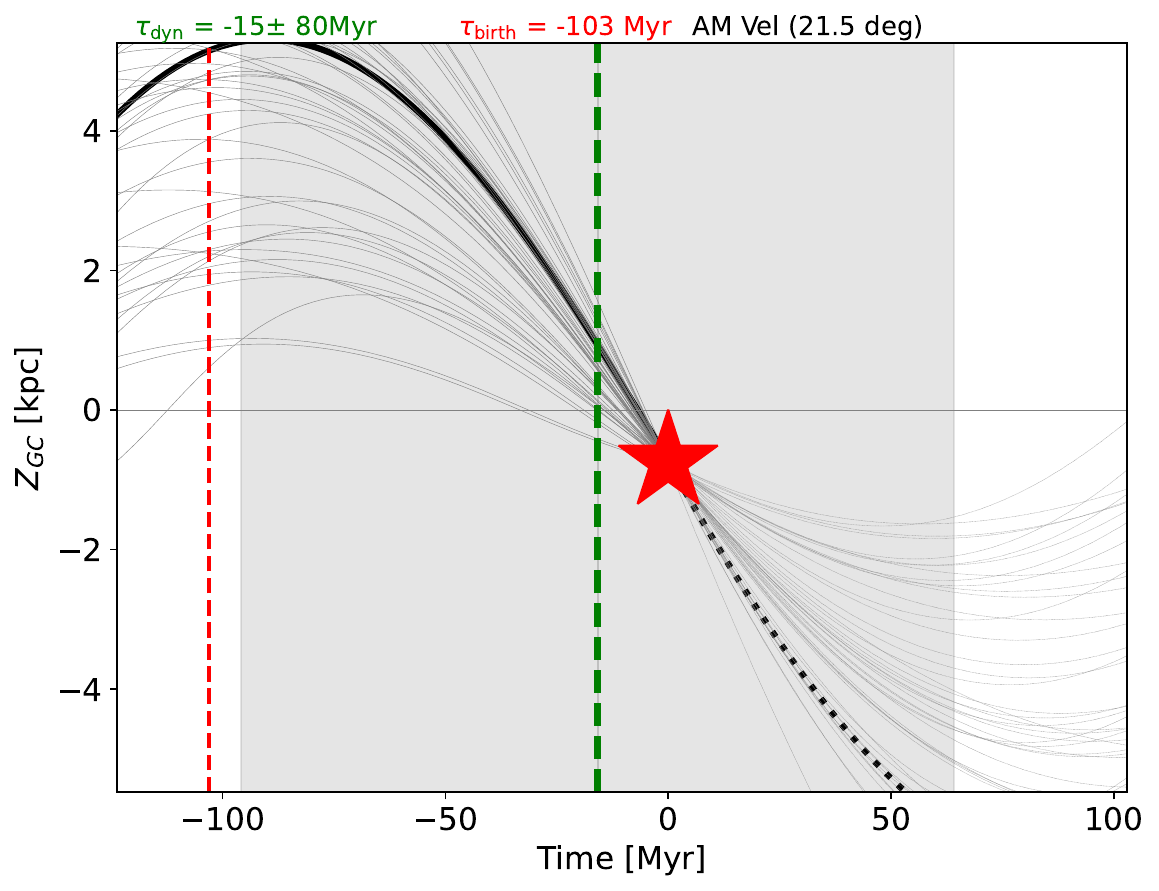}
\includegraphics[width=.5\columnwidth]{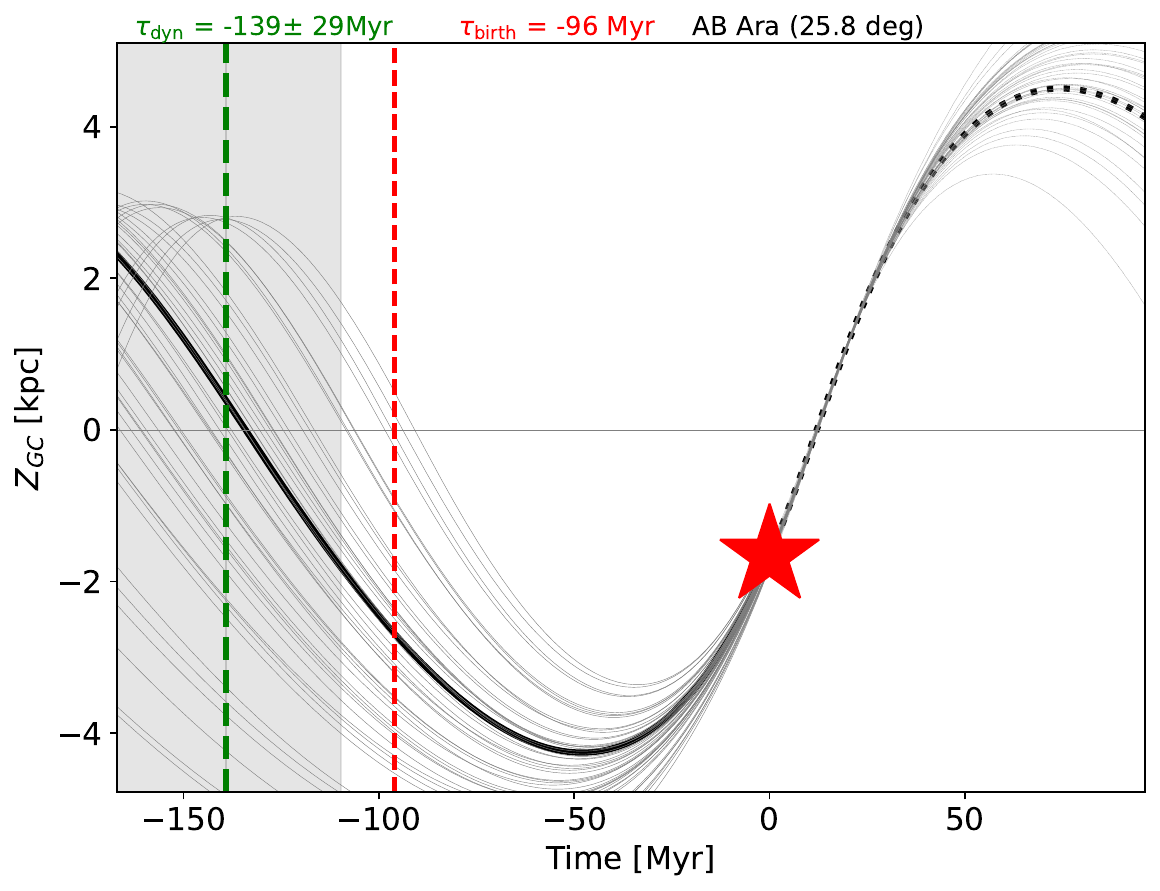}
\includegraphics[width=.5\columnwidth]{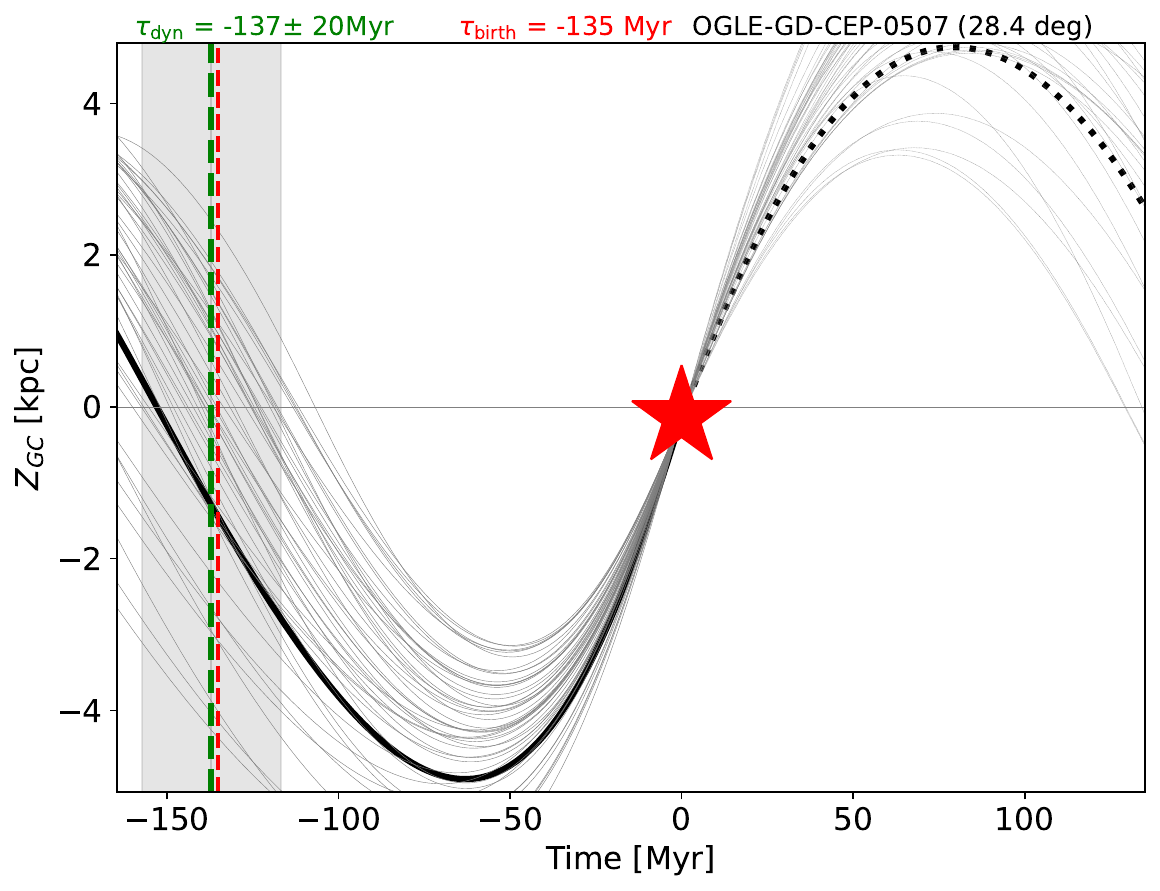}
\includegraphics[width=.5\columnwidth]{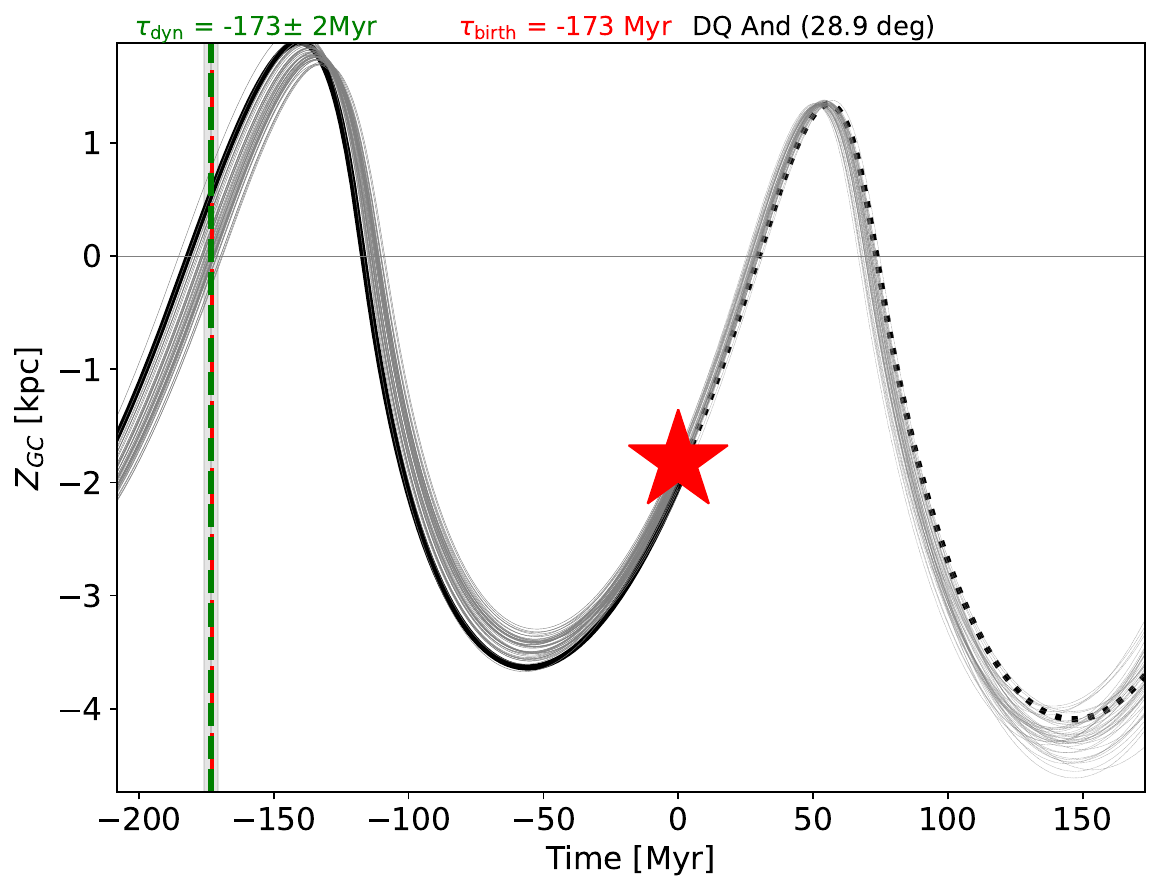}
\includegraphics[width=.5\columnwidth]{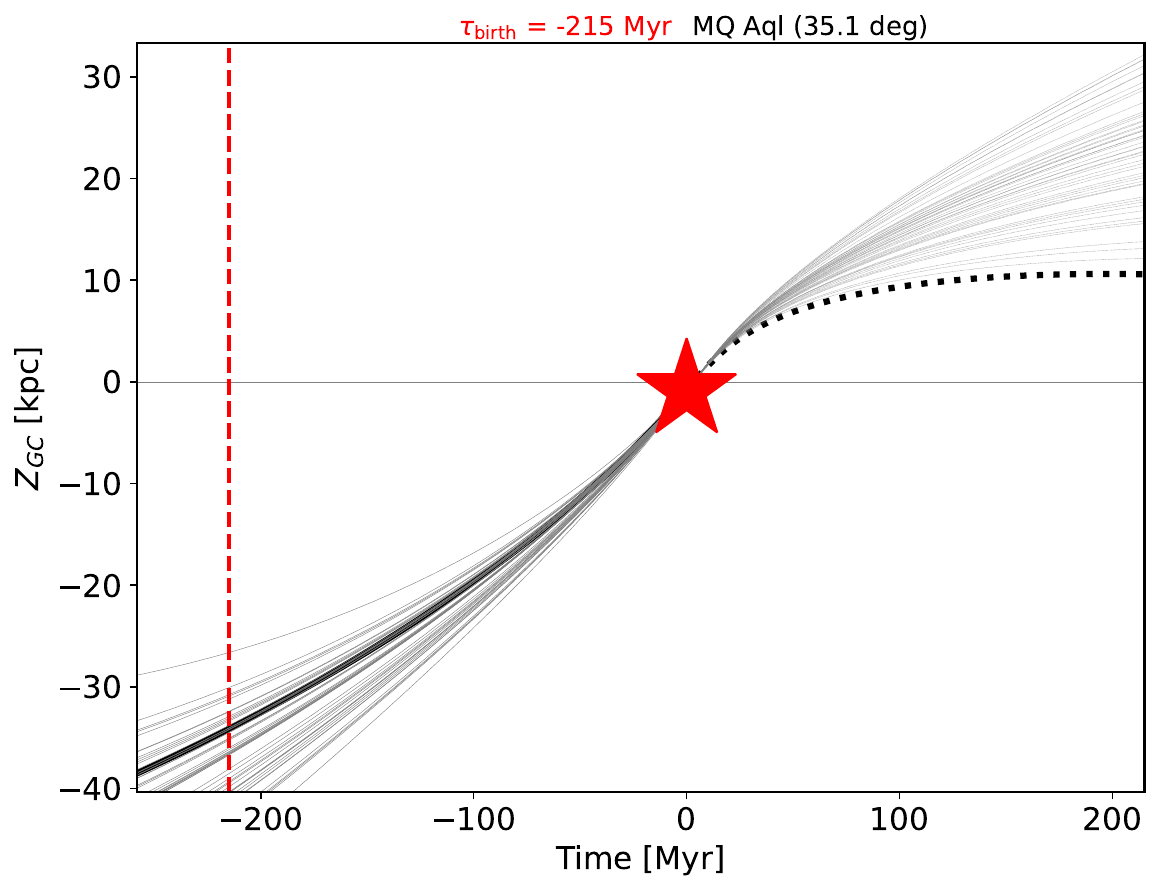}
\includegraphics[width=.5\columnwidth]{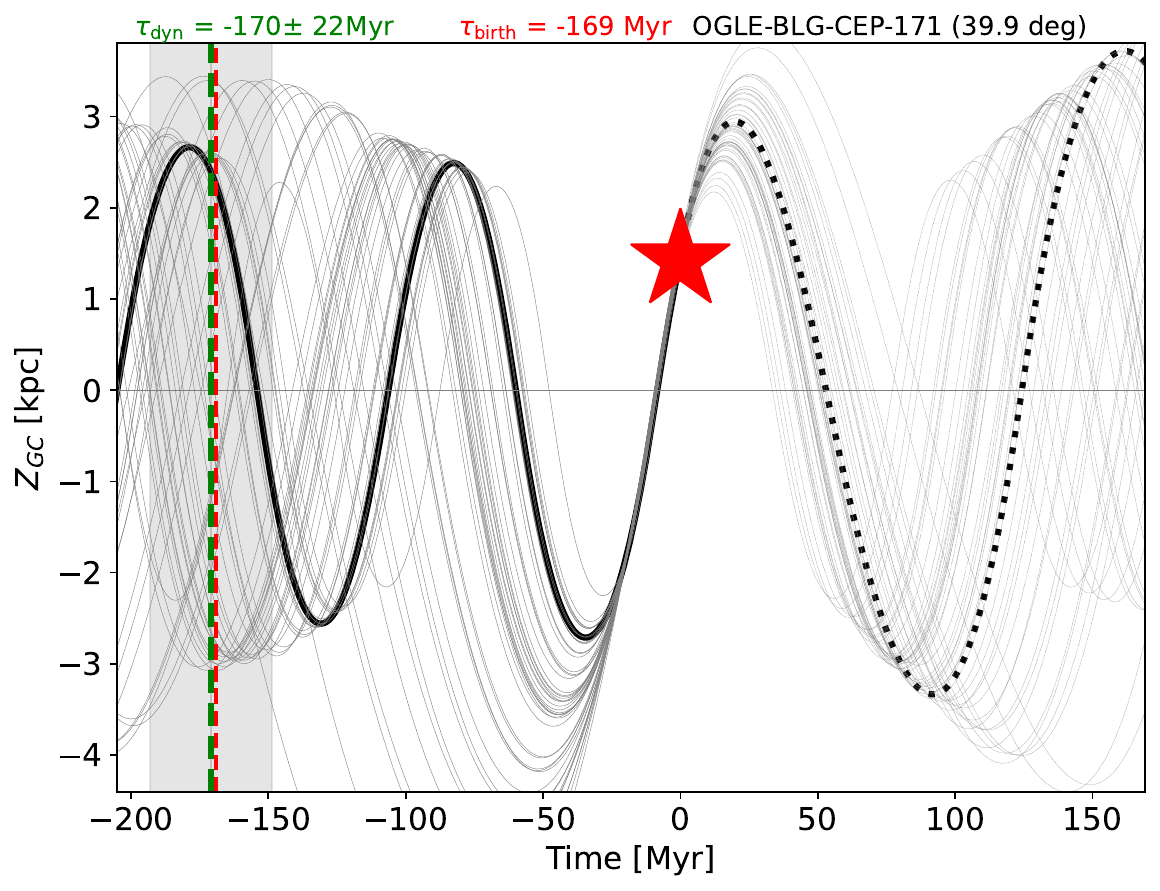}
\includegraphics[width=.5\columnwidth]{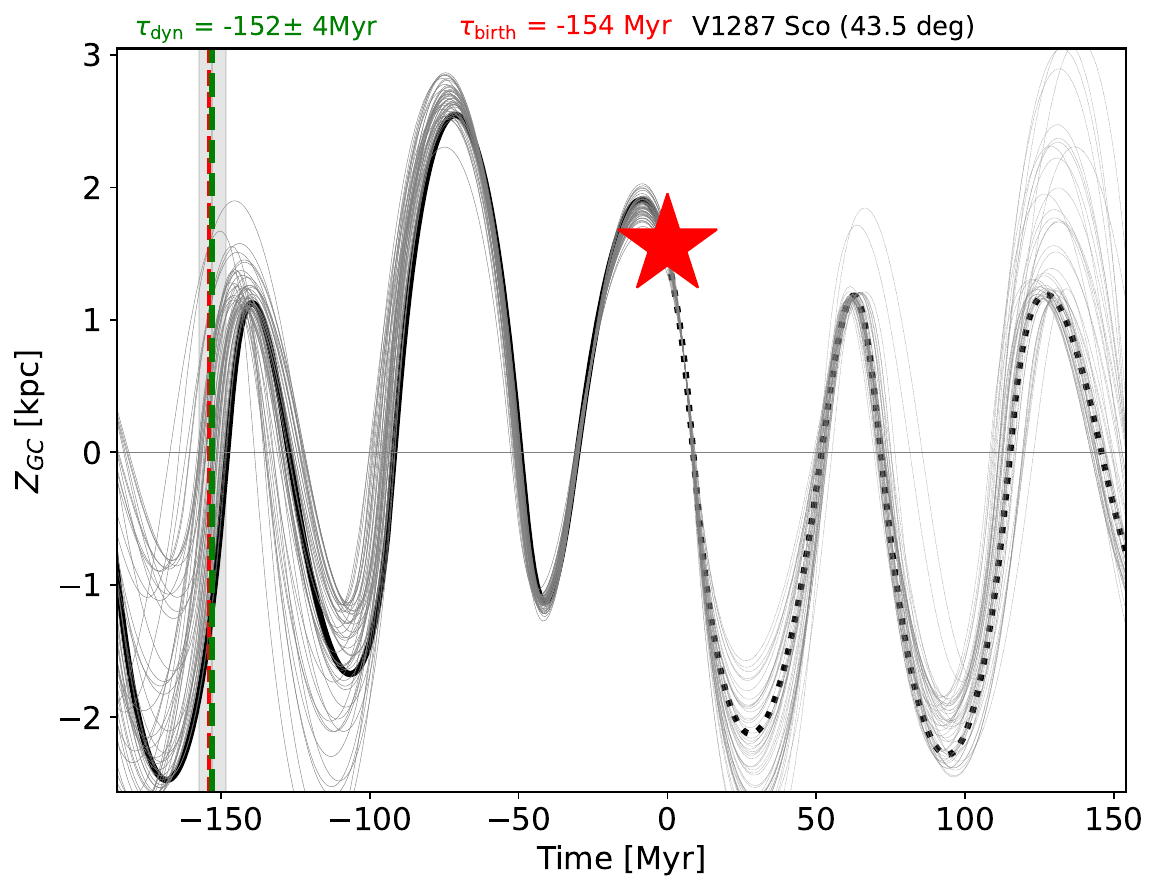}
\includegraphics[width=.5\columnwidth]{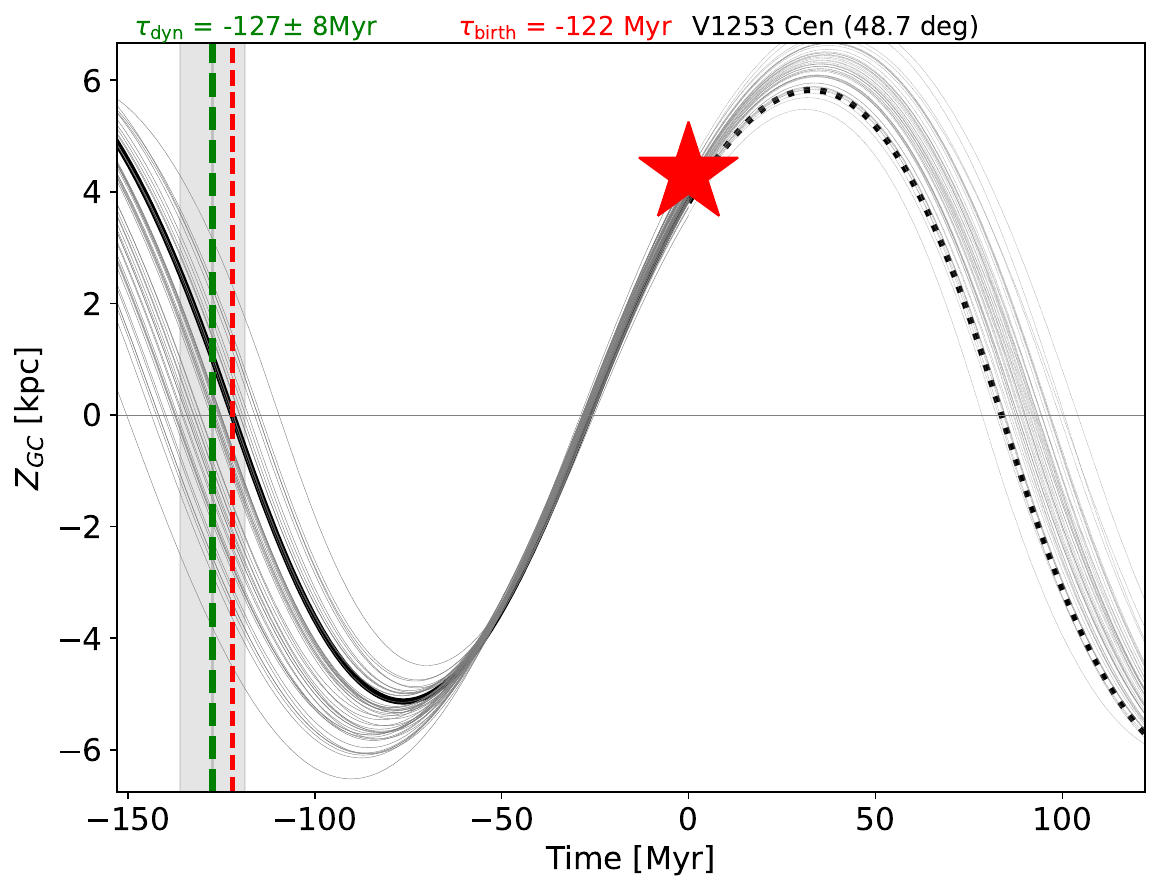}
\includegraphics[width=.5\columnwidth]{fig_tracks/tracks_16.pdf}
\includegraphics[width=.5\columnwidth]{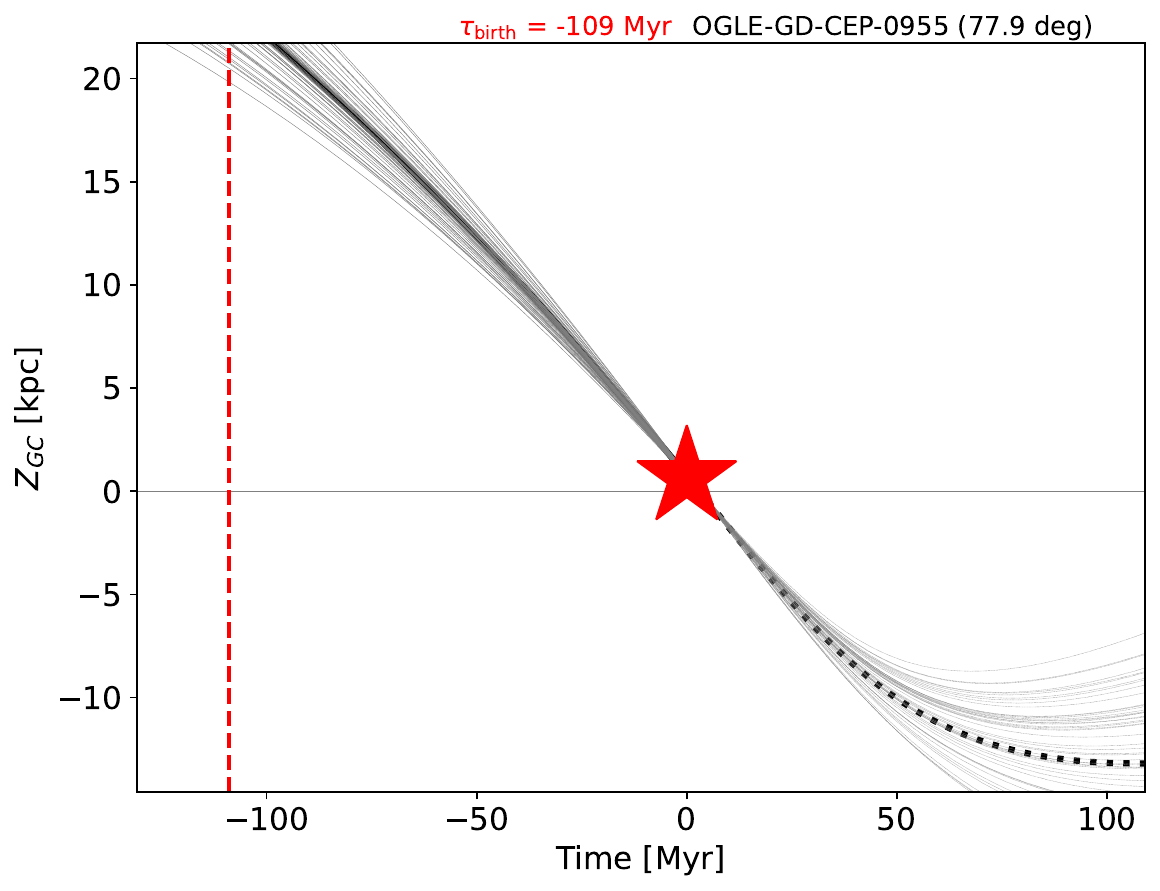}
\includegraphics[width=.5\columnwidth]{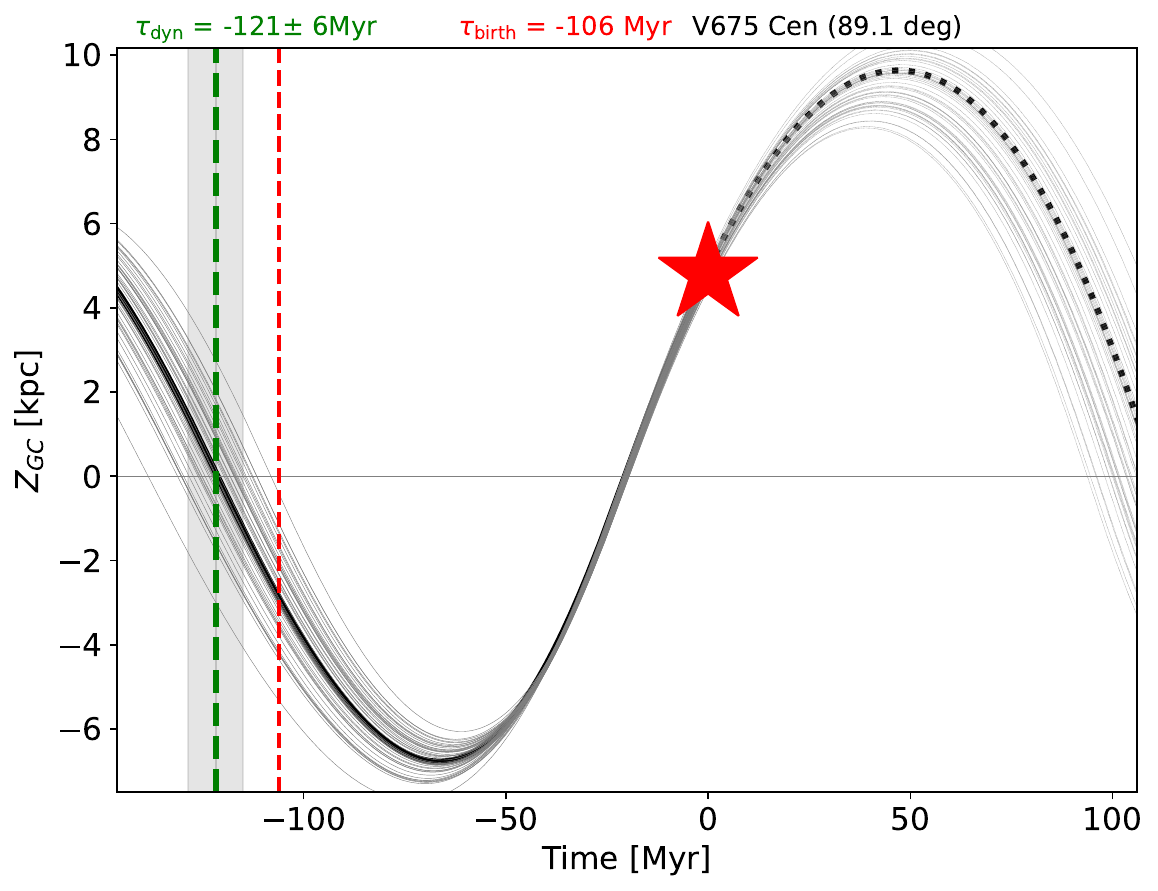}
\caption{Same as Figure~\ref{fig:example_ztrack} but for all the \rogue{} stars. In each case the title lists the Cepheid's assumed age \tbirth{}, its dynamical age estimated in this work \tdyn{}, as well as its survey source ID and orbital inclination from \autoref{table:cat}. In two cases there is either no disc crossing or the \tdyn{} exceeds 300 Myr, so we do not include their dynamical ages.} \label{fig:tracks}
\end{figure*}

Figure~\ref{fig:tracks} shows the orbital tracks charting the vertical displacement for all the \rogue{} stars. In each case the title lists the Cepheid's assumed age \tbirth{}, its dynamical age estimated in this work \tdyn{}, as well as its survey source ID and orbital inclination from \autoref{table:cat}. In two cases there is either no disc crossing (\texttt{MQ Aql})  or the \tdyn{} exceeds 300 Myr (\texttt{OGLE-GD-CEP-0955}), so we do not include their dynamical ages.

\end{appendix}

\appendix
\label{sec:app}

\end{document}